\documentclass[trackchanges]{aastex701}

\usepackage{ulem}
\usepackage{amsmath}
\usepackage{graphicx}
\usepackage{natbib}
\usepackage{longtable}
\usepackage{supertabular,booktabs}
\usepackage{sidecap}
\usepackage{enumitem}
\usepackage{txfonts}
\usepackage{epstopdf}
\usepackage{tabularx}
\usepackage{ltablex}
\newcommand{\Msun}{M$_{\odot}$}

\begin{document}

\title{Abundance of heavy r-process elements in CEMP-rs stars\\
The role of the i-process}

\author[orcid=0009-0000-1753-2212,gname='Akkara Muhammed',sname='Riyas']{A. M. Riyas}
\affiliation{Department of Physics, University of Calicut, Thenhipalam, Malappuram, Kerala-673635, India}
\email{riyasbinzaid@gmail.com}  

\author[orcid=0000-0002-3532-2793,gname=Drisya, sname='Karinkuzhi']{D. Karinkuzhi} 
\affiliation{Department of Physics, University of Calicut, Thenhipalam, Malappuram, Kerala-673635, India}
\affiliation{Institut d'Astronomie et d'Astrophysique, Universit\'e Libre de Bruxelles, ULB, Campus Plaine C.P. 226, Boulevard du Triomphe, 1050 Bruxelles, Belgium}
\email{drdrisyak@uoc.ac.in}

\author[orcid=0000-0003-0499-8608,gname=Sophie,sname= Van Eck]{S. Van Eck}
\affiliation{Institut d'Astronomie et d'Astrophysique, Universit\'e Libre de Bruxelles, ULB, Campus Plaine C.P. 226, Boulevard du Triomphe, 1050 Bruxelles, Belgium}
\affiliation{BLU-ULB, Brussels Laboratory of the Universe, blu.ulb.be}
\email{sophie.van.eck@ulb.be}

\author[orcid=0000-0001-6159-8470,gname=Arthur,sname= Choplin]{A. Choplin}
\affiliation{Institut d'Astronomie et d'Astrophysique, Universit\'e Libre de Bruxelles, ULB, Campus Plaine C.P. 226, Boulevard du Triomphe, 1050 Bruxelles, Belgium}
\affiliation{BLU-ULB, Brussels Laboratory of the Universe, blu.ulb.be}
\email{Arthur.Choplin@ulb.be}

\author[orcid=0000-0002-9110-941X,gname=Stephane,sname= Goriely]{S. Goriely}
\affiliation{Institut d'Astronomie et d'Astrophysique, Universit\'e Libre de Bruxelles, ULB, Campus Plaine C.P. 226, Boulevard du Triomphe, 1050 Bruxelles, Belgium}
\affiliation{BLU-ULB, Brussels Laboratory of the Universe, blu.ulb.be}
\email{stephane.goriely@ulb.be}

\author[orcid=0000-0001-6008-1103,gname=Lionel,sname=Siess]{L. Siess}
\affiliation{Institut d'Astronomie et d'Astrophysique, Universit\'e Libre de Bruxelles, ULB, Campus Plaine C.P. 226, Boulevard du Triomphe, 1050 Bruxelles, Belgium}
\affiliation{BLU-ULB, Brussels Laboratory of the Universe, blu.ulb.be}
\email{lionel.siess@ulb.be}

\author[orcid=0009-0000-6607-9258,sname= Keerthy]{ M. V. Keerthy}
\affiliation{Department of Physics, University of Calicut, Thenhipalam, Malappuram, Kerala-673635, India}
\email{keerthymv@gmail.com}

\author[orcid=0000-0002-1883-4578,gname=Alain,sname=Jorissen]{A. Jorissen}
\affiliation{Institut d'Astronomie et d'Astrophysique, Universit\'e Libre de Bruxelles, ULB, Campus Plaine C.P. 226, Boulevard du Triomphe, 1050 Bruxelles, Belgium}
\affiliation{BLU-ULB, Brussels Laboratory of the Universe, blu.ulb.be}
\email{Alain.Jorissen@ulb.be}

\author[orcid=0000-0001-8253-1603,gname=Thibault,sname=Merle]{T. Merle}
\affiliation{Institut d'Astronomie et d'Astrophysique, Universit\'e Libre de Bruxelles, ULB, Campus Plaine C.P. 226, Boulevard du Triomphe, 1050 Bruxelles, Belgium}
\affiliation{Royal Observatory of Belgium, Avenue Circulaire 3, 1180 Brussels, Belgium}
\email{Thibault.Merle@ulb.be}


\begin{abstract}

Carbon-enhanced metal-poor (CEMP) stars are ancient stars enriched in carbon and heavy elements.  Some of these stars exhibit enhanced s-process and/or r-process elements, hence are classified as CEMP-s, CEMP-rs, or CEMP-r. This classification is challenging due to the limited availability of heavy element abundances, particularly among r-process elements. Heavy r-process  elements such as terbium, holmium, thulium, ytterbium, lutetium, tantalum, and iridium have rarely been measured because their sensitive lines are located in the ultraviolet. 
However, they provide sensitive diagnostics of the s-, r-, and i- nucleosynthetic processes.
 In this work, we aim to obtain a secure classification of CEMP-s and -rs stars and investigate whether the i-process can account for the measured abundance patterns in CEMP-rs stars.
We derive the abundance profiles, notably for twelve heavy r-elements, including, in some cases, tantalum, using high-resolution UVES spectra of seventeen CEMP-s and -rs stars. 
Based on indicators such as the [s/r] abundance ratio or the model-independent `abundance distance’, nine stars are confirmed as CEMP-rs and six as CEMP-s. The classification of two objects remains uncertain.
The i-process satisfactorily reproduces the abundance patterns of CEMP-rs stars.
However, larger samples are needed to confirm trends with metallicity and clarify how CEMP-rs stars differ from CEMP-s stars.

\end{abstract}


\section{Introduction}
\label{Sect:Intro}

The chemical abundance profiles of metal-poor stars, which formed in the early history of the galaxy, offer valuable insights into the nature and efficiency of nucleosynthesis processes that occurred in the past, as well as insight concerning the chemical evolution of the galaxy.

 A significant proportion, estimated to be around 20\% at [Fe/H] $<$ $-$2.0 and increasing to about 30\% or more at [Fe/H] $<$ $-$3.0, 
 of the stars in the halo of the Milky way have been discovered to be metal-poor as well as heavily enriched in carbon ([C/Fe]$>{1.0}$) \citep{rossi1999,lucatello2006,placco2014,yoon2016,lee2017}.
 They are called carbon-enhanced metal-poor stars (CEMP) \citep{Beer2005}. The CEMP stars are 
 further
 classified into CEMP-r, CEMP-s, and CEMP-rs stars, depending on 
 whether they are enriched in elements produced by the rapid neutron-capture-process (r-process), or the slow neutron-capture process (s-process) or whether they display an hybrid r+s abundance profile.
The classification is mainly based on the abundances of two heavy elements, Ba (representative of the s-process) and Eu (representative of the r-process), however, the precise classification varies within authors \citep[][hereafter K21]{Beer2005,Jonsell2006,Masseron2010,Abate2016,Karinkuzhi2021}. 
 In the following, we denote as {\it mainly-s elements} or {\it mainly-r elements} those produced dominantly (i.e., by more than 50\%) by, respectively, the s- or the r-process in the solar system, as defined in \citet[][]{Goriely1999}. 

Though the astrophysical origin of s-process elements are rather well identified, primarily in AGB stars \citep{Busso1999,Kappeler2011}. In addition, rotating massive stars have been proposed as viable contributors to the s-process \citep{Frischknecht2016,Choplin2018,Limongi2018} and related chemical evolution models \citep{Cescutti2013,Prantzos2018,Rizzuti2021}, while the nucleosynthesis site(s) of the r-process is still debated. The very high neutron flux required for the r-process is identified to originate from astrophysical sites  
 such as mergers involving neutron stars and/or black holes
 \citep{Lattimer1977,Freiburghaus1999,Martin2015,goriely2011,Abbott2017,Drout2017,Watson2019}, core-collapse supernovae(CCSN) \citep{Mathews1990,Wheeler1998,Ishimaru2004,Farouqi2010}, collapsars \citep{Siegel2019} and magnetohydrodynamically driven supernovae \citep{Winteler2012,Nishimura2015,nishimura2017}. However, the direct observational evidence has been available only for the r-process site associated with the neutron star mergers. The enhancement of r-process elements among old metal-poor stars is a definite sign that the r-process took place 
early in the Galactic history \citep{Sneden2008,Thielemann2017}. The variations in abundances shown by different stellar populations, especially at lower metallicities, indicate the need for two or more astrophysical sites for the production of r-process elements \citep{Hansen2014}.

Two scenarios are still debated to explain the abundance pattern of CEMP-rs stars.
The measured abundances in CEMP-rs stars could be produced by 
the intermediate neutron capture process (i-process) occurring in low-to-intermediate mass asymptotic giant branch stars \citep[K21]{cowan1977,hampel2016,iwamoto2004,denissenkov2017,Denissenkov2018,Hampel2019,goswami2020,choplin2021}. 
Alternatively, CEMP-rs stars could also be the result of two independent pollutions, one by the s-process and another one by the r-process  \citep{lugaro2009,Abate2016,bisterzo2011,Jonsell2006}. These scenarios seem to account for 
some objects, such as the one presented in \citet{gull2018}. 

Though CEMP-rs stars are by definition enhanced in r-process elements, 
the overabundances are mostly measured 
from Eu, Gd, or Dy. This is mainly due to the difficulty of measuring heavy r-process elements using the optical spectra, as their most sensitive lines are in the UV. A major goal of this study is to determine whether heavy r-process elements are enhanced in CEMP-rs stars and to gain insights into their production mechanisms. Some of them were confirmed as CEMP-rs stars in our earlier study (K21), while others are taken from the literature \citep{behara2010,mcwilliam1995}. 
The study also includes a few CEMP-s stars for comparison.  Although sites such as neutron star mergers are potentially identified as possible sources of r-process elements, their exact contribution to the galactic chemical evolution and to the solar abundances remains to be clarified.  Hence, it is crucial to identify the possible source of these elements and the level of production at these sites.  

The structure of the paper is as follows. The sample selection of CEMP-s and CEMP-rs stars is outlined in Sect.~\ref{Sample selection}, followed in Sect.~\ref{Sect:parameters} by the derivation of the stellar parameters and abundances. 
Individual lines are discussed in Sect.~\ref{Sect:lines}. Sect.~\ref{Sect:Classification}  presents the classification of stars using different diagnostic indicators.
Sect.~\ref{Sect:nucleosynthesis} presents a comparison with nucleosynthesis predictions, , while Sect.~\ref{Sect:Classification discussion} provides a discussion on the classification scheme.  Sect.~\ref{Sect:Comparison of the heavy r-process} focuses on the comparison of heavy r-process abundances in different CEMP classes. Finally, Sect.~\ref{Sect: conclusion} lists the discussions and conclusions of this paper.

\section{Sample selection}
\label{Sample selection}
Our stellar sample consists in twelve stars previously analyzed in K21, for which we derive or, in some cases, update (using improved line selection) the abundances of Gd, Tb, Dy, Ho, Er, Tm, Yb, Lu, Hf, Ta, Os, and Ir. For several of these elements, particularly Tb, Ho, Tm, Yb, Lu, Ta, Os, and Ir, the abundances were not derived in K21 and are presented here for the first time. 
New UVES VLT spectra have been obtained (ESO proposals 105.20LJ.001 and 105.20LJ.002) for 5 objects of K21.
Additionally, we used the UVES archival spectra for 5 new objects, they are listed in Table ~\ref{Tab:Comparison_previous_parameters}.

 The approximate resolution is $R= \lambda/\Delta \lambda = 47,000$ and the spectral coverage is from 3280~\AA~ to 6835~\AA.

The sample thus consists of 8 CEMP-rs stars and 9 CEMP-s stars, according to previous literature classification, as referenced in Table~\ref{Tab:program_stars}.  Following the analysis in the present paper, this classification will in some instances be changed (see Table~\ref{Tab:Classification}). 

Concerning binarity, as listed in Table~\ref{Tab:program_stars}, most of the stars are confirmed binaries. Only CS 30322$-$023 and SDSS J0912+0216 lack binarity confirmation.

 \section{Derivation of the atmospheric parameters and elemental abundances}
\label{Sect:parameters}

\begin{table*}
\caption{
Atmospheric parameters of the programme stars.
}
\label{Tab:program_stars}
\begin{tabular}{llcccccccccccc}
\hline
\\
Name &  $T_{\rm eff}$&$\log g$           & $\xi$         &[Fe/H] & Bin & Ref. Stellar & Ref. Literature & Ref. Binarity\\
     &    (K)       & (cms$^{-2}$)    &  (km s$^{-1}$)&    &&Parameters &  Classification & \\
\hline\\
\textbf{CEMP-rs stars}\\
CS 22891$-$171  &   5215 $\pm$ 68 &  1.24 $\pm$ 0.09  &2.14 $\pm$ 0.14 &$-$2.50 $\pm$ 0.10& Y& 1 &1 & 12,15\\
CS 22947-187     &   5200 $\pm$ 62 &  1.50 $\pm$ 0.12  &1.70 $\pm$ 0.08  &$-$2.55 $\pm$ 0.10 &Y&  2&3 & 13,14\\
HD 145777     &   4443 $\pm$ 57  &  0.50 $\pm$ 0.10 &2.63 $\pm$ 0.10	&$-$2.32 $\pm$ 0.10  &Y&1 & 1 &8 \\
HD 187861     &   5000 $\pm$ 100 &  1.50 $\pm$ 0.25 &2.00 $\pm$ 0.20  &$-$2.60 $\pm$ 0.10& Y&1 & 1 & 11\\
HD 196944     &   5168 $\pm$ 48 &  1.28 $\pm$ 0.16  &1.68 $\pm$  0.11 &$-$2.50 $\pm$ 0.09& Y& 1& 1 & 1\\
HD 224959     &   4969 $\pm$ 64 &  1.26 $\pm$ 0.29  &1.63 $\pm$ 0.14  &$-$2.36 $\pm$ 0.09& Y&1&1 &10\\
SDSS J0912$+$0216     &   6140 $\pm$ 37 &  4.60 $\pm$ 0.21  & 1.19 $\pm$ 0.07  &$-$2.83 $\pm$ 0.07 & -&2&4&- \\
SDSS J1349$-$0229     &   6238 $\pm$ 95 &  4.41 $\pm$ 0.21  &1.45 $\pm$ 0.08 &$-$3.13 $\pm$ 0.07 &Y&2&4 & 4\\

\textbf{CEMP-s stars}\\
SDSS J1036+1212     &   5591 $\pm$ 99&  3.70 $\pm$ 0.08  & 0.87 $\pm$ 0.10  &$-$3.48 $\pm$ 0.09&Y&2&4 & 16,17 \\
CS 22887$-$048  &   6500 $\pm$ 50 &  3.20 $\pm$ 0.15  & 1.00 $\pm$ 0.05& $-$2.10 $\pm$0.09 &Y&  1 & 1 & 12,13\\
CS 22942$-$019  &   5100 $\pm$ 98 &  2.19 $\pm$ 0.20  &1.73 $\pm$ 0.10 & $-$2.50 $\pm$ 0.09&Y& 1& 1 & 6,7\\
CS 29512$-$073     &   5471 $\pm$ 82 &  2.78 $\pm$ 0.16  & 1.28 $\pm$ 0.08  & $-$2.35 $\pm$ 0.09 &Y&2 &5 & 12,13\\
CS 30322$-$023  &   4500 $\pm$ 100 &  1.00 $\pm$ 0.50  & 2.80 $\pm$ 0.10 &$-$3.35 $\pm$ 0.09  &-& 1& 1&-\\
HD 26         &   5169 $\pm$ 108 &  2.46 $\pm$ 0.18  & 1.46 $\pm$ 0.08 &$-$0.98 $\pm$  0.09  &  Y&1 & 1 & 8 \\
HD 55496      &   4642 $\pm$ 39 &  1.65 $\pm$ 0.14  &1.33 $\pm$ 0.08  &$-$2.10 $\pm$ 0.09&Y&1 & 1 & 7,9\\
HD 198269     &   4458 $\pm$ 15 &  0.83 $\pm$ 0.08  & 1.64 $\pm$ 0.09 &$-$2.10 $\pm$ 0.10 &Y&1 & 1 & 10\\
HD 206983     &   4200 $\pm$ 100 &  0.60 $\pm$ 0.20  & 1.50 $\pm$ 0.10  &$-$1.00 $\pm$ 0.10 &Y&1 & 1 & 7
 \\
\hline
\end{tabular}
\tablecomments{ The effective temperature $T_{\rm eff}$, surface gravity $\log g$, microturbulence $\xi$, and metallicity [Fe/H] are presented. The stars are grouped as CEMP-rs or CEMP-s stars according to their previous literature classification. The classification derived from the analysis of the present paper is listed in Table~\ref{Tab:Classification}.\\
References : (1) K21; (2) This work; (3) \citet{mcwilliam1995}; (4) \citet{behara2010}; (5) \citet{Roederer2014}; (6) \citet{Preston2001}; (7) \citet{Jorissen2005}; (8) \citet{Jorissen2016}; (9) \citet{Pereira2019}; (10) \citet{Mcclure1990}; (11) \citet{Gaia(3)2022}; (12) \citet{Gaia(1)2022}; (13) \citet{Beers2000}; (14) \citet{Buder2021}; (15) \citet{Beers1992}; (16) \citet{Bonifacio2021}; (17) \citet{Ahn2012} }
\end{table*}


 The atmospheric parameters, T$_{\rm eff}$, $\log g$, $\xi$, and [Fe/H]  are derived using the BACCHUS  code \citep{Masseron2016} in a semi-automated mode as explained by \citet{Karinkuzhi2018,Karinkuzhi2021} and are listed in Table~\ref{Tab:program_stars}. 
 As can be seen in Table ~\ref{Tab:Comparison_previous_parameters}, the agreement between the stellar parameters derived in the present work and those of previous studies is reasonable. The objects already discussed in K21 are not further mentioned here.
 

\begin{table}
\centering
\caption{
A comparison of the stellar parameters of the programme stars with literature}

\label{Tab:Comparison_previous_parameters}
\begin{tabular}{llllc}
\hline
\\
Name &  $T_{\rm eff}$&$\log g$                  &[Fe/H] & Ref.\\
     &    (K)       & (cms$^{-2}$)    & &   \\
\hline\\
CS 22947$-$187     &   5200 $\pm$ 62 &  1.50 $\pm$ 0.12    &$-$2.55 $\pm$ 0.10 &  1 \\
 & 5160 &1.30 $\pm$ 0.25 &  $-$2.49 & 2 \\
 & 5300  $\pm$ 52  &1.40 $\pm$ 0.37 &  $-$2.58 $\pm$ 0.06& 3 \\
 \vspace{1mm} \\
SDSS      &   6140 $\pm$ 37 &  4.60 $\pm$ 0.21   &$-$2.83 $\pm$ 0.07 &1 \\
J0912$+$0216 & 6500 &  4.5  &  $-$2.50 & 4 \\
& 6150 &  4.0   &  $-$2.68 & 5 \\
 \vspace{1mm} \\
SDSS      &   5591 $\pm$ 99&  3.70 $\pm$ 0.08   &$-$3.48 $\pm$ 0.09&1 \\
J1036t$+$1212 & 6000 &  4.0   &  $-$3.20 & 4 \\
& 5850 &  4.0  &  $-$3.47 & 5 \\
 \vspace{1mm} \\
SDSS      &   6238 $\pm$ 95 &  4.41 $\pm$ 0.21   &$-$3.13 $\pm$ 0.07 &1 \\
J1349$-$0229& 6200 &  4.0   &  $-$3.00 & 4 \\
& 6200 & 4.0 &  $-$3.40 & 5\\
 \vspace{1mm} \\
CS 29512$-$073     &   5471 $\pm$ 82 &  2.78 $\pm$ 0.16   & $-$2.35 $\pm$ 0.09 &1 \\
& 5650 $\pm$ 52& 3.60 $\pm$ 0.23 &  $-$1.93$\pm$ 0.07 & 3\\
& 5852 & 2.88 &  $-$2.15 & 6\\

 \\

\hline
\end{tabular}
\tablecomments{The stellar parameters (effective temperature, surface gravity, and metallicity) for the programme stars that were not discussed in K21 are compared with literature values.\\
References : (1) This work; (2) \citet{mcwilliam1995}; (3) \citet{Roederer2014};    (4) \citet{behara2010}; (5) \citet{Aoki2013};  (6) \citet{Limberg2021} }

\end{table}

 The abundances are derived by comparing the observed spectra with synthetic ones produced by the 1D local thermodynamic equilibrium (LTE)  TURBOSPECTRUM radiative transfer code \citep{Alvarez1998} using MARCS model atmospheres \citep{Gustafsson2008}. The \citet{Asplund2009} solar abundances were adopted.
Individual elemental abundances were calculated using line lists from \citet{Heiter2015} and \citet{Heiter2020}. The atomic lines used to derive the abundances of all the elements are presented in 
Table~\ref{Tab:Linelists}. The abundances were derived under LTE conditions, but non-LTE (NLTE) corrections were applied when they were available in the literature for stars of similar stellar parameters. 
Only the abundances of the heavy r elements are 
listed in Table~\ref{Tab:abundances}
for the stars already analyzed by K21. 
However, the 
full list of derived abundances 
is presented for the five objects not included in K21.


\section{Comments on individual abundances}
\label{Sect:lines}
We present the atomic and molecular lines used to derive the elemental abundances for the five objects that were not in K21. For the remaining objects, we only discuss the heavy r-process elements. Lithium could be a sensitive diagnostic of the i-process, since it is known to be produced during proton ingestion events \citep[e.g.][]{Iwamoto04, Cristallo-2009, Choplin2024b}, at least if not destroyed during subsequent AGB evolutionary phases.
Unfortunately it could not be measured in any of our targets.

\subsection{C, N, and O in the five new objects} 

\begin{figure}    
\centering
\plotone{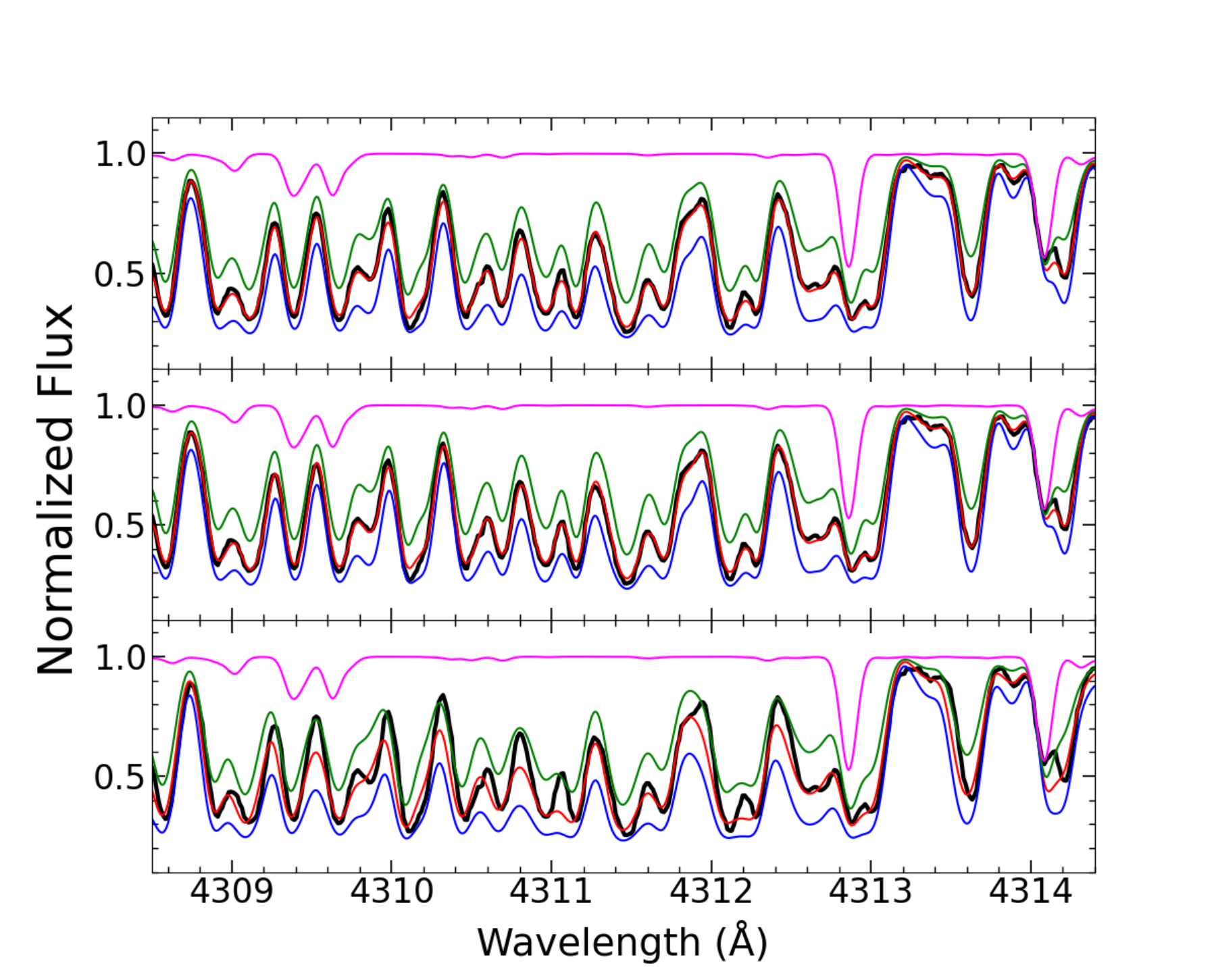}
\caption{Spectral fits for the determination of  $^{12}$C/$^{13}$C ratio using the CH G band at 4310 \AA\ in CS 29512$-$073. The upper (resp., middle and bottom) panel shows spectral synthesis with $^{12}$C/$^{13}$C$ =30$ (resp., $19$ and $1.5$). 
The red curve depicts the synthetic spectrum for an abundance of \(\log \epsilon(\mathrm{C}) = 7.45\), with the blue and green curves illustrating \(\pm 0.3\) dex variations. The black line represents the observed spectrum, and the magenta line corresponds to the spectral synthesis without carbon.
\label{fig:C4310}}
\end{figure}

\begin{figure}
\centering
\plotone{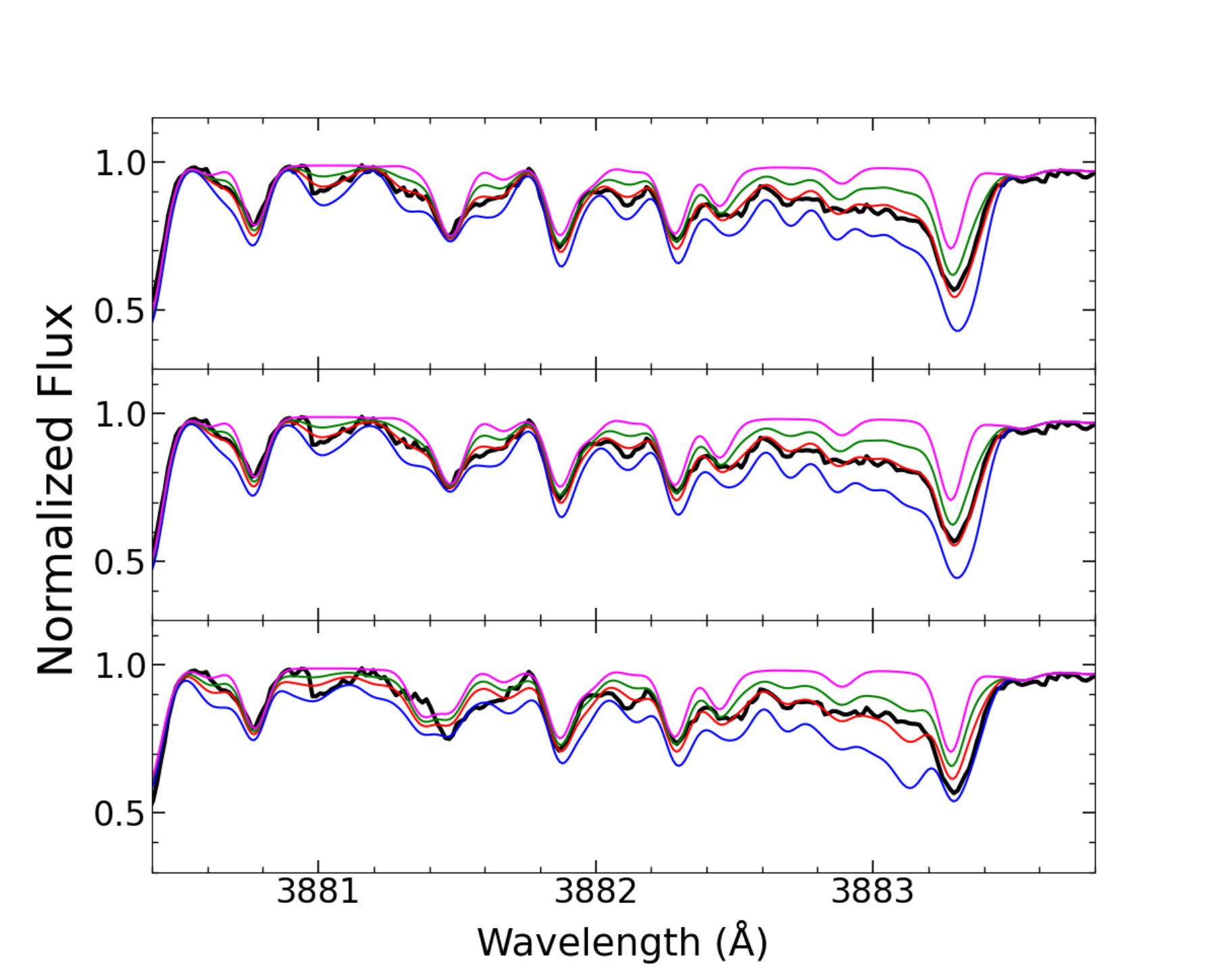}
\caption{Spectral fits for determining the $^{12}$C/$^{13}$C ratio using the CN band at 3883 \AA\ in CS 29512$-$073. The upper (resp., middle and bottom) panel shows spectral synthesis with $^{12}$C/$^{13}$C$= 30$ (resp., 11.5 and 1.5). The red curve depicts the synthetic spectrum for an abundance of \(\log \epsilon(\mathrm{N}) = 6.10\), with the blue and green curves indicating the synthesis with \(\pm 0.3\) dex variations. The black line represents the observed spectrum, and the magenta line corresponds to the spectral synthesis without nitrogen.
\label{fig:N3883}}
\end{figure}

The carbon (C) abundance is derived 
using only the CH G band at 4310~\AA\ 
because the C$_2$ bands 
are very weak due to their rather
high temperatures.  The oxygen abundance is measured 
using the forbidden line at 6300.303 \AA.  We could derive the nitrogen abundance in 
all five objects using the CN bands at 3883 \AA.

The $^{12}$C/$^{13}$C ratio is determined using $^{12}$CH and $^{13}$CH features in the G band region near 4300~\AA~ (Figure~\ref{fig:C4310}).
The $^{12}$CN and $^{13}$CN features 
around 3883 \AA\ were also used for 
confirmation (Figure~\ref{fig:N3883}).
The differences between the $^{12}$C/$^{13}$C ratios estimated from the G band and the 3883 \AA\ region are very small (of the order of $\pm 3.0$~units for all objects) except for CS 29512$-$073, where 
we find 
$^{12}$C/$^{13}$C = 19 
when derived from the CH band and 11.5 from the CN band. 

\subsection{Light elements: Na, Mg, Ca,  Sc, Ti, Cr, Mn, Co, Ni, Cu, and Zn}
 The abundances of the light elements are derived using the lines listed in Table A.1 of K21.
 For the five new objects,  the derived abundances are presented in Table~\ref{Tab:abundances}, while for the others, they are listed in Table~B.1 of K21. 
We could not measure the Na abundance in SDSS J1349$-$0229 because the Na lines were too strong and saturated.
The Mg abundance was derived exclusively from the \ion{Mg}{1} line at 5528.405~\AA. However, for the other two objects, both the  \ion{Mg}{1} line at 5528.405~\AA~and 5711.088~\AA~were used to derive the Mg abundance. \citet{Sofya2018} calculated a NLTE correction of 0.02 dex for the 5528.405~\AA~\ion{Mg}{1} line, corresponding to the following atmospheric parameters: $T_{\rm eff} = 6350$~K, $\log g = 4.09$, and [Fe/H]~$= -2.08$.
The stellar parameters of the stars in \citet{Sofya2018} are close but do not exactly match the ones of our objects; therefore, no NLTE correction was applied to our Mg abundances. However, from \citet{Sofya2018} we infer that the NLTE corrections for the \ion{Mg}{1} lines at 5528.405 and 5711.088~\AA~ would probably be very small.
The $\alpha$-elements show an average enrichment with [X/Fe] = 0.56, considering all the stars in our sample. 

The lines that were used for the abundance determination of light elements in addition to those listed in Table A.1 of K21, are listed in 
Table~\ref{Tab:Linelists}.

\subsection{s$-$process elements: Sr, Y, Zr, Nb, Mo, Ba, La, Ce, Pr, and Nd}

The abundances are listed in Table \ref{Tab:abundances}, some specific comments can be found below.

\noindent{\bf Strontium:} The Sr abundance is derived in the five new stars using \ion{Sr}{2} lines at 3464.453~\AA , 4077.707~\AA\ and 4215.520~\AA~. In K21, \ion{Sr}{1} lines at 4607.327, and 7070.070 \AA~ were also used, which are either absent or blended in these spectra, hence were not used in this analysis. 
 The NLTE corrections were taken from
 \citet{Bergemann2012,Mashonkina2008,Mashonkina2023} and, for the parameter range of all objects,
they are minor, ranging from 0.01 to $-$0.03 dex.
The only exception is SDSS J1036$+$1212, for which a NLTE correction of 0.14 dex for the 4077.707~\AA\ \ion{Sr}{2} line was applied, corresponding to a star with the following stellar parameters 
($T_{\rm eff} = 5600$~K, $\log g = 3.80$, and [Fe/H]~$= -$3.48) taken from \citet{Bergemann2012}.

\noindent{\bf Yttrium:} Many unblended or clean lines are available to measure the Y abundances.
For CS22947$-$187 and CS29512$-$073, a NLTE correction of 0.12 dex is adopted, corresponding to the average NLTE correction (for the lines that we used) computed in \citet{Sofya2023} for HD 122563 ($T_{\rm eff} = 4600$~K, $\log g$ = 1.43, and [Fe/H]~$= -$2.55). 

For SDSS J1036$+$1212, we applied corrections of 0.06 dex (resp., 0.07 dex), for the \ion{Y}{2} 4883.684~\AA~(resp., 4900.120~\AA) lines, since these values were computed by \citet{Sofya2023} for HD 140283 which has closely matching stellar parameters ($T_{\rm eff}$ = 5780 K, log g = 3.7, and [Fe/H] = $-$2.46).

In the same vein, since SDSS J0912$+$0216 and SDSS J1349$-$0229 have stellar parameters similar to those of HD~84937 ($T_{\rm eff} = 6350$~K, $\log g = 4.09$, and [Fe/H]~$= -2.12$), a turn-off very metal-poor star analyzed by \citet{Sofya2023}, we used their NLTE corrections of 0.06 dex for \ion{Y}{2} lines at  4883.684 and 4900.120 \AA~. 

We note that all our LTE abundances seem to be underestimated when compared to the model predictions (see Sect. \ref{Sect:nucleosynthesis}), and that NLTE corrections tend to reduce this discrepancy, since \citet{Sofya2023} measured positive NLTE corrections for all these lines with an average value of 0.21 dex for the parameter range corresponding to HD~122563. 


\noindent{\bf Zirconium:} In addition to the \ion{Zr}{2} lines listed in K21, we also used the \ion{Zr}{2} lines at 4208.977 \AA\ and 4359.720 \AA.

\noindent{\bf Niobium:} The \ion{Nb}{2} lines at 3425.425 \AA\ and 3426.531  \AA\ were used to derive Nb abundance in CS 29512$-$073  and SDSS J1349$-$0229. For SDSS J1036$+$1212, the \ion{Nb}{2} lines at 3651.187 \AA\ was used to estimate the Nb abundance. Niobium (Nb) could be reliably measured only in these three stars among the five new objects analyzed in this study.

\noindent{\bf Molybdenum:} The Mo abundance is measured in all the programme stars using the 
line at 3864.103 \AA\, as this is the only line available that is sensitive enough for abundance determination in our wavelength range.

\noindent{\bf Barium:} 
The  \ion{Ba}{2} lines at 5853.673 \AA, 6141.711 \AA~ and 6496.895~\AA\ were used to derive Ba abundance in SDSS J1349$-$0229 and SDSS J0912$+$0216. For the other three objects, we could also use the \ion{Ba}{2} line at 4166.000 \AA. 

\noindent{\bf Lanthanum:} The \ion{La}{2} lines at 4920 and 4921\AA, for which HF splitting is available, are primarily used to estimate the La abundance.  

The lines used to derive the abundances of Ce, Pr, and Nd are presented in the Table. \ref{Tab:Linelists}.

\subsection{r-process elements: Sm, Eu, Gd, Tb, Dy, Ho, Er, Tm, Yb, Lu, Hf, Ta, Os, and Ir}

\begin{figure}
\centering
\plotone{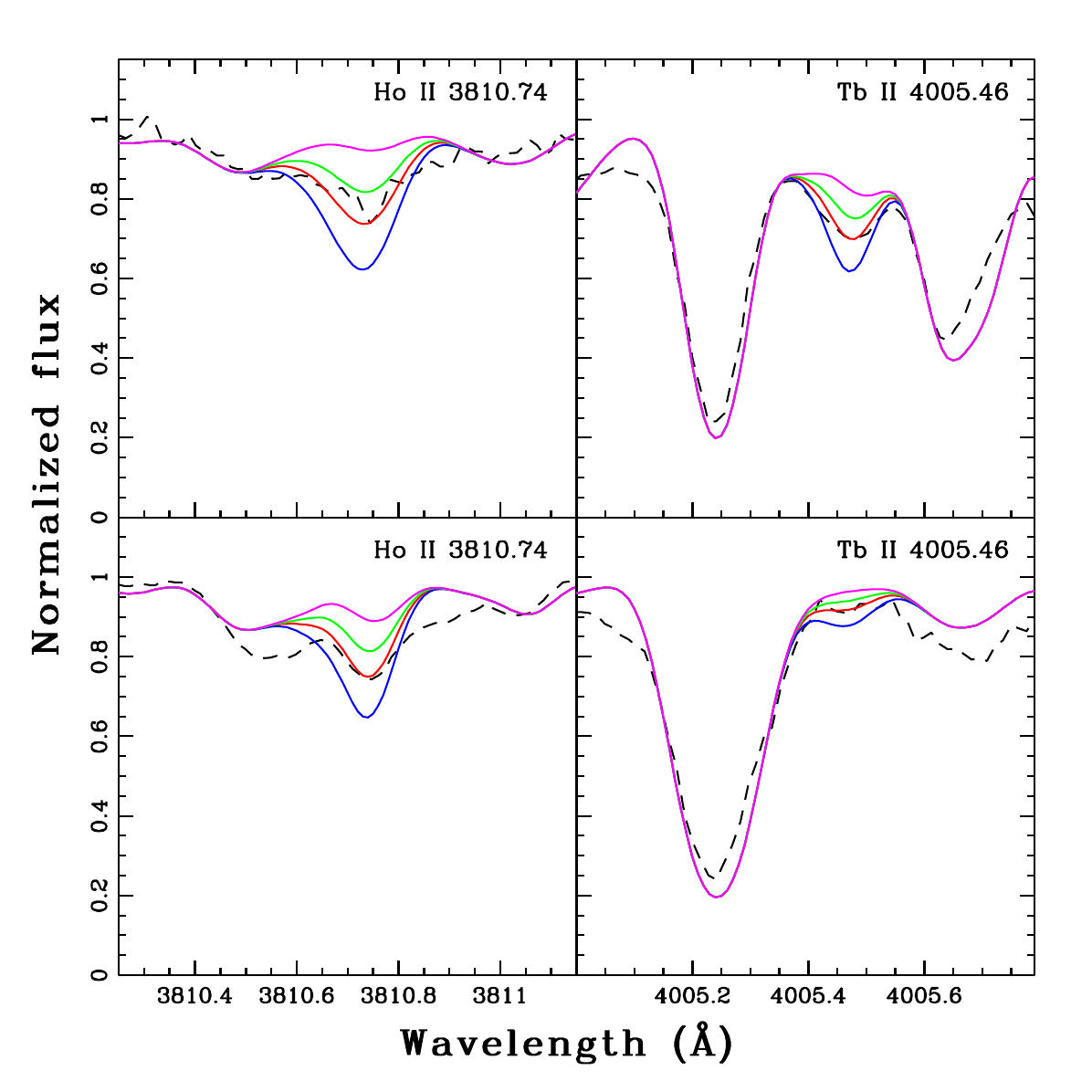}
\caption{The upper and lower left panels display the spectral fitting of the \ion{Ho}{2} lines for CS 22947$-$187 and HD 196944, while the upper and lower right panels display the \ion{Tb}{2} lines for HD 224959 and CS 30322$-$023.  Red lines correspond to spectral syntheses with the adopted \ion{Ho}{2} abundances of $-$1.6 dex, $-$2.0 dex for CS 22947$-$187 and HD 196944, and \ion{Tb}{2} abundances of $-$0.75 dex, $-$2.0 dex for HD 224959 and CS 30322$-$023  respectively.  Blue and green lines correspond to syntheses with abundances deviating by $\pm$0.3 dex from the adopted abundance. The black dashed line represents the observed spectrum. The magenta line corresponds to the synthesis with a null abundance for the corresponding element.}
\label{Fig:TbHo}
\end{figure}

In K21, the abundances of Sm, Eu, Gd, Dy, Er, Hf, and Os were determined for most stars. Abundances of Tb, Ho, Tm, and Yb for the five stars from K21 have been reported in our previous work \citep{riyas2024}.
In the present paper, we could add abundance determinations for five new objects and also complement these abundances with the ones of the heavy r-elements Tb, Ho, Tm, Yb, Lu, Ir, and Ta. A few fits are illustrated on Figure~\ref{Fig:TbHo}.

\noindent{\bf Samarium:} The  \ion{Sm}{2} lines at  
4318.926~\AA, 4390.854~\AA~ and 4420.520~\AA~were used to derive the Sm abundance for SDSS J0912$+$0216  and SDSS J1349$-$0229, while the line at 3941.876~\AA\ was used for SDSS J1036$+$1212. For CS 22947$-$187 and CS 29512$-$073, many Sm lines were accessible between 4000 and 5000~\AA\ as listed in Table~\ref{Tab:Linelists} of the Appendix.

\noindent{\bf Europium:} We determined the LTE and NLTE Eu abundances independently, as detailed in Sect. 4.6 of K21, and the results are shown in
Table~\ref{Tab:abundances}. 

\citet{Mashonkina2008} calculated the NLTE correction of 0.16  dex for the \ion{Eu}{2} line at 4129.680~\AA\ in  HD~122563 ($T_{\rm eff}$ = 4600~K, $\log g = 1.50$, [Fe/H] = $-$2.53). The  Eu abundance is derived using the same line for CS 22947$-$187 and CS~29512$-$073, so the same NLTE correction is applied to get the final Eu abundance since their parameters are similar to those of HD~122563.   The Eu$_{\rm LTE}$ in Table~\ref{Tab:abundances} is the average Eu abundance calculated without the 4129.680 \AA~line. 
The Eu abundance for SDSS J1349$-$0229 is obtained using the \ion{Eu}{2} line at 4205.065~\AA. We applied the NLTE correction of 0.11 dex from \citet{Mashonkina2014}, corresponding to the closest parameters $T_{\rm eff}$ = 5260~K, $\log g = 2.75$, [Fe/H] = $-$2.85. As these stellar parameters are different from the ones of SDSS J1349-0229, we list in Table~\ref{Tab:abundances} both the LTE and the (uncertain) non-LTE abundance.

For SDSSJ1036$+$1212, we used the two \ion{Eu}{2} lines at 3907.108 and 3930.506~\AA\ to derive the Eu abundance. \citet{Mashonkina2014} listed the NLTE corrections of 0.04 and 0.05 dex respectively for the 3907.108 and 3930.506~\AA~lines, corresponding to the parameters $T_{\rm eff}$ = 5010~K, $\log g = 4.80$, [Fe/H] = $-$3.40, which are close to those of SDSSJ 1036+1212. 

We calculated the Eu abundance for SDSSJ 0912$+$0216 using the three \ion{Eu}{2} lines at 3819.684, 3907.108, and 4205.065 ~\AA. \citet{Mashonkina2014} reported the NLTE corrections of 0.12, 0.15, and 0.11 dex respectively for these three lines, corresponding to the parameters $T_{\rm eff}$ = 5260~K, $\log g = 2.75$, [Fe/H] = $-$2.85, which are similar to those of SDSSJ 0912$+$0216. 

\noindent{\bf Gadolinium:}
 \ion{Gd}{2} lines at  3545.790 and 3768.396~\AA~are used in addition to the \ion{Gd}{2} line at 4251.731~\AA~ used by K21 for deriving the Gd abundance in SDSS J0912+0216 and SDSS J1036+1212. 

\noindent{\bf Terbium:}
A single \ion{Tb}{2} line at 3939.539~\AA\ is used to compute the Tb abundance in HD~145777.  
For HD 26, we used only the 
3658.888~\AA~line (leading to $\log \epsilon = 0.90$ dex), since the
3625.510~\AA\ line (leading to $\log \epsilon =0.30$ dex) appears to be affected by an unidentified feature.

\noindent{\bf Dysprosium:}
For HD~206983, Dy abundance is measured using the  \ion{Dy}{2} line at 4103.306~\AA~line, which is strong (20 \% of the local continuum) and blended by a Fe I line at 4103.298~\AA. For HD 26, this Dy line at 4103.306 \AA\ is also strong (40 \% of the local continuum), and it is marked as uncertain in the Table~\ref {Tab:abundances}.

\noindent{\bf Holmium:}
The \ion{Ho}{2} line at 3810.738~\AA\ could be used in CS~29512-073 and HD~196944, in addition to the other Ho lines as listed in Appendix \ref{Tab:Linelists}.




\begin{figure}
\centering
\plotone{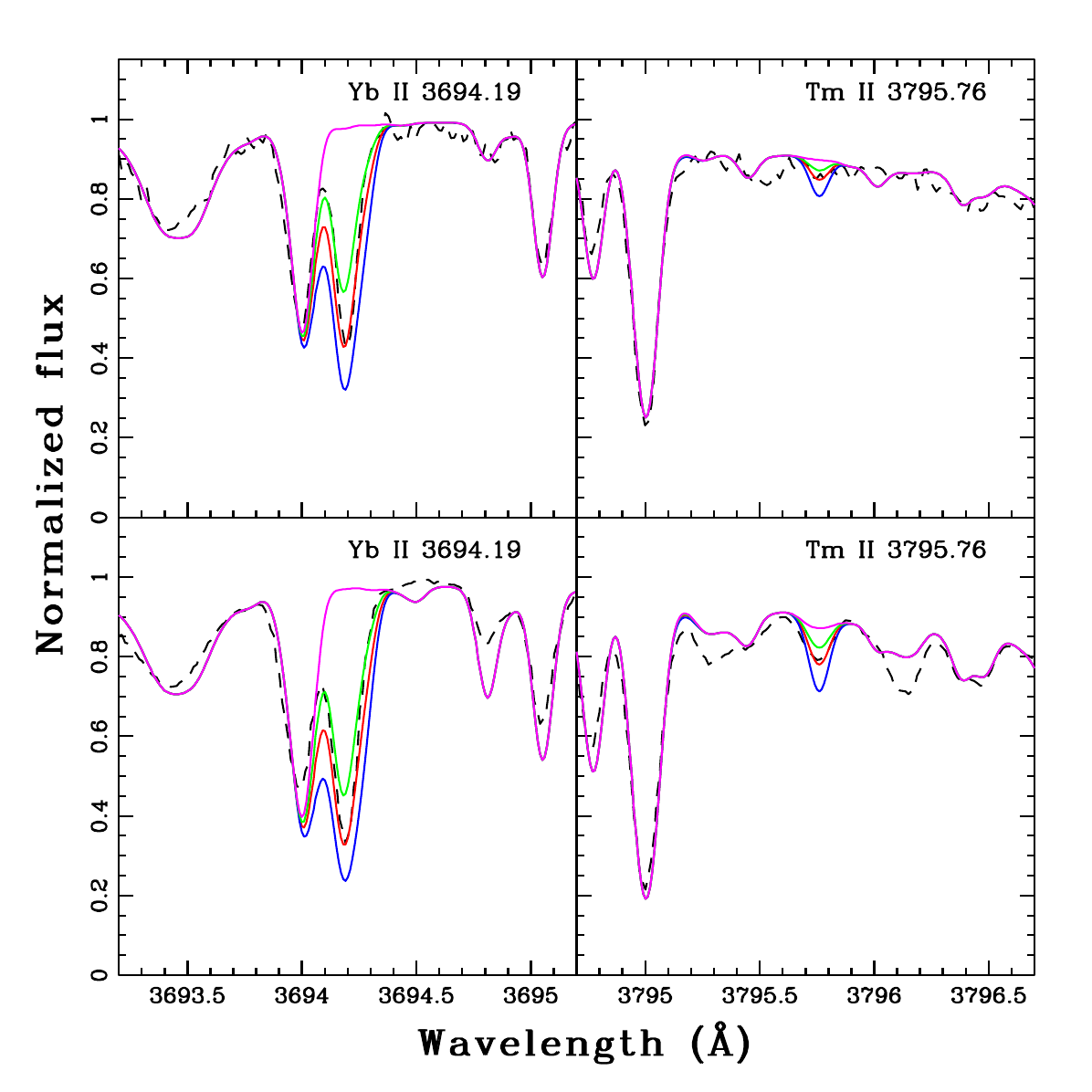}
\caption{Spectral fitting of the \ion{Yb}{2} and \ion{Tm}{2} lines is shown for two CEMP stars, CS~29512$-$073 and HD~196944 in the upper and lower panels, respectively.  Red lines correspond to spectral syntheses with the adopted \ion{Yb}{2}, \ion{Tm}{2} abundances of $-$0.55 dex, $-$1.20 dex for CS~29512$-$073, and $-$1.05 dex, $-$1.60 dex for HD~196944 respectively. The blue, green, magenta, and black curves have the same meaning as in Figure \ref{Fig:TbHo}.
\label{fig:TmYb}}
\end{figure}


\noindent{\bf Thulium:}
A total of seven \ion{Tm}{2} lines are used for deriving the Tm abundance in the programme stars. The spectral synthesis of the 3795.76~\AA~ Tm line is presented in Figure~\ref{fig:TmYb}. The \ion{Tm}{2} lines at 3700.255~\AA~ and 3701.362~\AA~ are used in all our programme stars, but in HD~198269, the Tm II line at 3700.255~\AA\ is slightly blended on the right wing. Although the abundance from this line is consistent with the one from the line at 3701.362~\AA\, it is not included in the analysis. \\
\noindent{\bf Ytterbium:}
The Yb abundance is measured in all the programme stars using the \ion{Yb}{2} line at 3694.192~\AA~  (Figure~\ref{fig:TmYb}), as it is the only line sufficiently sensitive that we could identify for abundance determination. However, in some stars (HD~26, HD~198269, HD~206983, and HD~224959) it is very strong ($\le $ 30 \% of the local continuum), and therefore rejected, because the core of the line must form in the upper layers of the photosphere, where the LTE approximation vanishes.

\noindent{\bf Lutetium:}
In HD 198269, the \ion{Lu}{2} line at 3507.395~\AA, which is measurable in most other stars, is absent. Therefore, the Lu abundance is determined using the other two \ion{Lu}{2} lines: 5983.701~\AA~ and 6221.592~\AA, leading to $\log \epsilon = -0.10$ dex and $ -0.80$ dex, respectively. Given the discrepancy between these two values, the Lu abundance (computed as their average) is considered highly uncertain.

\noindent{\bf Hafnium:}
The \ion{Hf}{2} lines at 3918.090~\AA~ and 4093.150~\AA~ yield consistent abundances in all the stars where they are detectable. When they could not be measured, we used the other Hf lines listed in Appendix \ref{Tab:Linelists}.

\begin{figure}
\centering
\plotone{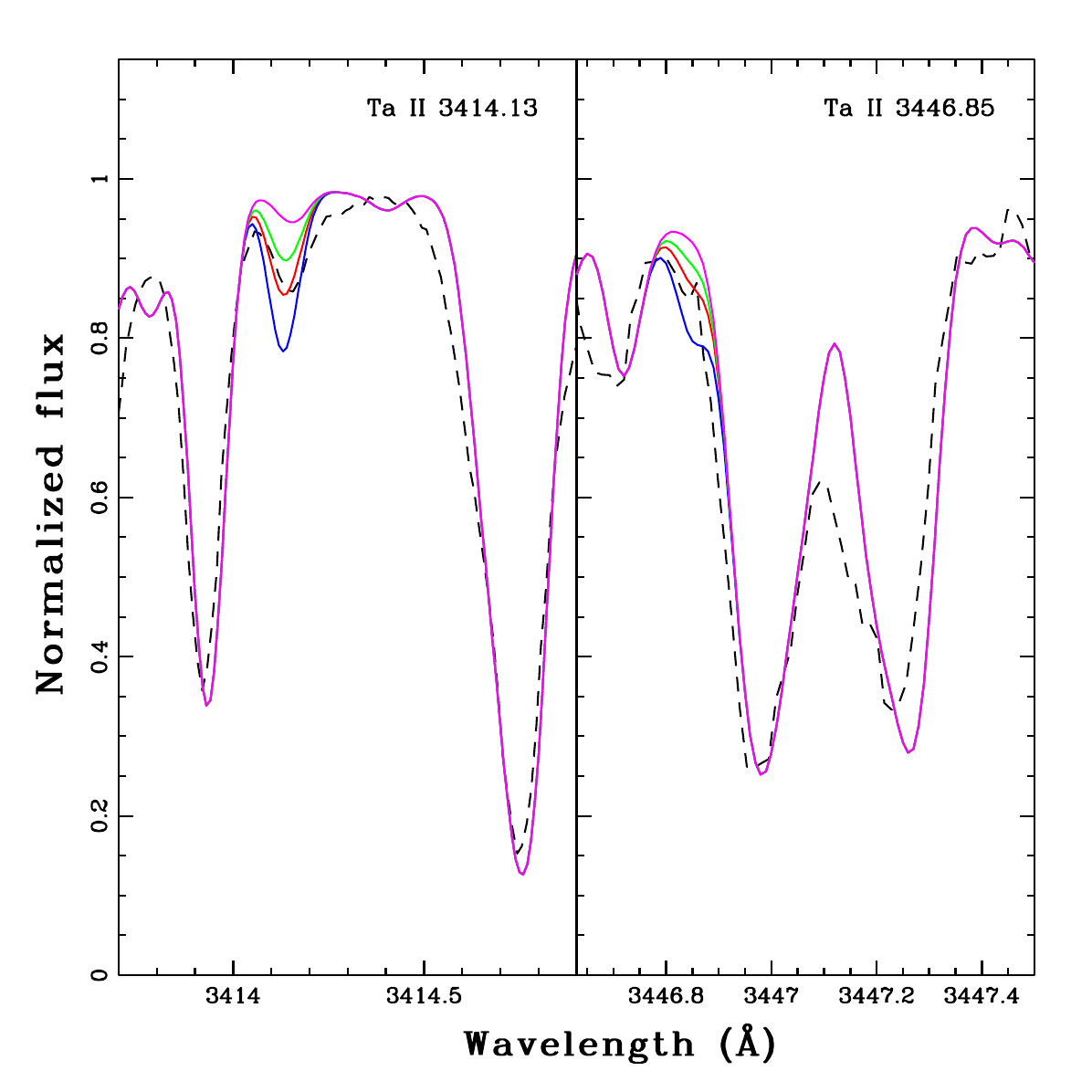}
\caption{ The spectral fits for the \ion{Ta}{2} lines at 3414.13 \AA\ and 3446.85 \AA\ are presented for HD~196944 in the left and for HD~224959 in the right panels, respectively. The red curve represents the synthesis with the adopted abundances ($-$0.6 dex for the left panel and $-$0.25 dex for the right panel). The blue, green, magenta, and black curves have the same meaning as in Figure \ref{Fig:TbHo}.}
\label{Fig:Ta}
\end{figure}

\noindent{\bf Tantalum:}
Eight lines of \ion{Ta}{2}  could be identified in the spectra, as listed in Table~\ref{Tab:Linelists}. Two lines are illustrated on Figure~\ref{Fig:Ta}. However, because of a poor agreement between synthetic and observed spectra, only upper limits or uncertain abundances could be derived for all the stars except for HD~196944. For HD~196944, the best fit was obtained from the 3414.128~\AA~ line ($\log \epsilon = -0.60$ dex), providing an abundance roughly in agreement with the (clearly less well-fitted) 3057.232 and 3056.603~\AA~lines ($\log \epsilon \sim -0.40$ dex), but in conflict with the 3440.312~\AA~line ($\log \epsilon = 0.30$ dex). 
Table~\ref{Tab:abundances} lists the Ta abundance of HD~196944 derived from the sole 3414.128~\AA\ line.

\noindent{\bf Osmium:}
We investigated three Os lines: 4135.781 \AA, 4260.849 \AA, and  4420.477~\AA. It turns out that the 4135.781~\AA~line gives higher abundance (by 0.60 dex) than the 4260.849~\AA~ and 4420.477~\AA~ lines in HD~55496 and CS~22891$-$171. 
In addition, the abundance derived from the 
4135.781~\AA\ line is much larger than the model predictions.
Therefore, this line was rejected, and the Os abundance  (or upper limit) as listed in Table \ref{Tab:abundances} was only derived from the 4260.849~\AA~ and 4420.477~\AA~lines.

\noindent{\bf Iridium:}
The \ion{Ir}{1} line at 3992.121~\AA\ is sensitive, but the overall agreement in this spectral region is not satisfactory. 
Some additional Ir lines are sometimes visible, but we only provide upper limits or uncertain Ir abundances in Table \ref{Tab:abundances}.

\subsection{The third peak s-process element Pb}

The \ion{Pb}{1} line at 4057.807~\AA~is the only suitable one to measure the Pb abundance in our spectra. The hyperfine splitting is taken into account. 
NLTE corrections of +0.37, +0.62, and +0.52 dex were adopted for 
CS~22947$-$187, CS~29512$-$073, and SDSS~J1349$-$0229, respectively, based on the values derived by \citet{Mashonkina2012} for stars with comparable stellar parameters.

\subsection{Uncertainties in abundances}

\begin{table}
\centering
\caption{Uncertainties in abundances. }
\label{Tab:uncertainties}
\begin{tabular}{crrrr}
\hline
       & \multicolumn{4}{c}{$\Delta \log \epsilon_{X}$} \\
        \cline{2-5}
 Element&   $\Delta T_{\rm eff}$ & $\Delta \log g$ & $\Delta \xi_t$ &$\Delta$ [Fe/H]   \\
&   ($+$100 K) & ($+$0.5) & ($+$0.5 km~s$^{-1}$) & ($+$0.5 dex)\\
\hline
C  &  0.20  &  $-$0.15   &    $-$0.05  &   0.00 \\
N  &  0.15  &  0.05   &     0.10  &    0.00 \\
O  &  0.10  &  0.20   &    0.05  &   0.05 \\
Na &  0.05  &  0.00   &     0.00  &    0.00 \\
Mg &  0.05  &  0.00   &     $-$0.05  &    0.00 \\
Ca &  0.06  &  $-$0.03  &     $-$0.09  &    $-$0.01 \\
Sc &  0.05  &  0.18   &     $-$0.02  &    0.00 \\
Ti &  0.09  &  0.10   &     0.00  &    0.06 \\
Cr &  0.08  &  0.00   &     $-$0.06  &    $-$0.06 \\
Mn &  0.10  &  0.00   &     0.00  &    0.00 \\
Fe &  0.14  &  0.06   &    $-$0.10  &    0.01 \\
Co &  0.13  &  0.00   &     $-$0.10  &    0.00 \\
Ni &  0.08  &  0.00   &     $-$0.02  &    0.01 \\
Cu &  0.10  &  $-$0.05   &     0.00  &    0.00 \\
Zn &  0.05  &  0.05   &     $-$0.05  &    0.00 \\
Sr &  0.13  & 0.20   &    $-$0.47  &    $-$0.05  \\
Y  &  0.05  &  0.16   &     0.00  &   0.01 \\
Zr &  0.00  &  0.05   &    0.00  &   0.00 \\
Mo &  0.05  &  $-$0.10   &     $-$0.10  &    0.00 \\
Ba &  0.03  &  0.18   &     $-$0.13  &    0.02 \\
La &  0.07  &  0.15   &    0.00  &   0.02 \\
Ce &  0.05  &  0.15   &   0.00  &   0.02\\
Pr &  0.05  &  0.10   &    0.00  &   0.02 \\
Nd &  0.05  &  0.15   &    $-$0.03  &   0.02  \\
Sm & 0.07  &  0.14   &   $-$0.02  &   0.02 \\
Eu & 0.03  &  0.10   &  $-$0.02  &   $-$0.02 \\
Gd &  0.07  &  0.17   &   0.00  &   0.02\\
Tb &  0.10  &  0.15   &    $-$0.03  &   0.03 \\
Dy &  $-$0.05  &  0.05   &     $-$0.08  &    0.00 \\
Ho &  0.05  &  0.12   &     0.00  &    0.00 \\
Er &  0.08  &  0.22   &    $-$0.02  &   0.02 \\
Tm &  0.00  &  0.13   &     0.02  &    0.03 \\
Yb &  0.05  &  0.15   &     $-$0.10  &    0.00 \\
Lu &  0.06  & 0.22    &  0.00  & 0.02 \\
Hf &  0.07  &  0.22   &   0.00  &   0.05 \\
Ta &  0.07 &  $-$0.03   &    $-$0.11  &   0.02 \\
Os &  0.15  &  0.05   &    0.00  &   0.00 \\
Ir &  0.06  & 0.10   & 0.00  &0.00 \\
Pb &  0.10  &  0.00   &     0.00  &    0.00\\
\hline
\end{tabular}
\tablecomments{Abundance variations ($\Delta \log \epsilon_{X}$) with variations of the atmospheric parameters in CS~22947$-$187. The terbium and osmium lines are extremely close to the continuum in CS~22947$-$187; we thus used a star with similar stellar parameters, HD~224959 (see Table~\ref{Tab:program_stars})
to estimate the abundance sensitivity to parameter changes for Tb and Os.}
\end{table}

The uncertainties on the elemental abundances 
were obtained from  Equation.~2 of \citet{Johnson2002}, following the procedure 
explained 
in \citet[K21]{Karinkuzhi2018}:
\begin{equation}
\begin{split}
\label{Eq:Johnson}
\sigma^{2}_{\rm tot}=\sigma^{2}_{\rm ran}
\;+\; \left(\frac{\partial \log\epsilon}{\partial T}\right)^{2}\sigma^{2}_{T}  \;+\; \left(\frac{\partial \log \epsilon}{\partial \log g}\right)^{2}\;\sigma^{2}_{\log g} 
\;+\; \left(\frac{\partial \log \epsilon}{\partial \xi }\right)^{2}\;\sigma^{2}_{\xi}\;+\;
\left(\frac{\partial \log\epsilon}{\partial  \mathrm{[Fe/H]}}\right)^{2}\sigma^{2}_{\mathrm{[Fe/H]}} \;+\; \\ 2\bigg [\left(\frac{\partial \log\epsilon}{\partial T}\right) \left(\frac{\partial \log \epsilon}{\partial \log g}\right) \sigma_{T \log g}
\;+\; \left(\frac{\partial \log\epsilon}{\partial \xi}\right) \left(\frac{\partial \log \epsilon}{\partial \log g}\right) \sigma_{\log g \xi} 
\;+\;\left(\frac{\partial \log\epsilon}{\partial \xi}\right) \left(\frac{\partial \log \epsilon}{\partial T}\right) \sigma_{ \xi T}\Bigg].
\end{split}
\end{equation}

In Equation~\ref{Eq:Johnson}, $\log \epsilon$ represents an elemental abundance, while $\sigma_{T}$, $\sigma_{\log g}$, and $\sigma_{\xi}$  are the typical uncertainties on the atmospheric parameters, which are estimated to be
$\sigma_{T}$ = 75~K, $\sigma_{\log g}$ = 0.2~dex, $\sigma_{\xi}$ = 0.05~km/s, and $\sigma_{\mathrm{[Fe/H]}}$ = 0.10~dex. The partial derivatives appearing in Equation~\ref{Eq:Johnson}
were determined in the particular cases of CS 22947-187  and
HD 224959, varying the atmospheric parameters $T_{\rm eff}$, $\log g$, microturbulence $\xi$, and [Fe/H] by 100~K, 0.5, 0.5~km/s, and 0.5 dex, respectively and the corresponding abundance changes are provided in Table ~\ref{Tab:uncertainties}. We adopt the covariances  $\sigma_{T \log g}$ = $-$0.5, $\sigma_{\log g \xi}$ = 0.02 and $\sigma_{\xi T}$ = 3, as measured by K21, since these cases are specifically found for HD 196944 and HD 198269. Similarly, here the objects, CS 22947$-$187 and HD 224959, also exhibit stellar parameters that are almost comparable to those of HD 196944 and HD 198269.  

We assumed a random error  $\sigma_{\rm ran} = 0.1$~dex if only one line is used to derive the abundance.  Otherwise we have calculated $\sigma_{\rm ran}$ 
as $\sigma_\mathrm{ran} = \sigma_{l}/N^{1/2}$, where $\sigma_{l}$ is the standard deviation of the abundances derived from the $N$ lines of the considered element. The final uncertainties are then calculated using Equation \ref{Eq:[X/Fe]} and are presented in Table ~\ref{Tab:abundances}. 
The final error in abundance ratios ($\sigma_{\rm [X,Fe]}$) is calculated using Equation 6 of \citet{Johnson2002}.
  \begin{equation}
\label{Eq:[X/Fe]}
\sigma^{2}_{\rm [X/Fe]} =\sigma^{2}_{\rm X} + \sigma^{2}_{\rm Fe} - 2\;\sigma_{\rm X,Fe},
\end{equation}
 Where $\sigma_{\rm X,Fe}$ is the covariance between two abundances (X and Fe), and  is given by: 
\begin{center}
\begin{equation}
\begin{split}
\label{Eq:covariance}
\sigma_{\rm X,Fe} = \left(\frac{\partial \log\epsilon_{X}}{\partial T}\right)\left(\frac{\partial \log\epsilon_{Fe}}{\partial T}\right)\sigma^{2}_{T} \;+\; 
\left(\frac{\partial \log\epsilon_{X}}{\partial \log g}\right)\left(\frac{\partial \log\epsilon_{Fe}}{\partial \log g}\right)\sigma^{2}_{\log g}  \;+\;\left(\frac{\partial \log\epsilon_{X}}{\partial \xi}\right)\left(\frac{\partial \log\epsilon_{Fe}}{\partial \xi}\right)\sigma^{2}_{\xi}  \;+\; \\
 \Bigg [\left(\frac{\partial \log\epsilon_{X}}{\partial T}\right) \left(\frac{\partial \log \epsilon_{Fe}}{\partial \log g}\right) \;+\; \left(\frac{\partial \log\epsilon_{X}}{\partial \log g}\right) \left(\frac{\partial \log \epsilon_{Fe}}{\partial T}\right) \Bigg] \sigma_{T \log g}
\;+\; 
 \Bigg [\left(\frac{\partial \log\epsilon_{X}}{\partial \xi}\right) \left(\frac{\partial \log \epsilon_{Fe}}{\partial \log g}\right) \;+\; \left(\frac{\partial \log\epsilon_{X}}{\partial \log g}\right) \left(\frac{\partial \log \epsilon_{Fe}}{\partial \xi}\right) \Bigg] \sigma_{\xi \log g}. \\
\end{split}
\end{equation}
\end{center}

{\bf \section{Classification} 
\label{Sect:Classification} }

Distinguishing between CEMP-s and CEMP-rs stars is challenging and may even be unrealistic, particularly if CEMP-rs stars originate from an i-process capable of producing a wide range of abundance profiles, from s-type to rs-type. However, we describe various indicators below and discuss their consistency, as well as cases where they yield ambiguous results.

\begin{table*}\small
\centering
\caption{Classification based on s- and r-elements.}

\label{Tab:Classification}
\begin{tabular}{l r r c c r c r c c r r c c }
\toprule
 Object& \multicolumn{3}{c}{[s/r]} && \multicolumn{4}{c}{Distance} & &  \multicolumn{3}{c}{$\chi^2$} & Final Class \\
        \cline{2-4} 
        \cline{6-9} 
        \cline{11-13} \rule{0pt}{2.5ex}
   &  [La/Eu] & [Ba/Eu] & Class$_{sr}$ && $d_s$ & Class$_{d_s}$& $d_{rms}$ & Class$_{d_{rms}}$ &&  $\chi^2_{\text{s}}$  & $\chi^2_{\text{i}}$   & Class$_{\chi^2}$&  \\
\midrule
CS 22891$-$171 & 0.55& 0.37 & s/rs      &&0.48 & rs& 0.73      &s/rs && 3.81 & 1.66 & rs & rs\\
CS 22947$-$187 & 0.33&0.46 & rs       && 0.43  &rs&  0.64   &rs && 5.78 &0.86 &rs & rs\\
HD 145777 & 0.10& 0.02 & rs         &&0.47 &rs& 0.57      & rs&& 2.01 &2.58&  s/rs & rs\\
HD 187861 & $-$0.05 &$-$0.16& r/rs     && 0.04 &r & 0.48     &rs &&6.54 & 1.78 & rs & rs\\
HD 196944 & $-$0.01 & 0.24& rs      && 0.39 &rs& 0.52     & rs&& 5.04 &1.60 & rs & rs\\
HD 224959 & 0.17& 0.19 & rs         && 0.35 &rs& 0.56     &rs&& 3.29 & 1.61 &  rs & rs\\
SDSS J0912$-$0216  &0.24 & $-$0.27& rs&& 0.52 &rs&  0.68    &s/rs&&19.87 & 8.29 & rs?  &  rs\\
SDSS J1349$-$0229 & 0.01 & $-$0.29& rs&& 0.40  &rs&  0.59   &rs && 25.81 & 13.63 & rs? & rs \\
SDSS J1036$+$1212 & 0.17 & $-$0.98& rs&& 0.38 &rs&  0.72    &s/rs&&27.79 & 8.17 & rs? &  rs\\
CS 22887$-$048 & 0.50& 0.42& s        && 0.63 &s/rs& 0.75     &s&& 3.69 & 1.30 &  rs & s or rs\\
CS 22942$-$019& 0.51& 0.79& s         && 0.91 &s& 1.07     &s&& 2.41 & 1.08 &  rs & s\\
CS 29512$-$073  & 0.58& 0.68& s       &&  0.77 &s& 0.90   &s&& 3.38 & 0.88 & rs & s\\
CS 30322$-$023 & 0.82& 0.64& s        && 0.83 &s& 0.97     &s&& 2.60 & 1.87 &  rs & s\\
HD 26 &  0.92& 0.94      & s        && 1.18 &s& 1.28     &s&& 2.09   & 1.88 &  s/rs &s \\ 
HD 55496 &... &...& ...                   && 1.00 &s& 1.22     & s   && 3.87 & 6.32 &  s & s\\
HD 198269 &0.55& 0.55 & s           && 0.83 &s&0.95      &s&& 1.10 &1.17 & s/rs & s  \\
HD 206983 & 0.30 & 0.27& rs         && 0.60 &s/rs&0.70      & s/rs&& 1.24 & 0.87 &  s/rs & s or rs\\
\bottomrule
\end{tabular}
\medskip\\
\tablecomments{The three [s/r] columns use both [La/Eu] and [Ba/Eu] to classify the star as s or rs. The four Distance columns refer to the distance to the r-process, as defined in the text. The star is classified as rs if $d_s \le 0.6$ (resp., $d_{rms} \le 0.7$).
The  $\chi^2_{\text{s}}$ and $\chi^2_{\text{i}}$ indicators evaluate how closely the measured abundances of elements ($Z > 30$) align with the best-fitting theoretical predictions produced by s-process and i-process AGB models, respectively. When the $\chi^2_{\rm s}$ and $\chi^2_{\rm i}$ are similar and reasonably low, the star is classified as `s/rs' (i.e., the class cannot be securely determined from comparisons to models). A question mark in the class$_{\chi^2}$ column highlights problematic stars with high $\chi^2_{\rm s}$  or $\chi^2_{\rm i}$ values (see text for details). 
The `Final Class' column represents the tentative best classification based on the values of various indicators used in this study ($d_{s} $, $d_{rms}$, $\chi^2_{\rm s}$ and $\chi^2_{\rm i}$).}

\end{table*}

\subsection{The [La/Eu] indicator}
 In Table~\ref{Tab:Classification}, the 
 Class$_{sr}$ column is based on the [La/Eu] (or [Ba/Eu]) abundance ratio. 
 With $X$ standing for La or Ba,
 stars with [X/Eu] $>$ 0.5 are categorized as CEMP-s, those with $0.0 \le$ [X/Eu] $\le 0.5$ as CEMP-rs, and those with [X/Eu] 
 clearly negative
 as CEMP-r. 
 Difficulties arise because the Ba and La diagnostics are sometimes conflicting (e.g., CS~22891$-$171).
 HD~196944, with [La/Eu]=-0.01 but [Ba/Eu]=0.24, is too close to the borderline to be considered as a CEMP-r star and is classified as CEMP-rs. 
 Such indicators are prone to errors because they are based on a small number of elemental abundances, sometimes only on two line measurements. In the following, indicators based on a larger number of lines and elements are considered. 

\subsection{The ``distance" indicator}
In an attempt to establish a model-independent classification, we compute the ``distance'' of the measured abundance distribution to the canonical Solar System r-process distribution. Although the r-process is not strictly universal, it is found to exhibit less variability (e.g., with metallicity) than the s-process and therefore serves as a more reliable reference point for quantifying abundance anomalies.
The  $d_s$  and $d_{rms}$  distances are detailed in Sect. 5 of K21 and are calculated using the element set: Y, Zr, Ba, La, Ce, Nd, Sm, and Eu. 
The limit between CEMP-s and CEMP-rs stars is, somewhat arbitrarily, placed at $d_{s} = 0.6$ ($d_{rms}=0.7$) to give consistent results.
The classification based on both distance indicators is listed in columns Class$_{d_s}$ and Class$_{d_{rms}}$ in Table~\ref{Tab:Classification}. 

\subsection{The $^{12}$C/$^{13}$C ratio}
In s-process AGB models (i.e., without PIE), a surface $^{12}$C/$^{13}$C ratio larger than typically $20-30$ is found during the AGB phase \citep[e.g. Figures~1 and 2 in][]{Choplin2024letter}. 
Following a PIE, this ratio is predicted to range between 3 and 9 \citep{iwamoto2004, Choplin2024b} with a potential increase up to around 20 if additional thermal pulses occur.

The nine programme stars classified as rs (last column of Table~\ref{Tab:Classification}) have\footnote{We considered the $^{12}$C/$^{13}$C ratios derived from the CN band.} $2.3<$~$^{12}$C/$^{13}$C~$<16$, with an average of 8.3. The two stars with an unclear classification (s or rs) have $^{12}$C/$^{13}$C$=12$ and $24$. Finally, the six stars classified as s have $5<$~$^{12}$C/$^{13}$C~$<16$, with a mean of 11.6.  
The low ratios found in rs stars are consistent with i-process model predictions.
Also, on average, rs-stars exhibit lower $^{12}$C/$^{13}$C ratios than s-stars, consistent with the lower values predicted by i-process models. 
Nevertheless, the ratios measured in s-stars are relatively low compared to those expected from s-process AGB models.
However, the companion star that accreted material from the AGB donor may have subsequently undergone the first dredge-up, a process known to decrease the $^{12}$C/$^{13}$C ratio \citep[e.g.][]{Charbonnel1994}. Among our programme stars, 13 out of 17 have $\log g < 3$, suggesting they may have experienced the first dredge-up. 
Due to the possible surface effects related to the first dredge-up, we do not consider here the measured $^{12}$C/$^{13}$C ratios as a robust indicator for distinguishing s from rs stars in our sample.

\section{Comparison with nucleosynthesis predictions}
\label{Sect:nucleosynthesis}

\subsection{AGB models and fitting procedure} 
\label{Sect:AGB-fitting}
The enrichment in heavy elements measured in CEMP-s stars is thought to originate from a now-extinct AGB companion that transferred material to the secondary star via stellar winds (e.g. \citet{Masseron2010} and references therein). The same scenario has been invoked to explain the heavy element enrichment in CEMP-rs stars \citep[e.g.][]{Lugaro2012, Bisterzo2012, Abate2013, Abate2015, Karinkuzhi2021,Karinkuzhi2023,Choplin2024}. 
To test this hypothesis, the current observations are compared with AGB nucleosynthesis predictions computed using the STAREVOL code \citep{Siess2000,Siess2008}. 

In our AGB models, the s-process (i-process) is followed by a network comprising 414 (1160) nuclei and linked by 639 (2123) reactions.  
During proton ingestion episodes (PIE), the nuclear burning and convective turnover timescales become comparable, so that the nucleosynthesis and transport equations are coupled. 
We considered AGB models with $M_{\rm ini} = $ 1 and 2~\Msun\ and $-3\leq$~[Fe/H]~$\leq-1$.  
Among these are: (1) s-process models, representing AGB stars that undergo only s-process nucleosynthesis (2~\Msun\ at [Fe/H]~$=-1$, $-2$, $-2.5$ and $-3$), and (2) i-process models, representing AGB stars that experience i-process nucleosynthesis (1 and 2~\Msun\ at $-3\leq$~[Fe/H]~$\leq-1$). While the i-process is thought to occur preferentially in low-mass, low-metallicity stars \citep[e.g. Figure~3 in][]{Choplin2022}, it remains unclear which AGB stars undergo s-process, i-process, or both types of nucleosynthesis.
In some cases, varying the mixing parameters within the same stellar model can lead to either s- or i-process nucleosynthesis.
Therefore, both s- and i-process models were sometimes considered at a given mass and metallicity.
Different overshooting strengths $f_{\rm top}$ above the convective thermal pulse were explored in i-process models. As discussed in Sect.3.2 of \cite{Choplin2024} for a 1 \Msun\ AGB model at [Fe/H]~$=-2.5$, increasing the value of $f_{\rm top}$ enhances the abundances of elements with $36 \leq Z \leq 56$ by up to $\sim$~1 dex, while simultaneously reducing the abundances of heavier elements with $Z \geq 82$ by a similar amount \citep[see also Fig.~4a in][]{Choplin2024}.
Further details about the AGB models are provided in \cite{Goriely18c,Karinkuzhi2021,choplin2021,Choplin2022,Choplin2024}.

\begin{figure}
\centering
\plotone{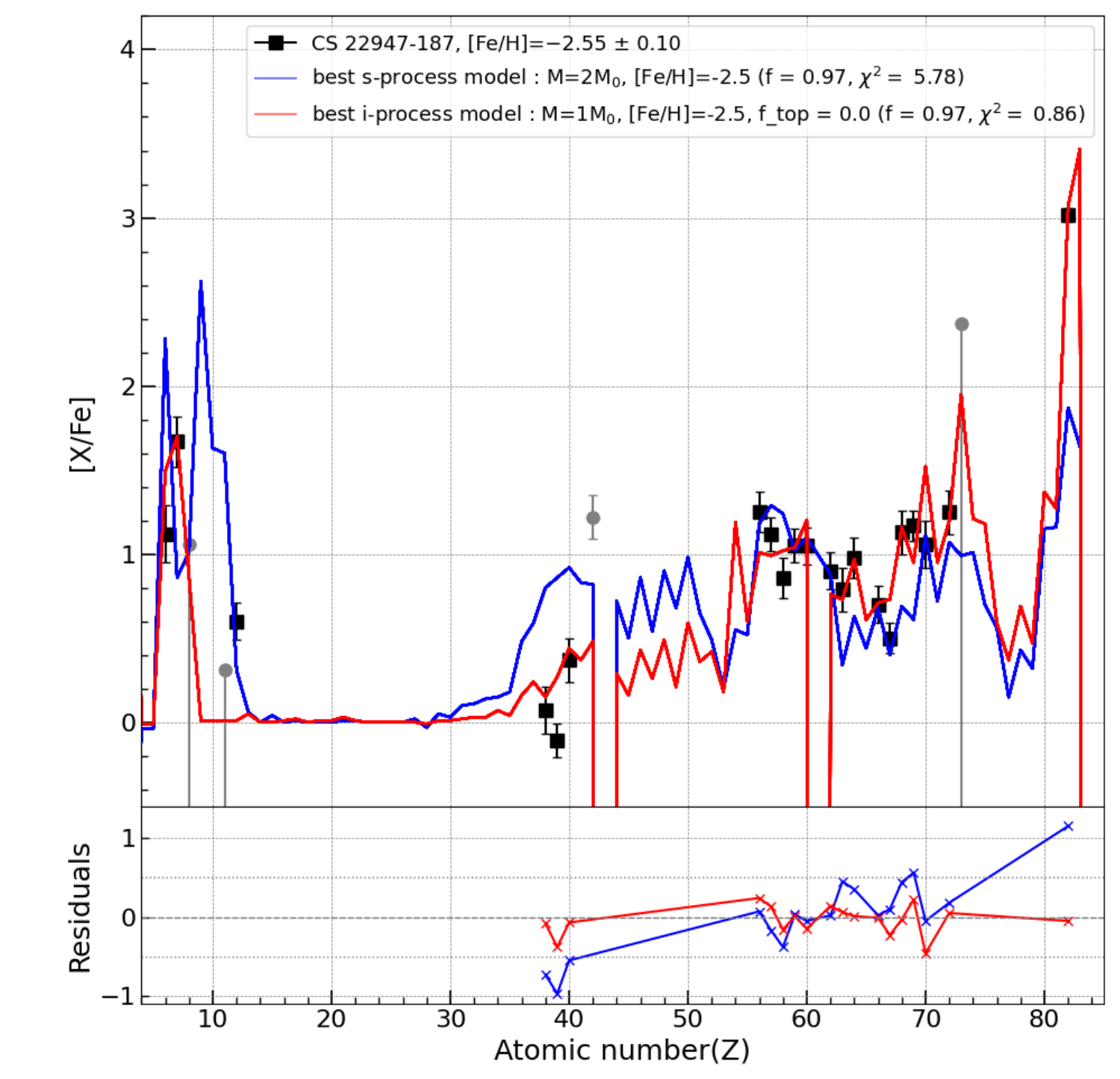}
\caption{The abundance patterns of the 17 CEMP stars are compared with nucleosynthesis predictions from the STAREVOL code. The measured abundances are indicated by black squares, uncertain abundances by grey circles, and upper limits by grey circles with a downward arrow. In all cases, the best-fitting theoretical predictions for both the s-process (blue) and i-process (red) are displayed. Only one star is displayed here for the reference; the rest are in Appendix B. Models are described in Sect.~\ref{Sect:nucleosynthesis}.}
\label{fig:pattern}
\end{figure}

To fit the chemical abundances of our sample stars with our AGB models, we follow the procedure outlined in Sect.~6.2 of \cite{choplin2021}. For each sample star, the best-fitting AGB nucleosynthesis prediction is identified by minimizing the $\chi^2$ value, representing the quality of the fit \cite[see Equation~7 in][]{choplin2021}. The minimal $\chi^2$ value (representing the difference between the measured and theoretical chemical abundances) is searched by mixing some of the material ejected by our AGB models in the envelope of the companion. During the fitting procedure, the [X/Fe] ratios in our models are obtained by summing the isotopic abundances. The abundance $X_i$ of each isotope $i$ is computed according to the relation
\begin{equation}
X_{i} =  (1-f) \, X_{s} + f \, X_{\rm ini}
\label{eq_abdil}
\end{equation}
where $X_s$ and $X_{\rm ini}$ are the surface and initial mass fractions of isotope $i$, respectively (available since individual isotope abundances are followed in our AGB models). The dilution factor $f$ is a free parameter, varied between 0 and 1 so as to minimize the $\chi^2$ value. It controls the amount of AGB ejecta mixed into the companion envelope.

Only elements heavier than Zn ($Z=30$) are considered in the $\chi^2$ calculation. 
Elements with uncertain abundance determinations, including upper limits (listed in Table~\ref{Tab:abundances} and reported in Figure~\ref{fig:pattern} by grey symbols) are excluded from the $\chi^2$ computation. 
We note that the number of reliably determined chemical abundances varies from star to star, consequently affecting the number of elements included in the $\chi^2$ calculation.
The best fitting s-process (blue pattern) and i-process  (red pattern) models are illustrated in Figure~\ref{fig:pattern} for one representative star and most extensively in Figure~\ref{fig:appendix-figure} for the full sample.

\section{Classification discussion}
\label{Sect:Classification discussion}

 We now compare the measured abundance profile with the one predicted from a given nucleosynthesis model (s- or i-process), for example with a $\chi^2$ statistics.

 First of all, we make the assumption that the secondary star initially has a solar-scaled abundance pattern, i.e., is not enriched in heavy elements, although chemical inhomogeneities are expected in the Galactic halo \citep[e.g. ][]{Cescutti-2008, Wehmeyer-2015}. Actually, there is no requirement for the secondary star to follow a solar-scaled abundance pattern at the metallicities of our targets. 
However, this assumption is legitimate for two reasons.
First, we focus on polluted stars that typically exhibit significant heavy element overabundances (+1 to +2 dex). We therefore assume that the pollution dominates any pre-existing, possibly non-solar, abundance patterns.
Second, and more importantly, we fit the abundances of the mainly s-process elements using either an s- or i-process model. Since the onset of s-process enrichment in the Galaxy occurs relatively late, it is reasonable to assume that the secondary star was not initially enriched in s-process elements. We then examine how well this fit reproduces the measured abundances of the mainly r-process elements.

$\chi_s^2$ (resp., $\chi_i^2$) values are computed from the best-fitting s- or i-process models (see Sect.~\ref{Sect:AGB-fitting}), from all available chemical element abundances above Zn after excluding the abundances which are uncertain and have only upper limits, and listed in Table~\ref{Tab:Classification}. 
The s- or i-processes in our AGB models provide a reasonable description of the abundance distributions for most of the programme stars (Figures~\ref{fig:pattern} and \ref{fig:appendix-figure}).
The resulting classification is listed in the Class$_{\chi^2}$ column in Table~\ref{Tab:Classification}.
Among the 17 CEMP stars analyzed, 6, 12 and 14 stars have $\chi_s^2$ or $\chi_i^2 \leq 1.5$, 2, and 4 respectively (Table~\ref{Tab:Classification}).

Except for the two objects CS 22887$-$048 and HD~206983, all objects in common with K21 remain in the same category.
In other words, the newly determined elemental abundances for most stars confirm their K21 CEMP classification. HD 206983 was classified as s in K21; however its low [La/Eu] and [Ba/Eu] are the main indications of a possible rs assignation. The distance and the $\chi^2$ indicators are rather inconclusive for this star. Similar to the case for CS22887$-$048, it was also classified as s in K21 based on the distance measurements. In the present analysis, the d$_s$ and $\chi^2$ measurements put it in the rs class while [La/Eu] and d$_{rms}$ values 
classify it among the s stars. 

 One star in particular shows a large dispersion in its light s element abundances: SDSS J1036+1212 has [Sr/Ba] = -1.77 (NLTE) or -1.63 (LTE). Together with [Ba/Fe]=1.28, this locates it close to the locus of CEMP-rs stars in the diagram [ls/hs] (light s over heavy s) as a function of [hs/Fe]: for example, [Sr/Ba] = f([Ba/Fe]) diagram \citep{Hansen-2019}. Sr is low in SDSS J1036+1212, as well as other light s elements (e.g. [Y/Fe]= 0.87(LTE) to 0.94 (NLTE), though uncertain). However, another ls element shows a much larger abundance: [Zr/Fe] = 1.60, which would place the star among CEMP-s stars and might explain why this star was originally categorized as CEMP-s. However, the other diagnostics, like the [Ba/Eu] or [La/Eu] ratio, and the $d_{\rm s}$ distance (using 8 different heavy element abundances) favor the CEMP-rs classification (the d$_{\rm rms}$ distance is intermediate and inconclusive).  Finally, the $\chi^2$ is very large for the i-process, but even  larger for the s-process.
Therefore we decided to change this object from the CEMP-s class to the CEMP-rs class.

The 10 stars classified as CEMP-rs with the $d_s$ distance are consistently better fitted (i.e. lower $\chi^2$) by the i-process models, except HD~145777 which is nearly equally well fitted by the s- ($\chi_{\rm s}^2 = 2.01$) and i-process ($\chi_{\rm i}^2 = 2.58$) models (Table~\ref{Tab:Classification}).
Their residuals (measured-predicted abundances) typically fall within the range of  $0.5 - 1$~dex (Figures~\ref{fig:pattern} and \ref{fig:appendix-figure}). This is acceptable given that nuclear uncertainties in i-process AGB models are of the same order, i.e. $0.5 - 1$~dex \citep{Goriely2021, Martinet2024}.  However, the three SDSS stars exhibit a few elements with very high abundances that are challenging to reproduce with our i-process models, the most extreme case being Pr ($Z=59$), which shows an overabundance of 3.8~dex in SDSS~J1349$-$0229.
Higher [X/Fe] ratios can be achieved if the PIE occurs when the remaining AGB convective envelope is smaller. As a test, we allowed for dilution in a smaller AGB envelope mass for SDSS~J1349$-$0229 to mimic a late PIE, possibly taking place near the end of the AGB phase or during the post-AGB phase \citep[e.g.,][]{Herwig11, DeSmedt2012}.
In this case, the best-fitting i-process model yields $\chi_i^2 = 9.2$ (compared to 13.6 when $f=0$) and successfully reproduces most elemental abundances, with a few notable exceptions such as Ba ($Z=56$) and Pr ($Z=59$, Figure~\ref{fig:appendix-figure}, red dashed line labeled ``late TP"). Notably, the model accounts for the high Ta ($Z=73$) abundance of approximately 4~dex. Ta is a crucial element for distinguishing between the s- and i-processes, as its production is significantly higher during i-process nucleosynthesis.

 In the present study, most of the best-fitting models yield high dilution factors $f$ (cf. Eq.~\ref{eq_abdil}). However, a few stars\footnote{CS22887-048, HD26, HD224959 and the three SDSS stars for s-process; SDSS J1036 and SDSS J1349 for i-process} are better matched by models with low $f$ values. 
As discussed in \cite{Choplin21} and \cite{Choplin22cor}, low dilution factors sometimes imply unrealistically high accreted masses, which may be difficult to reconcile with binary evolution models.
These low-dilution cases may therefore raise concerns regarding the required accreted mass. A detailed analysis of this issue is beyond the scope of this paper, but we plan to investigate it in future work.

Surprisingly, 3 out of the 6 stars (CS~22942$-$019, CS~29512$-$073, and CS~30322$-$023) classified as s with the $d_s$ distance are better fitted by i-process models 
(we exclude HD 26 because its $\chi^2_s$ and $\chi^2_i$ are very close, making its $\chi^2$ classification uncertain). These stars have $0.77 \le d_s \le 0.91$, close to the threshold of 0.6 that separates CEMP-rs from CEMP-s stars.
For instance, the abundance pattern of two of these stars (CS~29512$-$073 and CS~22942$-$019) is better reproduced by our i-process models, particularly for elements with 
$38 \leq Z \leq 42$ and $Z=82$ for CS~29512$-$073. 
CS22887$-$048 is another borderline case, with the $d_s$ and $d_{rms}$ indicators in the borderline region. However, its significant Ta overabundance of $\sim 3$~dex (excluded from the $\chi^2$ calculation due to its uncertainty) is reproduced by the i-process model within the error limits but markedly underproduced by the s-process model. A more detailed analysis of Ta in this star could provide a clearer classification. We note that considering additional s-process AGB models with different initial assumptions (e.g., initial mass) could also potentially reconcile these stars with s-process predictions.

\section{Comparison of the r-process abundances in different CEMP classes}
\label{Sect:Comparison of the heavy r-process}

It is important to emphasize that the designation ``r-process elements" is inherited from their origin in the solar system, where they are predominantly attributed to r-process nucleosynthesis. However, in the case of the present paper, as far as CEMP-s and CEMP-rs stars are concerned, these elements are more accurately interpreted as originating from the tail of the s- or i-process abundance distributions. Indeed, even a standard radiative s-process can produce up to ~5\% of europium, with similarly small yet non-negligible contributions to other elements typically associated with the r-process. In the following, the r-process elements produced by the s (resp., i) process should be understood as ``s- (resp., i-) process-made r-elements" (and denoted as $r_s$ or $r_i$ elements for brevity).
This nuance is critical when interpreting the measured abundance patterns in these stars.

We now discuss the abundance trends displayed by the different elements.

\begin{figure}
\plotone{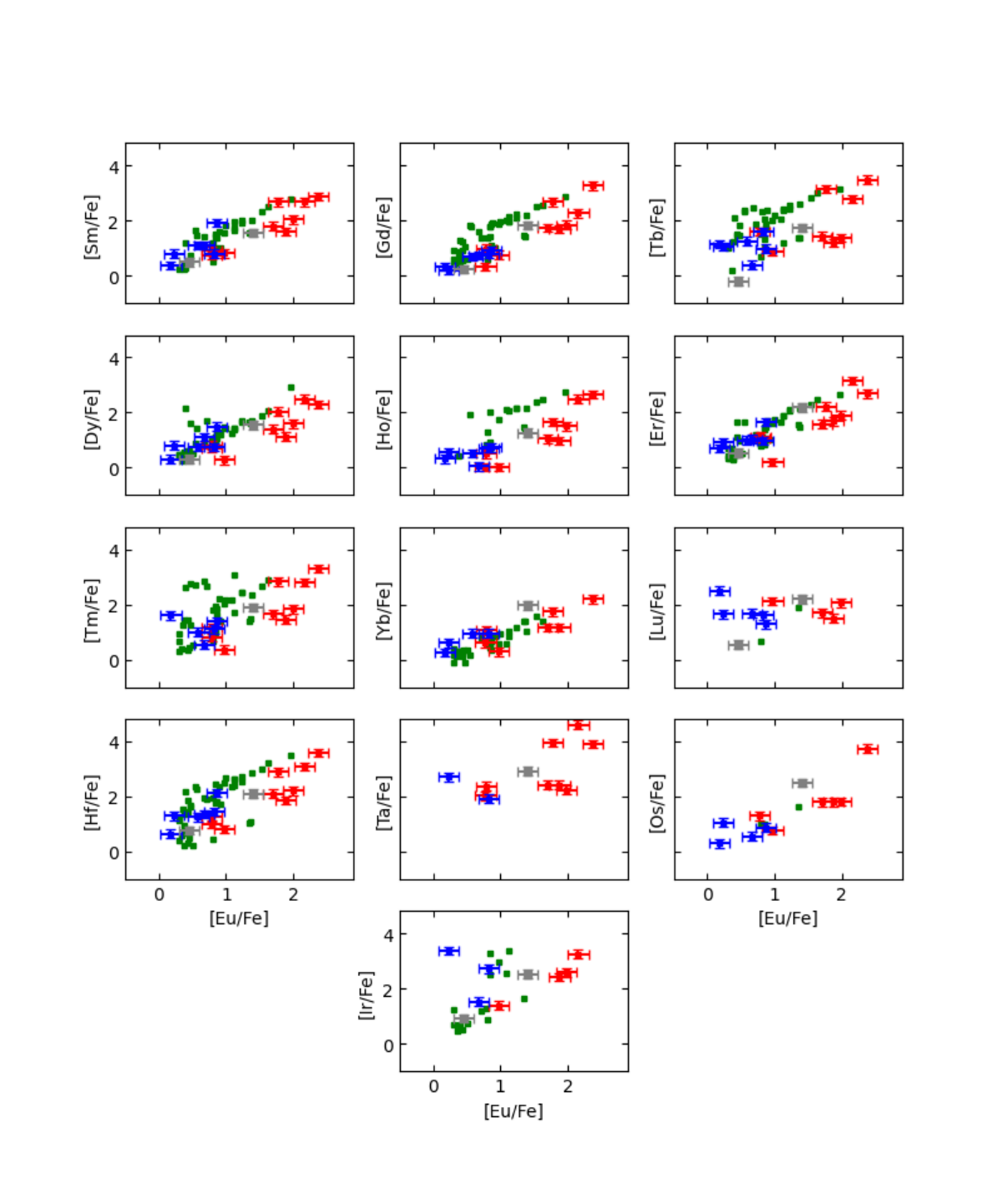}
\caption{[X/Fe] as a function of [Eu/Fe] for CEMP-rs stars (red squares), CEMP-s stars (blue squares), 
CEMP stars of uncertain (r or rs) classification (grey squares), and 47 r-process enriched stars (some of them are carbon-enriched) (green squares) from \cite{Roederer2014,Ivans2006,Westin2000,Sneden2003}. All abundances, including upper limits and uncertain values, were considered. }
\label{Fig:Eu/Fe}
\end{figure}

In Figure~\ref{Fig:Eu/Fe}, we examine the correlations among r-elements (i.e., predominantly attributed to r-process nucleosynthesis in the solar system) in CEMP-rs stars, in CEMP-s stars as well as in a sample of r-process enriched stars from the literature \citep{Roederer2014,Ivans2006,Westin2000,Sneden2003}. The r-process enriched stars considered here are r-I and r-II stars, classified following \citet{Beer2005}, with [Eu/Fe] between +0.3 and +1.0 and [Ba/Eu] $<$ 0 for r-I, and [Eu/Fe] $>$ +1.0 and [Ba/Eu] $<$ 0 for r-II. Their metallicities ($-4\leq\mathrm{[Fe/H]}\leq-1$) are similar to those of our CEMP-s and CEMP-rs stars.
As expected, most r-elements exhibit a strong correlation with europium, consistent with the hypothesis that they originate from the same astrophysical site. 
The observed scatter may primarily reflect abundance measurement uncertainties. Notably, the correlation appears weaker for Lu and Ir, which could indicate larger abundance errors for these elements. 

We note that in this figure, it is expected to observe an overlap between CEMP-rs and CEMP-s stars. This is because the level of enrichment ([r/Fe]) is independent of the star’s classification as CEMP-s or CEMP-rs, which is originally based on abundance ratios such as [La/Eu] or [Ba/Eu] (i.e., [s/r]). For instance, in Fig.~\ref{Fig:Eu/Fe}, two CEMP stars are found at very low [r/Fe] enrichment: CS 22942$-$019 and HD 145777. Nevertheless, CS 22942$-$019 is classified as a CEMP-s star, as together with its relatively low [Eu/Fe] = 0.83, it exhibits a high [Ba/Fe] = 1.62. In contrast, HD 145777 is categorized as a CEMP-rs star, with [Eu/Fe] = 0.97 and [Ba/Fe] = 0.99.  

\begin{figure}
\plotone{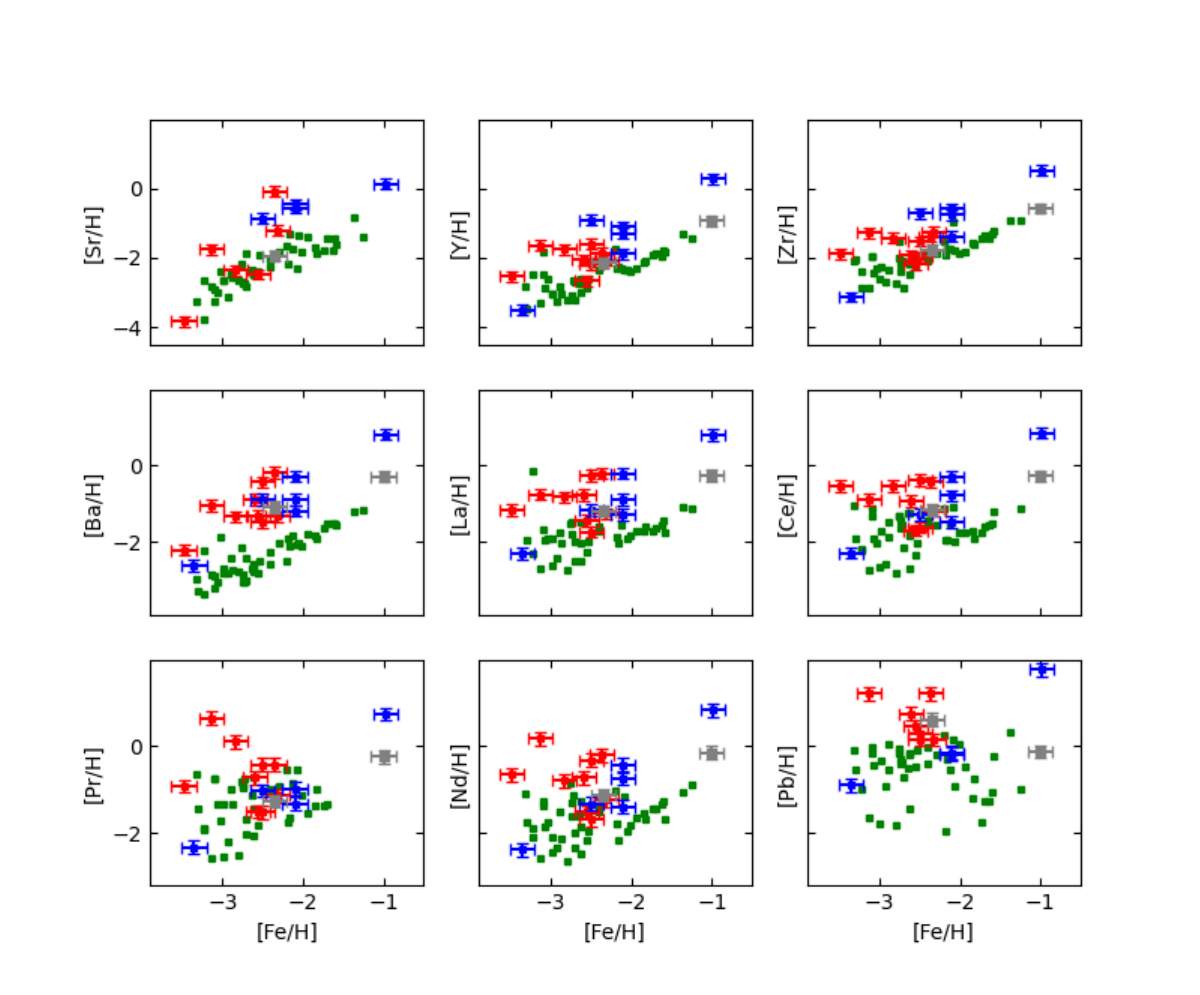}
\caption{The Enrichment in mainly-s elements [s/H] as a function of [Fe/H]. The red, blue, grey, and green squares have the same meaning as in Figure~\ref{Fig:Eu/Fe}.}
\label{Fig:s/H}
\end{figure}

\begin{figure}
\plotone{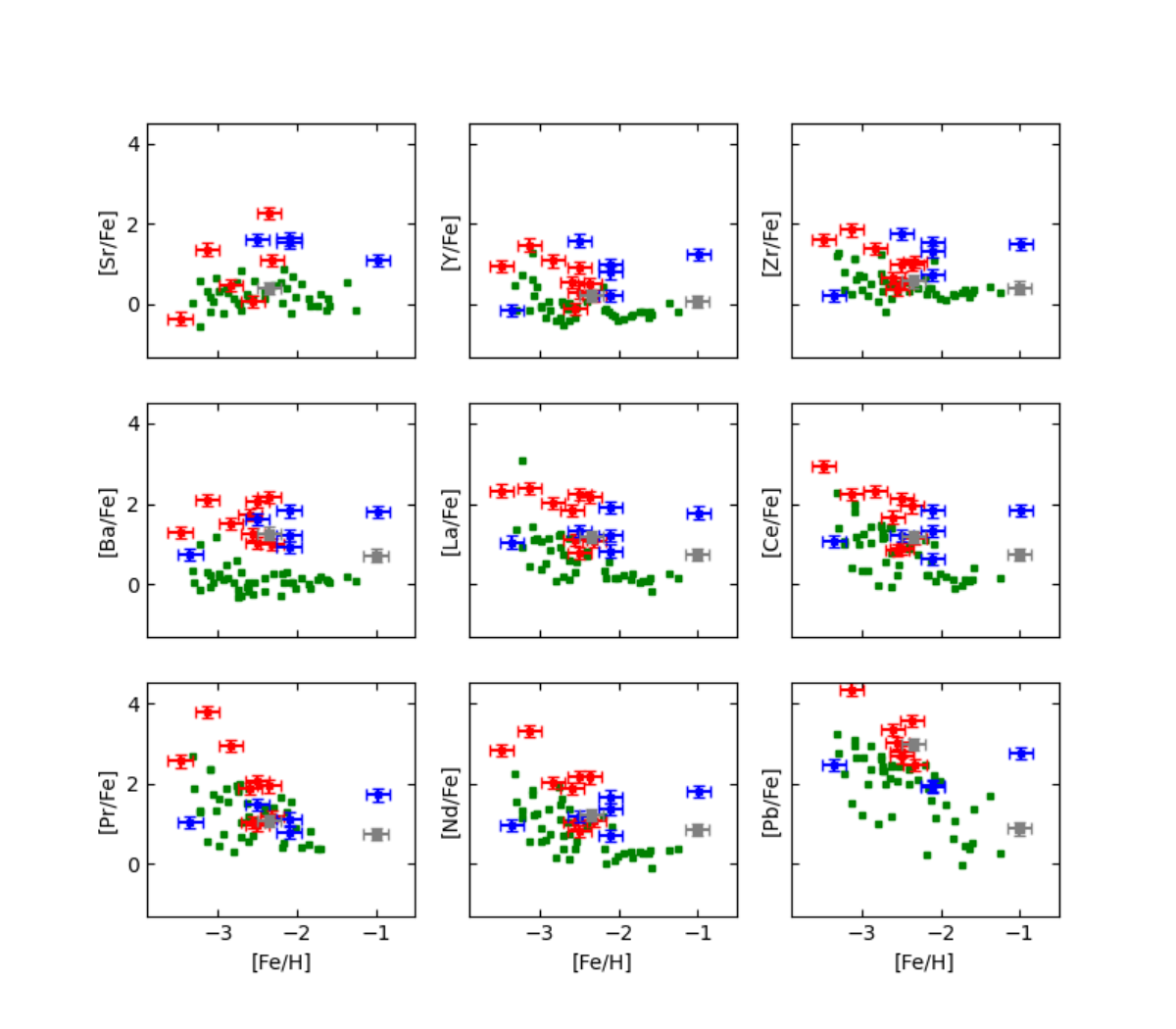}
\caption{The Enrichment in mainly-s elements [s/Fe] as a function of [Fe/H]. The red, blue, grey, and green squares have the same meaning as in Figure~\ref{Fig:Eu/Fe}.}
\label{Fig:s/Fe}
\end{figure}

\begin{figure}
\plotone{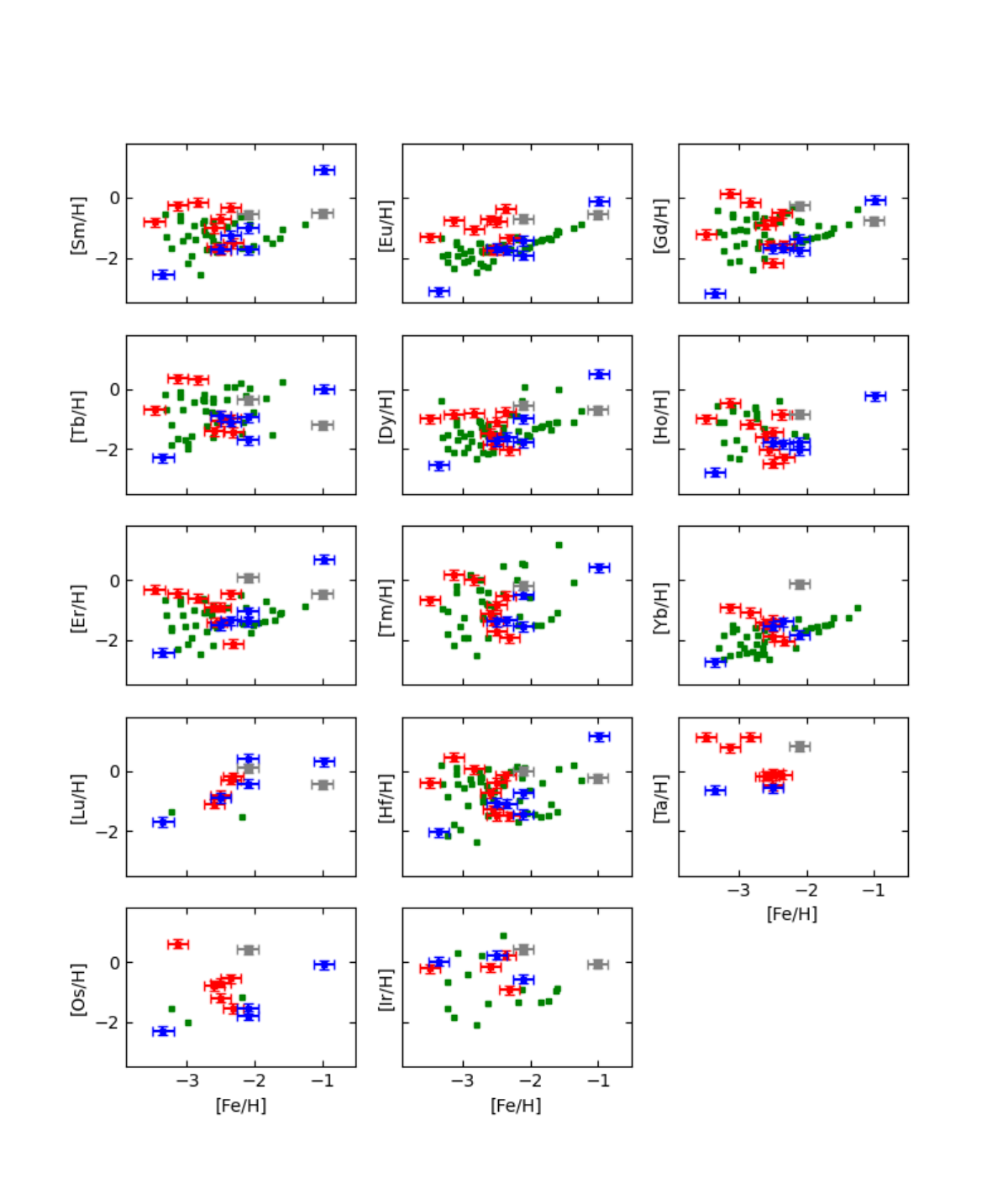}
\caption{[X/H] as a function of [Fe/H] for mainly-r elements. The red, blue, grey, and green squares have the same meaning as in Figure~\ref{Fig:Eu/Fe}.}
\label{Fig:r/H}
\end{figure}

\begin{figure}
\plotone{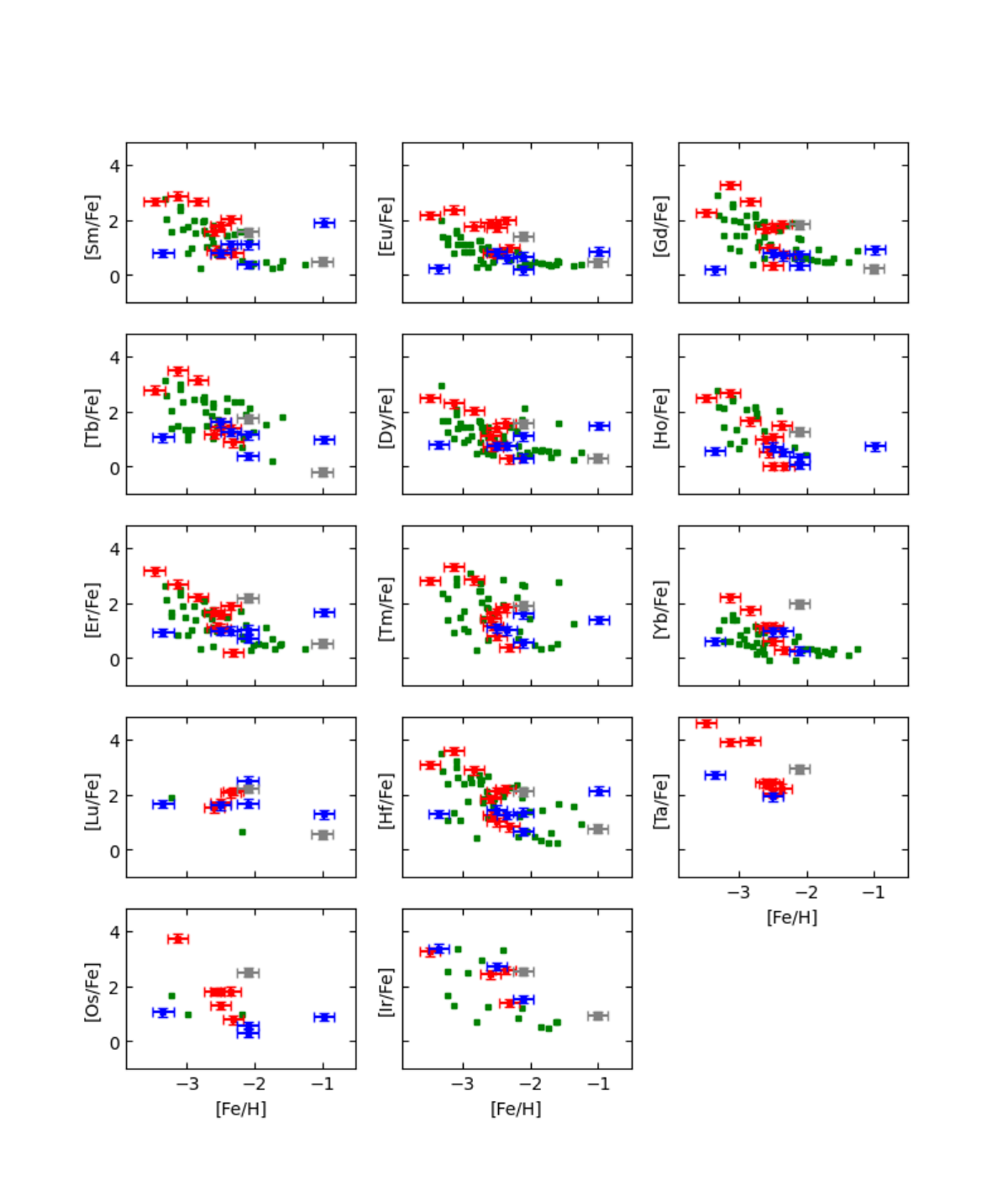}
\caption{[X/Fe] as a function of [Fe/H] for mainly-r  elements. The red, blue, grey, and green squares have the same meaning as in Figure~\ref{Fig:Eu/Fe}.}
\label{Fig:r/Fe}
\end{figure}

We now examine the evolution of the mainly-s and mainly heavy-r abundances as a function of metallicity
(Figures~\ref{Fig:s/H} and \ref{Fig:r/H}).  
For mainly s-elements (Figure \ref{Fig:s/H}), there is a nice trend of increasing [X/H] with increasing metallicity for CEMP-s and r-process enriched (some of them are carbon-enriched also)
stars. For the r-process enriched stars, this either reflects the contribution of AGB stars,
or the (small) r-process contribution to s-process elements. Rotating massive stars are also potential contributors to the s-process, particularly at low metallicity \citep[e.g.][]{Frischknecht2016}. In our sample however, lead, when measured, is found to be highly overabundant. This is difficult to reconcile with yields from rotating massive stars, which are generally not expected to produce significant amounts of Pb. That said, in stars where lead could not be measured, a contribution from rotating massive stars cannot be ruled out. In CEMP-r stars, the mainly-s elements should therefore reflect the galactic chemical evolution. Indeed, in Figure~\ref{Fig:s/Fe}, it can be seen that [X/Fe] (in particular when X = Ce) stays roughly solar from metallicity values between [Fe/H] = $-$2.0 and $-$1.0. This is well in line with \citet{Cunha-2017}, who find a constant [Ce/Fe] in field stars from metallicity values between [Fe/H]$=-1.0 $ and 0.0.

CEMP-s stars, obviously, have higher [s/H] abundances than CEMP-r stars. This can be attributed to the nature of the polluting star and to a smaller dilution, since CEMP-s stars have been polluted (after their formation) by a nearby AGB, whereas CEMP-r stars owe their heavy elements to protostellar clouds enriched by one or several events. 

However, for CEMP-rs stars, the situation is not as clear.
This is further supported by very dispersed Spearman correlation coefficient values between $-$0.68 (for Pb) and 0.58 (for Sr). The main conclusion is that, based on their dominant s-process elements, CEMP-rs stars form a group distinct from both r-process enriched and CEMP-s stars. They occupy a region of higher [s/H] than r-process enriched stars and lower metallicity than CEMP-s stars. However, the extent to which selection biases (i.e., the identification of a star as CEMP-rs) influence this distinction remains to be assessed.

Similar trends are visible for mainly-r elements (Figure \ref{Fig:r/H}),  in CEMP-s, -rs, and -r stars. 
For r-process enriched and CEMP-s stars, there is again, for mostly-r elements, a nice trend of increasing [X/H] with increasing [Fe/H].
This reflects the contribution of r-process production events (kilonovae, collapsars and MHD SNe) to r-process enriched stars. For CEMP-s stars, it might reflect the r-process contribution to r-process elements, which are present with small abundances in CEMP-s stars.

Again, for CEMP-rs stars, the situation is not that clear: they might show a trend similar to that of r-process enriched stars (which would favour an r+s explanation for CEMP-rs stars) or define a different sequence (in which case, their abundance pattern would be due to an event different from the one causing the s- or i abundance patterns, and also distinct from the one causing the r-pattern).
The Spearman correlation coefficient values are similar for CEMP-s and CEMP-r
stars, while the negative values in CEMP-rs stars indicate an anti-correlation with metallicity. 

In summary, in CEMP-rs stars, the [r/H] ratio appears to increase as metallicity decreases.
In contrast, the increase in  [r/H] values with metallicity in CEMP-s and r-process enriched stars could be due to the galactic chemical evolution. 

Figure~\ref{Fig:r/Fe} is well in line with Figure~14 of \cite{Sneden2008} and Figure~5 of \cite{Hansen-2018}, displaying [Eu/Fe] as a function of [Fe/H].
A large scatter in [Eu/Fe] abundances is observed at low metallicities. \citet[][and references therein]{Sneden2008}  suggested it reflects a chemically inhomogeneous early Galaxy. At those early times, enrichment was likely dominated by individual nucleosynthetic events (e.g., supernovae, kilonovae), whose yields were unevenly distributed. As metallicity increases, corresponding to later stages of Galactic evolution, this scatter diminishes, presumably due to the progressive chemical mixing of the interstellar medium.
\begin{figure}
\plotone{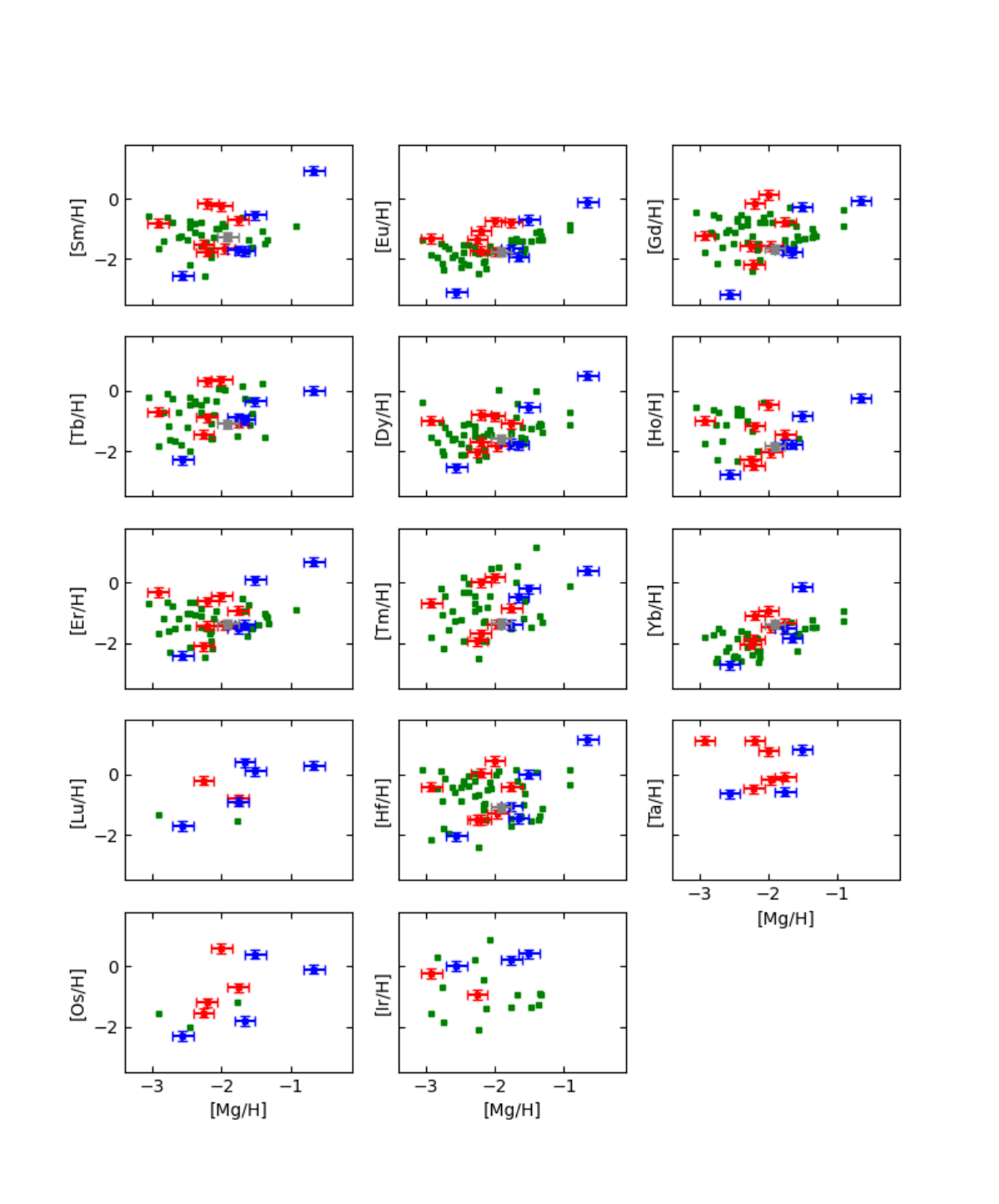}
\caption{[X/H] as a function of [Mg/H] for r-process elements. The red, blue, grey, and green squares have the same meaning as in Figure~\ref{Fig:Eu/Fe}.}
\label{Fig:Mg/H}
\end{figure}
In Figure~\ref{Fig:Mg/H}, we present the abundances of elements traditionally classified as r-process products as a function of [Mg/H].  The models predict that magnesium is produced in AGB stars with initial masses of 2–3~M$_\odot$, but not in stars of 1~M$_\odot$, at the metallicities considered. In contrast, the i-process is found to operate efficiently in 1~M$_\odot$ models, but not in higher-mass stars. Thus, the predictions suggest that low-mass (1~M$_\odot$) AGB stars can undergo i-process nucleosynthesis without producing significant amounts of Mg, whereas higher-mass stars primarily undergo s-process nucleosynthesis, accompanied by Mg production. This trend is consistent with the observational data shown in Figure~\ref{Fig:Mg/H}: among CEMP-s stars, a positive correlation is observed between [Mg/H] and the abundances of heavy neutron-capture elements (produced by the s-process). In contrast, no such correlation is observed in the presumably lower-mass CEMP-rs stars, which are expected to have experienced the i-process but not Mg production. 
This explanation has to be nuanced by the fact that the stellar evolution and nucleosynthesis models do not take into account any $\alpha$-enrichment at low metallicity. But at a given metallicity, such an $\alpha$ enrichment should affect CEMP-s and -rs stars in the same way. Since in our sample the metallicity of CEMP-rs stars is, on average, lower than that of CEMP-s stars,  this should increase the Mg abundance in CEMP-rs stars, which is not observed.
So, neglecting the $\alpha$ enrichment cannot be the cause of the positive correlation (resp. the absence of correlation) between Mg and heavy neutron capture elements in CEMP-s stars (resp., CEMP-rs stars). 
\section{Conclusions}
\label{Sect: conclusion}
We used high-resolution UVES spectra for seventeen CEMP stars to derive abundances for n-capture elements, focusing on heavy r-elements. 

The stars are classified into different CEMP classes with updated abundances, including the NLTE corrections. Several classification indicators were used, in addition to the classical [La/Eu].
In particular, in an attempt to base the classification on a larger number of chemical elements, the model-independent `abundance distances' $d_s$ and $d_{\rm rms}$ suggested by K21 were computed.
The agreement with s- and i-process nucleosynthesis models was also quantified ($\chi^2_s$ and $\chi^2_i$).

All of the 8 stars initially classified as CEMP-rs stars are confirmed to belong to the CEMP-rs category. Among the 9 stars initially classified as CEMP-s stars, 6 are confirmed and 2 are classified as ``s or rs", essentially because of 
$d_s$ and  $d_{rms}$ distances close to the threshold values $0.6$ and $0.7$, resp.
In a few stars Tantalum could be tentatively measured, these measurements constitute strong constraints on the operation of the i-process. Additionally, the star SDSS J1036+1212, which was initially classified as a CEMP-s star, is now reassigned to the CEMP-rs category, based on its lower [s/r] ratio and $d_s$ value.
The systematic comparison of 
 measured abundances with nucleosynthesis model predictions, supported by model-independent abundance-distance diagnostics, provides a more nuanced classification and highlights the need for a multi-element approach in disentangling the complex chemical signatures of these stars. 

Despite this progress, several open questions remain. What are the precise stellar conditions, such as neutron densities, mixing mechanisms, and progenitor masses, that lead to efficient i-process nucleosynthesis in AGB stars? How does the i-process varies with metallicity? Furthermore, the detection of elements like Ta, which are rarely measured but sensitive to neutron-capture conditions, raises the question of whether additional, presently unmeasured elements could offer even more definitive diagnostics of the i-process.

To address these questions, further high-resolution spectroscopic observations of CEMP stars of various metallicities are crucial. In parallel, advancements in i-process modeling, including hydrodynamical simulations, are needed to fully capture the complexity of the nucleosynthetic environments involved. Expanding the sample of well-characterized CEMP stars with accurate NLTE corrections and full n-capture element coverage will be essential for refining our understanding of the i-process and its astrophysical sites.


\begin{acknowledgments}
We thank the referee for the insightful remarks, which have greatly contributed to the improvement of the paper.
MR acknowledges the financial support from the UGC, Govt. of India through the UGC-JRF (NTA Ref.No:201610156431/CSIR-UGC NET NOVEMBER 2020). DK acknowledges the financial support from ANRF through the SURE grant with the file number (SUR/2022/000748). MR and DK gratefully acknowledge financial support from Belgium - India project on Precision Astronomical spectroscopy for Stellar and Solar system bodies (BIPASS), approved by the International Division, Department of Science and Technology (DST, Govt. of India; DST/INT/BELG/P-01/2021 (G)) and the Belgian Federal Science Policy Office (BELSPO, Govt. of Belgium; BL/33/IN22-BIPASS).  
A.C. is a Postdoctoral Researcher of the Fonds de la Recherche Scientifique – FNRS. This work is based on observations collected at the European Southern Observatory under ESO programme(s) 076.D-0451 and 078.D-0217, obtained from the ESO Science Archive Facility.
\end{acknowledgments}

\facilities{VLT(UVES)}


\appendix


\begin{figure*}
\centering

\hfill
\begin{minipage}[t]{0.20\textwidth}
\section{Line list}
\label{Tab:Linelists}
\begin{tabular}{l@{\hspace{8pt}}l@{\hspace{2pt}}r@{\hspace{2pt}}}

\toprule
$\lambda$ (\AA) & $\chi_{\rm low}$  & $\log gf$ \\
 & (eV) &      \\
\midrule 
Na I & & \\
5682.633 & 2.102 & -0.706 \\
5688.205 & 2.104 & -0.450 \\
5889.951 & 0.000 &  0.108\\
5895.924 & 0.000 & -0.144\\
\hline
Mg I & & \\
5528.405 & 4.346 & -0.620 \\
5711.088 & 4.346 & -1.833 \\
\hline
Ca I & & \\
5581.965 & 2.523 & -0.555\\
5588.749 & 2.526 & 0.358\\
5590.114 & 2.521 & -0.571\\
5594.462 & 2.523 & 0.097\\
5598.480 & 2.521 & -0.087\\
6102.723 & 1.879 & -0.793\\
6122.217 & 1.886 & -0.316\\
6162.173 & 1.899 & -0.090\\
6169.563 & 2.526 & -0.478\\
\hline
Sc  II & & \\
3576.340 & 0.008 &  0.007\\
3613.193 & 7.877 & -2.174\\
3630.844 & 7.877 & -1.402\\
4246.822 & 0.315  & 0.242\\
5526.790 & 1.768 &  0.024\\
5641.001 & 1.500 & -1.131\\
5657.896 & 1.507 & -0.603\\
\hline
Ti  I & & \\
5210.385 & 0.048 & -0.884\\
Ti  II & & \\
4411.925 & 1.224 &-2.520\\
4417.714 & 1.165 & -1.190\\
4418.330 & 1.237 & -1.970\\
4441.729 & 1.180 & -2.330\\
4450.482 & 1.084 & -1.520\\
4464.449 & 1.161 & -1.810\\
4468.507 & 1.131 & -0.600\\
4488.325 & 3.124 & -0.510\\
4501.270 & 1.116 & -0.770\\
5211.530 & 2.590 & -1.160\\
5226.538 & 1.566 & -1.260\\
5336.771 & 1.582 & -1.590\\
5381.015 & 1.566 & -1.920\\
\hline
Cr  I & & \\
5206.037 & 0.941 &  0.020\\
5208.425 & 0.941 &  0.170\\
5296.691 & 0.983 & -1.360\\
5298.272 & 0.983 & -1.140\\
5345.796 & 1.004 & -0.980\\
5348.315 & 1.004 & -1.290\\  
\hline\\

\end{tabular}
\end{minipage}
\hfill
\begin{minipage}[t]{0.20\textwidth}
\vspace{10pt}
\begin{tabular}{l@{\hspace{8pt}}l@{\hspace{8pt}}r@{\hspace{2pt}}}
\toprule
$\lambda$ (\AA) & $\chi_{\rm low}$ & $\log gf$ \\
 & (eV) &      \\
\midrule
Cr  II & & \\
5237.329 & 10.760 & -0.740\\
Mn I & & \\
4030.753 & 0.000 & -0.470 \\
4033.062 & 0.000 & -0.618 \\
4034.483 & 0.000 & -0.811\\
4041.355 & 2.114 &  0.285\\ 
4783.427 & 2.298  & 0.044\\
4823.524 & 2.319 & 0.144\\
\hline
Co I &&\\
4118.767 & 1.049  &-0.490\\
4121.311 & 0.923  &-0.320\\
\hline
Ni  I & & \\
5017.568 & 3.539 & -0.020\\
5035.357 & 3.635 &  0.290\\
5080.528 & 3.655  & 0.330\\
5081.107 & 3.847  & 0.462\\
5084.089 & 3.679 &  0.030\\
5137.070 & 1.676 & -1.990\\
5168.656 & 3.699 & -0.430\\
5476.900  &1.826 & -0.890\\
\hline
Cu  I &&\\
5105.537 & 1.389 & -1.516\\
5218.197 & 3.817  & 0.264\\
\hline
Zn I &&\\
4810.528 & 4.078 & -0.160\\
\hline
Sr II  &&\\
3464.453 & 3.040 &  0.530\\
4077.719  & 0.000 & 0.170\\
4215.519 & 0.000 & -0.170\\
\hline
Y II  &&\\
4854.863 & 0.992 & -0.111\\
4883.684 & 1.084  & 0.265\\
4900.120 & 1.033  & 0.103\\
5087.416 & 1.084 & -0.170\\
5200.406 & 0.992 & -0.570\\
5205.724 & 1.033 & -0.193\\ 
\hline
Zr II &&\\
3614.765 & 0.359 & -0.252\\
3751.590 & 0.972  & 0.000\\
3766.795 & 0.409 & -0.830\\
3836.761 & 0.559 & -0.120\\
3958.220 & 0.527 & -0.320\\
3991.127 & 0.758 & -0.310\\
3998.954 & 0.559 & -0.520\\
4149.198 & 0.802 & -0.040\\
4161.200 & 0.713 & -0.590\\
4208.977 & 0.713 & -0.510\\
4359.720 & 1.236 & -0.510\\
5112.270 & 1.665 & -0.850\\  
\hline\\

\end{tabular}
\end{minipage}
\hfill
\begin{minipage}[t]{0.20\textwidth}
\vspace{10pt}
\begin{tabular}{l@{\hspace{8pt}}l@{\hspace{2pt}}r@{\hspace{2pt}}}
\toprule
$\lambda$ (\AA) & $\chi_{\rm low}$  & $\log gf$ \\
 & (eV) &      \\
\midrule
5350.089 & 1.827 & -1.240\\
5350.350 & 1.773 & -1.160\\
Nb II & & \\
3425.425 & 1.354 & -0.140 \\
3426.531 & 1.315 & -0.420 \\
3651.187 & 0.931 & -0.400\\
\hline
Mo  I   &    &      \\
3864.103 & 0.000 & -0.010 \\
\hline
Ba II &&\\
4130.649 & 2.722 &  0.525\\
4166.000 & 2.722 & -0.433\\
5853.673 & 0.604 & -0.909 \\
6141.711 & 0.704 & -0.030\\
6496.895 & 0.604 & -0.406\\
\hline
La II &&\\
4920.965 &0.126 &-2.261\\     
4920.965 &0.126 &-2.407\\     
4920.966 &0.126 &-2.065\\     
4920.966 &0.126 &-2.078\\     
4920.966 &0.126 &-2.738\\      
4920.968 &0.126 &-1.831\\     
4920.968 &0.126 &-1.956\\      
4920.968 &0.126 &-2.629\\     
4920.971 &0.126 &-1.646\\      
4920.971 &0.126 &-1.895\\    
4920.971 &0.126 &-2.650\\      
4920.975 &0.126 &-1.490\\     
4920.975 &0.126 &-1.891\\      
4920.975 &0.126 &-2.760\\     
4920.979 &0.126 &-1.354 \\     
4920.979 &0.126 &-1.957\\      
4920.979 &0.126 &-2.972\\     
4920.985 &0.126 &-1.233\\      
4920.985 &0.126 -&2.162\\      
4920.985 &0.126 &-3.375\\     
4921.774 &0.244 &-1.139\\      
4921.774 &0.244 &-2.220\\      
4921.774 &0.244 &-3.601\\     
4921.775 &0.244 &-1.233\\     
4921.775 &0.244 &-2.005\\      
4921.775 &0.244 &-3.207\\     
4921.776 &0.244 &-1.334\\     
4921.776 &0.244 &-1.445\\      
4921.776 &0.244 &-1.915\\      
4921.776 &0.244 &-1.927\\     
4921.776 &0.244 &-2.923\\      
4921.776 &0.244 &-3.010\\     
4921.777 &0.244 &-1.566\\      
4921.777 &0.244 &-1.955\\      
4921.777 &0.244 &-2.939\\       
\hline\\

\end{tabular}
\end{minipage}
\hfill
\begin{minipage}[t]{0.20\textwidth}
\vspace{10pt}
\begin{tabular}{l@{\hspace{8pt}}l@{\hspace{8pt}}r@{\hspace{2pt}}}
\toprule
$\lambda$ (\AA) & $\chi_{\rm low}$  & $\log gf$ \\
 & (eV) &      \\
\midrule 
4921.778 &0.244 &-1.700\\     
4921.778 &0.244 &-1.848\\     
4921.778 &0.244 &-2.006\\      
4921.778 &0.244 &-2.053\\      
4921.778 &0.244 &-2.258\\
4921.778 &0.244 &-3.123\\
4970.386  &0.321 & -1.16\\
5290.818  &0.000 & -1.65\\
\hline
Ce  II  &&\\
4073.474 & 0.478 &  0.230\\
4081.219 & 0.478  & 0.010\\
4083.222 & 0.701  & 0.270\\
4118.143 & 0.696  & 0.190\\
4127.363 & 0.684  & 0.350\\
4137.645 & 0.516  & 0.440\\
4165.599 & 0.910  & 0.530 \\
4222.597 & 0.122  & 0.020\\
4943.840 & 0.956 & -1.000\\
5274.229 & 1.044  & 0.130\\
5330.556 & 0.869  &-0.400\\
5353.524 & 0.879  & 0.110\\
6043.373 & 1.206  &-0.480\\
\hline
Pr  II &&\\
4000.173 & 0.204 & -0.401\\
4004.702 & 0.216 & -0.337\\
4033.830 & 0.372 & -0.128\\
4044.813 & 0.000 & -0.311\\
4056.534 & 0.630 &  0.353\\
4062.805 & 0.422 &  0.330\\
4100.717 & 0.550  & 0.572\\
5219.045 & 0.795 & -0.053\\
5220.108 & 0.796  & 0.298\\
5259.728 & 0.633  & 0.114\\
5322.772 & 0.483  &-0.141\\ 
\hline
Nd  II &&\\
4446.380 & 0.205 & -0.350\\
4451.560 & 0.380 &  0.070\\
4797.150 & 0.559  &-0.690\\
4914.380 & 0.380 & -0.700\\
4959.115 & 0.064 & -0.800\\
5089.832 & 0.205  &-1.160\\
5092.788 & 0.380  &-0.610\\
5123.779 & 0.380 & -0.610\\
5130.586 & 1.304 &  0.450\\
5132.328 & 0.559 & -0.710\\
5192.610 & 1.136 &  0.270\\
5249.576 & 0.976 &  0.200\\
5255.502 & 0.205 & -0.670\\
5273.427 & 0.680 & -0.180 \\
5276.869 &  0.859 & -0.440\\
\hline\\
\end{tabular}
\end{minipage}
\hfill

\end{figure*}

\newpage


\begin{figure*}
\centering

\hfill
\begin{minipage}[t]{0.20\textwidth}
\begin{tabular}{l@{\hspace{8pt}}l@{\hspace{8pt}}r@{\hspace{2pt}}}
\toprule
$\lambda$ (\AA) & $\chi_{\rm low}$  & $\log gf$ \\
 & (eV) &      \\
\midrule
5293.160 & 0.823  & 0.100\\
5311.450 & 0.986 & -0.420\\
5319.810 & 0.550  &-0.140\\
5361.467 & 0.680  &-0.370\\
\hline
Sm II &&\\
3941.876 & 0.000 & -0.860\\
4064.579 & 0.248 & -0.676\\
4318.926 & 0.277 & -0.250\\
4390.854 & 0.185 & -0.450\\
4420.520 & 0.333 & -0.430\\
4424.337 & 0.485  & 0.140\\
4433.890 & 0.434  &-0.190\\
4434.318 & 0.378  &-0.070\\
4452.720 & 0.277  &-0.410 \\
4467.340 & 0.659  & 0.150\\
4815.800 & 0.185  &-0.820\\
\hline
Eu II &&\\
3819.577 & 0.000 & -0.620\\  
3819.594 & 0.000 & -0.511\\  
3819.596 & 0.000 & -1.289\\   
3819.618 & 0.000 & -0.402\\    
3819.620 & 0.000 & -1.099\\    
3819.622 & 0.000 & -2.507 \\ 
3819.648 & 0.000 & -0.297\\  
3819.651 & 0.000 & -1.045\\  
3819.654 & 0.000 & -2.361\\  
3819.684 & 0.000 & -0.198\\ 
3819.689 & 0.000 & -1.087\\
3819.693 & 0.000 & -2.448\\   
3819.727 & 0.000 & -0.105\\   
3819.733 & 0.000 & -1.277\\   
3819.738 & 0.000 & -2.776\\ 
3907.046 & 0.207 & -0.374\\   
3907.080 & 0.207 & -0.542\\   
3907.093 & 0.207 & -1.186\\   
3907.108 & 0.207 & -0.742\\ 
3907.119 & 0.207 & -1.020\\    
3907.131 & 0.207 & -0.994\\   
3907.132 & 0.207 & -2.283\\  
3907.138 & 0.207 & -1.010\\  
3907.147 & 0.207 & -1.358\\    
3907.149 & 0.207 & -1.918\\   
3907.152 & 0.207 & -1.096\\   
3907.159 & 0.207 & -1.772\\   
3907.160 & 0.207 & -1.261\\   
3907.165 & 0.207 & -1.805\\
3930.424 & 0.207 & -1.219\\   
3930.429 & 0.207 & -0.326\\   
3930.469 & 0.207 & -1.044\\    
3930.472 & 0.207 & -0.534\\  
3930.477 & 0.207 & -1.219\\    
3930.506 & 0.207 & -1.022\\    
\hline\\

\end{tabular}
\end{minipage}
\hfill
\begin{minipage}[t]{0.20\textwidth}
\begin{tabular}{l@{\hspace{8pt}}l@{\hspace{8pt}}r@{\hspace{2pt}}}
\toprule
$\lambda$ (\AA) & $\chi_{\rm low}$ & $\log gf$ \\
 & (eV) &      \\
\midrule
3930.507 & 0.207 & -0.771\\  
3930.510 & 0.207 & -1.044\\   
3930.535 & 0.207 & -1.098\\   
3930.536 & 0.207 & -1.040\\   
3930.538 & 0.207 & -1.022\\   
3930.556 & 0.207 & -1.307\\ 
3930.557 & 0.207 & -1.316\\  
3930.558 & 0.207 & -1.098\\    
3930.569 & 0.207 & -1.404\\   
3930.570 & 0.207 & -1.307\\ 
3971.892 & 0.207 & -0.316\\  
3971.894 & 0.207 & -1.328\\   
3971.896 & 0.207 & -2.650\\   
3971.942 & 0.207 & -0.436\\   
3971.944 & 0.207 & -1.145\\   
3971.945 & 0.207 & -2.307\\    
3971.984 & 0.207 & -0.567\\ 
3971.985 & 0.207 & -1.111\\  
3971.986 & 0.207 & -2.198\\ 
3972.016 & 0.207 & -0.713\\   
3972.017 & 0.207 & -1.170\\  
3972.017 & 0.207 & -2.307\\    
3972.039 & 0.207 & -0.876\\    
3972.039 & 0.207 & -1.353\\   
3972.052 & 0.207 & -1.052\\
4129.600 & 0.000 & -1.512\\  
4129.604 & 0.000 & -1.035\\   
4129.617 & 0.000 & -1.316\\   
4129.622 & 0.000 & -0.977\\   
4129.626 & 0.000 & -1.512\\    
4129.642 & 0.000 & -1.257\\   
4129.648 & 0.000 & -0.847\\   
4129.653 & 0.000 & -1.316\\   
4129.675 & 0.000 & -1.294\\   
4129.682 & 0.000 & -0.696\\    
4129.688 & 0.000 & -1.257\\   
4129.717 & 0.000 & -1.480\\   
4129.724 & 0.000 & -0.545\\   
4129.730 & 0.000 & -1.294\\    
4129.774 & 0.000 & -0.401\\  
4129.781 & 0.000 & -1.480\\
4204.896 & 0.000  &-1.112\\ 
4204.899 & 0.000  &-1.413\\   
4204.904 & 0.000 & -2.367\\   
4204.921 & 0.000 & -0.936\\
4204.927 & 0.000 & -1.230\\    
4204.934 & 0.000 & -2.258\\    
4204.958 & 0.000 & -0.773\\  
4204.965 & 0.000 & -1.171\\   
4204.974 & 0.000 & -2.367\\    
4205.006 & 0.000 & -0.627\\  
4205.015 & 0.000 & -1.205\\  
\hline

\end{tabular}
\end{minipage}
\hfill
\begin{minipage}[t]{0.20\textwidth}
\begin{tabular}{l@{\hspace{8pt}}l@{\hspace{8pt}}r@{\hspace{2pt}}}
\toprule
$\lambda$ (\AA) & $\chi_{\rm low}$  & $\log gf$ \\
 & (eV) &      \\
\midrule
4205.026 & 0.000 & -2.710\\    
4205.065 & 0.000 & -0.496\\  
4205.076 & 0.000 & -1.388\\    
4205.135 & 0.000 & -0.376\\
\hline
Gd II  &&\\
3362.239 & 0.079 &  0.430\\
3545.790 & 0.144  & 0.190\\
3549.359 & 0.240  & 0.290\\
3646.195 & 0.240  & 0.320\\
3671.205 & 0.079 & -0.220\\ 
3719.452 & 1.233  & 0.460\\
3768.396 & 0.079  & 0.360\\
3844.578  &0.144 & -0.400\\
4077.966 & 0.103 & -0.040\\
4215.022 & 0.427  &-0.440\\
4251.731 & 0.382  &-0.220\\
4342.181 & 0.600 & -0.270\\
3845.633 & 0.790 &  0.252 \\
3899.188 & 0.373 &  0.330\\
3939.539 & 0.000 & -0.270\\
4002.566 & 0.641  & 0.100\\
4005.467 & 0.126 & -0.020\\
\hline
Tb  II   &    &      \\
3625.510 & 0.401 & -0.060\\ 
3658.888 & 0.126 &-4.083\\
3702.869 & 0.126 & -1.671\\
3775.268 & 0.790 & -0.570\\
\hline
Dy  II  &&\\
3407.796 & 0.000 &  0.180\\
3445.570 & 0.000 & -0.150\\ 
3538.519 & 0.000 & -0.020\\
3645.398 & 0.103  & 0.340\\
3944.680 & 0.000  & 0.110\\
4000.450 & 0.103  & 0.040\\
4073.120 & 0.538 & -0.320\\
4077.966 & 0.103 & -0.040\\
4103.306 & 0.103 & -0.380\\
\hline
Ho  II   &    &      \\
3453.123 & 0.079 & -0.930\\
3456.020 & 0.000 & -2.828 \\
3484.816 & 0.079 & -0.636\\
3676.358 & 0.126 &  0.480\\
3747.380 & 0.401 &  0.130\\
3796.754 & 0.000 & -1.722\\
3810.738 & 0.000 &  0.142 \\
4045.470 & 0.000 & -0.918\\
4152.586 & 0.079 &-2.733 \\
\hline
Er  II  && \\
3524.913 & 0.000 & -0.887\\ 
3692.649 & 0.055 &  0.138\\
3729.524 & 0.000 & -0.488\\
3830.481 & 0.000 & -0.365\\

\hline\\

\end{tabular}
\end{minipage}
\hfill
\begin{minipage}[t]{0.20\textwidth}
\begin{tabular}{l@{\hspace{8pt}}l@{\hspace{8pt}}r@{\hspace{2pt}}}
\toprule
$\lambda$ (\AA) & $\chi_{\rm low}$  & $\log gf$ \\
 & (eV) &      \\
\midrule
3896.233 & 0.055 & -0.241\\
3906.311 & 0.000 & -0.052\\
\hline
Tm  II   &    &      \\
3700.255 & 0.029 & -0.380 \\
3701.362 & 0.000  & -0.540 \\
3734.123 & 0.029 & -0.710\\
3761.958 & 3.328 & -2.850\\
3795.759 & 0.029 & -0.230 \\
3848.019 & 0.000 & -0.140 \\ 
3958.097 & 0.000 & -1.120\\
\hline
Yb  II   &    &      \\
3694.192 & 0.000 & -0.300 \\
\hline
Lu II  & &\\
3507.395 & 0.000 &-1.637\\
5983.701 &1.462 &-1.952\\
6221.592 &1.541 &-2.471\\
\hline
Hf II  &&\\
3399.790 & 0.000  &-0.570\\
3505.219 & 1.037 & -0.140\\
3719.276 & 0.608  &-0.810\\ 
3793.379  &0.378 & -1.110\\
3918.090 & 0.452 & -1.140\\
4093.150 & 0.452 & -1.150\\
Ta  II   &    &      \\
3379.507 & 0.511 & -0.520\\
3406.666 & 0.847 & -0.110\\
3414.128 & 0.847  &-0.210\\
3440.312 & 1.797&   0.540\\
3446.851 & 1.575 &  0.240\\
3541.882 & 0.128 & -1.140\\
3573.407  &1.459 &  0.340\\
3694.501 & 1.797  & 0.150\\ 
\hline
Os I  &&\\
4260.849 & 0.000 &-1.434\\
4420.477 & 0.000 & -1.590\\
\hline
Ir I &&\\
3725.392 & 2.363 & -0.430\\
3738.530 & 0.784 & -2.070\\
3747.205 & 0.717 & -1.480\\
3768.673 & 1.467 & -1.600\\
3800.120 & 0.000 & -1.450\\
3951.948 & 1.515 & -1.570\\
3992.121 & 1.225 & -1.220\\
\hline
Pb I & & \\
4057.832 & 1.320 & -0.220\\
\hline\\
\end{tabular}
\end{minipage}
\hfill

\end{figure*}


\setcounter{table}{4} 

{\footnotesize
\begin{table*}[h!]
\small
\section{Elemental abundances for the programme stars}
\caption{Elemental abundances}
\label{Tab:abundances}
\hspace*{-2cm}
\begin{tabular}{lclrrrrrrrrrr}
\hline
\multicolumn{2}{c}{}& \multicolumn{5}{c}{CS 22947$-$187} && \multicolumn{4}{c}{CS29512$-$073} \\
\cline{4-7}\cline{9-12}
 &    Z  &    log$_{\odot}{\epsilon}^a$ & log${\epsilon}$&$\sigma_{l}$ (N)& [X/Fe] $\pm~\sigma_{t}$ & [X/Fe]$^b$& & log${\epsilon}$&$\sigma_{l}$(N)& [X/Fe]~$\pm~ \sigma_{t}$  &[X/Fe]$^b$ \\
\hline
C  & 6  & 8.43 &   7.00  &   0.10(1)  &   1.12 $\pm$ 0.17 &1.26& & 7.45 &  0.10(1) &  1.37 $\pm$ 0.17  & 1.05& \\
$^{12}$C/$^{13}$C$^c$& & &  &       &  1.5 $\pm$ 0.3    &  & &         &           &  19 $\pm$ 8.8        & & \\
$^{12}$C/$^{13}$C$^d$& & &  &       &  2.3 $\pm$ 0.7     &  & &         &           &  11.5  $\pm$1.9        & & \\
N  & 7  & 7.83 &   6.95  & 	 0.10(1)  &   1.67 $\pm$ 0.15 &1.71& &    6.10  & 	 0.10(1) &   0.62 $\pm$ 0.15 &0.40&  \\
O  & 8  & 8.69 &    7.20*    &   0.10(1)    &    1.06* $\pm$ 0.16 &0.62& &     8.60:  &   0.10(1)  &  2.26: $\pm$ 0.16 &0.42\\
Na & 11 & 6.24 &   4.00*  &   0.10(1)  &  0.31* $\pm$ 0.13  & 0.16&&   4.05::&  0.05(2)  &  0.16:: $\pm$ 0.09  &  & \\
Mg & 12 & 7.66 &   5.70  &   0.10(2)  &  0.59 $\pm$ 0.11  & 0.56&&    5.75 &   0.10(2) & 0.44 $\pm$ 0.11 & 0.01  \\

Ca & 20 & 6.34 &   4.25  &   0.03(9)  &  0.46 $\pm$ 0.08  & 0.51&&     4.33  &   0.05(6)  & 0.34 $\pm$ 0.08  & 0.25&  \\
Sc & 21 & 3.15 &   0.75  &   0.05(2)  &  0.15 $\pm$ 0.11  & 0.11&&  0.71 &   0.10(2)  &  $-$0.09 $\pm$ 0.13  & $-$0.21    \\
Ti & 22 & 4.95 &   2.74  &   0.09(13)  &  0.34 $\pm$ 0.10  & 0.32&&     2.83  &   0.04(6)  &  0.23 $\pm$ 0.10 &0.06   \\
Cr & 24 & 5.64 &   3.06  &   0.07(7)  &  $-$0.03 $\pm$ 0.09  & $-$0.06& &   3.15 &   0.05(4)  & $-$0.14 $\pm$ 0.09 &  $-$0.14 \\
Mn & 25 & 5.43 &   2.60  &   0.10(1)  &  $-$0.28 $\pm$ 0.13 &$-$0.21& &     2.74 &   0.10(5) & $-$0.34 $\pm$ 0.10   & $-$0.44  \\
Fe & 26 & 7.50 &   4.95  &   0.07(81) &   & & &      5.15 &  0.07(91)  &    & & \\
Co & 27 & 4.99 &   2.60 &   0.10(2)  &  0.16 $\pm$ 0.12  &$-$0.05&  &     2.60&   0.10(2) & $-$0.04 $\pm$ 0.12  &$-$0.24   \\
Ni & 28 & 6.22 &   3.80  &   0.06(6)  &  0.13 $\pm$ 0.09  & 0.17&&  3.88 &  0.07(7) & 0.01 $\pm$ 0.09  & 0.06  \\
Zn & 30 & 4.56 &   2.30 &   0.10(1)  &  0.29 $\pm$ 0.13  &0.27& &    2.50&   0.10(1) &  0.29 $\pm$ 0.13   & 0.11  \\
Sr$_{\rm LTE}$ & 38 & 2.87 &   0.40    &   0.10(1)      &  0.08 $\pm$ 0.14     &0.29& &   0.95   & 0.10(1)     &  0.43 $\pm$ 0.15 & 0.31  \\
Sr$_{\rm NLTE}$ & 38 & 2.87 &  0.39  &   0.10(1)     &   0.07 $\pm$ 0.14      && &  0.92  &  0.10(1)    &   0.40 $\pm$ 0.15    &   \\
Y$_{\rm LTE}$  & 39 & 2.21 &   $-$0.57  &   0.05(6)  &  $-$0.23 $\pm$0.10  & $-$0.14&&   $-$0.05  &   0.06(5)  &  0.09 $\pm$ 0.11 &0.21 \\
Y$_{\rm NLTE}$  & 39 & 2.21 &   $-$0.45  &   0.05(6)  &  $-$0.11 $\pm$0.10  & &&   0.07  &   0.06(5)  &  0.21 $\pm$ 0.11 & &\\
Zr & 40 & 2.58 &   0.40  &   0.10(7)  &  0.37 $\pm$ 0.13  &0.34& &   0.80 &  0.10(2)  &  0.57 $\pm$ 0.11  &0.50& \\
Nb & 41 & 1.46 &  ...   &      ...     &  ...   && &   0.70:    & 0.10(1)       &  1.59: $\pm$ 0.12   &  \\
Mo & 42 & 1.88 &   0.55:  &   0.10(1)  &  1.22: $\pm$ 0.13  &1.45& &   0.40  &  0.10(1) & 0.87 $\pm$ 0.13  &1.18 \\
Ba & 56 & 2.18 &   0.88  &   0.08(3)  &  1.25 $\pm$ 0.12  & 1.53&&   1.10 &  0.13(3)  &  1.27 $\pm$ 0.13 & 1.44\\
La & 57 & 1.10 &  $-$0.33  &   0.04(4) &  1.12 $\pm$ 0.10  & 0.96&&     $-$0.08 & 0.03(2) & 1.17$\pm$ 0.11  &1.04 \\
Ce & 58 & 1.58 &  $-$0.11 &   0.07(11) &  0.86 $\pm$ 0.12  & 0.89&&    0.42 &  0.03(9) &  1.19 $\pm$ 0.10 & 1.26 \\
Pr & 59 & 0.72 &   $-$0.78  &   0.08(5)  & 1.05 $\pm$ 0.10 & 0.95&&     $-$0.54&   0.05(7)  & 1.09 $\pm$ 0.09  & 1.09\\
Nd & 60 & 1.42 &   $-$0.08  &   0.05(6) &  1.05 $\pm$ 0.11  &1.04& &   0.30  &  0.05(9) & 1.23 $\pm$ 0.10 &1.21   \\
Sm & 62 & 0.96 &  $-$0.69  &   0.08(4)  &  0.90 $\pm$ 0.11  & 0.73&&   $-$ 0.30 &   0.06(5) &  1.09 $\pm$ 0.10 &  0.86 \\
Eu$_{\rm LTE}$ & 63 & 0.52 &  $-$1.36  &   0.15(4)  &  0.67 $\pm$ 0.12  &0.56& &  $-$1.32 &  0.14(5)  &  0.51 $\pm$ 0.11 & 0.36  \\
Eu$_{\rm NLTE}$ & 63 & 0.52 &  $-$1.24  &   0.10(1)  & 0.79 $\pm$ 0.13  &&& $-$1.24 & 0.14(1)   & 0.59 $\pm$ 0.11   &  & \\
Gd & 64 & 1.07 &   $-$0.50 &   0.10(2)  & 0.98 $\pm$ 0.12  & 0.91&&   $-$0.59 &   0.12(5)  & 0.69 $\pm$ 0.15  & &  \\
   
Tb & 65 & 0.30 & ...   &   ...  &  ...&& &  $-$0.80::  & 0.10(1) & 1.25:: $\pm$ 0.11  & 1.83  \\
Dy & 66 & 1.10 &   $-$0.75 &   0.10(2)  & 0.70 $\pm$ 0.11  &0.61&&  $-$0.50 &   0.10(4)  & 0.75 $\pm$ 0.15 &0.88\\
Ho & 67 & 0.48 &   $-$1.57 &   0.03(2)  & 0.50 $\pm$ 0.09  & 0.79&&  $-$1.36: & 0.05(3)  & 0.51: $\pm$ 0.09 && \\
Er & 68 & 0.92 &  $-$0.50 &   0.09(3)  &  1.13 $\pm$ 0.13    &1.00& &    $-$0.44&  0.09(4) &  0.99 $\pm$ 0.13 & 0.89  \\
Tm & 69 & 0.10 &   $-$1.28 &   0.04(3)  & 1.17 $\pm$ 0.09  & &&    $-$1.25:: &  0.05(2)  & 1.00::$\pm$ 0.10 &1.98\\
Yb & 70 & 0.84 &   $-$0.65 &   0.10(1)  & 1.06 $\pm$ 0.14  & 1.38&&   $-$0.55: &   0.10(1)  & 0.96: $\pm$ 0.14&1.05 \\
Hf & 72 & 0.85 &   $-$0.45  &   0.05(2)  & 1.25$\pm$ 0.13 & 1.28&&  $-$0.25* &  0.05(2)  & 1.25* $\pm$ 0.13 & 1.56&\\
Ta & 73 & $-$0.12 &  $-$0.30*  &  0.10(3)  & 2.37*$\pm$ 0.11 & &&  ...  &  ...    &  ...  & &\\
Pb$_{\rm LTE}$ & 82 & 1.75 &   1.85 &   0.10(1)  &  2.65 $\pm$ 0.04 &2.33& &  1.85&  0.10(1)  &  2.45 $\pm$ 0.04   &  2.07&\\
Pb$_{\rm NLTE}$ & 82 & 1.75 &   2.22 &   0.10(1)  &  3.02 $\pm$ 0.04 & &&  2.37 &   0.10(1)  &  2.97 $\pm$ 0.04  &  \\
\hline\\
\end{tabular}
\\

\tablecomments{
$^{a}$ \citet{Asplund2009} \\
$^{b}$ \citet{Roederer2014}\\
$:$ Uncertain abundances due to noisy/blended region\\
$::$ Very Uncertain abundances due to noisy/blended region\\
$*$ Upper limit and uncertain\\
$^{c}$ $^{12}$C/$^{13}$C ratio fom CH G band\\
$^{d}$ $^{12}$C/$^{13}$C ratio fom CN band\\}
\end{table*}
}

\setcounter{table}{4}
{\footnotesize
\begin{table*}
\caption{Continued.}
\hspace*{-3cm}
\small
\renewcommand{\arraystretch}{1} 
\begin{tabular}{l@{\hspace{2pt}}c@{\hspace{8pt}}l@{\hspace{1pt}}r@{\hspace{4pt}}r@{\hspace{2pt}}r@{\hspace{3pt}}r@{\hspace{1pt}}r@{\hspace{2pt}}r@{\hspace{4pt}}r@{\hspace{2pt}}r@{\hspace{2pt}}r@{\hspace{2pt}}r@{\hspace{2pt}}r@{\hspace{4pt}}r@{\hspace{2pt}}r@{\hspace{2pt}}r}
\hline
\multicolumn{2}{c}{}& \multicolumn{5}{c}{SDSSJ0912$+$0216} && \multicolumn{4}{c}{SDSSJ1036$+$1212}  && \multicolumn{4}{c}{SDSSJ1349$-$0229}\\
\cline{4-7}\cline{9-12}\cline{14-16}
 &    Z  &    log$_{\odot}{\epsilon}^a$ & log${\epsilon}$&$\sigma_{l}$ (N)& [X/Fe]~$\pm~ \sigma_{t}$ &[X/Fe]$^b$ & &log${\epsilon}$&$\sigma_{l}$(N)  & [X/Fe]~$\pm~ \sigma_{t}$ &[X/Fe]$^b$ & &log${\epsilon}$&$\sigma_{l}$(N)  & [X/Fe]~$\pm~ \sigma_{t}$ &[X/Fe]$^b$  \\
\hline
C  & 6  & 8.43 &   7.70  &   0.10(1)  &   2.10 $\pm$ 0.17 & 2.17&& 6.30  &  0.10(1)   &   1.35 $\pm$ 0.17  & 1.47 & & 8.20  &  0.10(1)   &   2.90 $\pm$ 0.17 & 2.82 \\
$^{12}$C/$^{13}$C$^c$& & &  &       &  11.5$\pm$1.9    &  & &         &    &  4$\pm$1.5         &  & &  &       &  5.6 $\pm$1.4   & \\
$^{12}$C/$^{13}$C$^d$& & &  &       &  9$\pm$3.1    &  & &         &           &  5.6 $\pm$1.4        &  & &  &       &  5.6 $\pm$1.4   &\\
N  & 7  & 7.83 &   6.85  & 	 0.10(1)  &   1.85 $\pm$ 0.15 &1.75& &     6.40:  &  0.10(1)   &   2.05: $\pm$ 0.15  & 1.29 &&   6.80  &  0.10(1)   &   2.10 $\pm$ 0.15& 1.60 \\
Na & 11 & 6.24 &  4.40* &   0.10(2)  &  0.99* $\pm$ 0.13  & 0.38&&      3.40   &   0.10(2)  &   0.64 $\pm$0.19    &0.43 &&&&&\\
Mg& 12 & 7.60 &   5.40  &   0.10(1)  &  0.63 $\pm$ 0.13  &0.21& &     4.70: &  0.10(1)   &  0.56: $\pm$ 0.13  & 0.00 &&   5.60  &  0.10(1)   &  1.13 $\pm$ 0.13  & 0.57  \\

Ca & 20 & 6.34 &   4.00  &   0.10(3)  &  0.49 $\pm$ 0.11  & 0.42&&     3.33  &  0.05(3)   &  0.47 $\pm$ 0.09  & 0.38 &&     3.80:  &  0.10(2)   &  0.59: $\pm$ 0.11 &0.40   \\
Sc & 21 & 3.15 & 0.78 &  0.13(3)   & 0.46 $\pm$ 0.13 &0.28 &&  -0.30  &  0.11(3)   &    0.03 $\pm$ 0.13 &0.11 & & 0.59 &0.12(4)  &  0.57 $\pm$ 0.13  &  \\
Ti & 22 & 4.95 &   2.70  &   0.04(3)  &  0.58 $\pm$ 0.10  & 0.51&&     2.10  &  0.05(2)   &   0.63$\pm$ 0.11  & 0.75 &&     2.50  &  0.04(3)   &   0.68$\pm$ 0.10  & 0.55 \\
Cr & 24 & 5.64 &  2.70 &   0.08(3)  &  $-$0.11 $\pm$ 0.09  &$-$0.16 &&     2.10  &  0.10(2)   &   $-$0.06 $\pm$ 0.11  & $-$0.12 &&   2.80  &  0.08(3)   &   0.29 $\pm$ 0.09  & $-$0.02 \\
Mn & 25 & 5.43 &   2.10  &   0.10(2)  &  $-$0.50 $\pm$ 0.12  & $-$0.55& &  1.27  & 0.05(3)   &  $-$0.68 $\pm$ 0.09  &  & &   2.20  &  0.10(2)   &  $-$0.10 $\pm$ 0.13  & $-$0.68    \\
Fe & 26 & 7.50 &   4.67  & 0.07(67)  & & & &     4.02  &  0.07(52) &   & &&4.37  & 0.07(52) &   & \\
Co & 27 & 4.99 &   2.65 &   0.10(2)  &  0.49 $\pm$ 0.12  &0.31 &&   1.90  &  0.10(1)  &   0.39 $\pm$ 0.14  & 0.57 &&...&...&...& \\
Ni & 28 & 6.22 &  4.20* &  0.18(4)  & 0.81* $\pm$ 0.12  &0.07 &&   3.60  &  0.05(2)   &  0.86 $\pm$ 0.09 &0.28 &&...&...&...&\\
Cu & 29 & 4.19 &  ...  & ... & ...  & & &   1.60 &  0.10(1)   &  0.89 $\pm$ 0.14  &  & & 3.00: &  0.10(1)   &  1.94: $\pm$ 0.14  &    \\
Zn & 30 & 4.56 &   2.30* &   0.10(1)  &  0.57* $\pm$ 0.13  & & &   1.70* & 0.10(1)  &  0.62$\pm$0.13  &  & &  2.70: &  0.10(1)  &  1.27:$\pm$0.13  &   \\
Sr$_{\rm LTE}$ & 38 & 2.87 &   0.50    &   0.10(2)      &  0.46 $\pm$ 0.14     & 0.57& &  $-$1.10  &0.10(1)   &  $-$0.49 $\pm$ 0.14 & $-$0.56 &&  1.10 & 0.10(2)  &  1.36$\pm$ 0.15 &1.30   \\
Sr$_{\rm NLTE}$ & 38 & 2.87 & 0.51  & 0.10(2)  &   0.47 $\pm$ 0.14  & &&  $-$0.96 & 0.10(1)  &  $-$0.35 $\pm$0.14  & & &  1.11 & 0.10(2)  &  1.37$\pm$ 0.15  &        \\
Y$_{\rm LTE}$  & 39 & 2.21 &  0.35  &   0.05(2)  &  0.97 $\pm$ 0.14  &0.61& &    $-$0.40* & 0.10(2)  & 0.87* $\pm$ 0.12 & 0.24 &&    0.50 & 0.17(4)  &  1.42 $\pm$ 0.13 & 1.29\\
Y$_{\rm NLTE}$  & 39 & 2.21 &  0.46  &   0.05(2)  &  1.08 $\pm$ 0.14  && &    $-$0.33* & 0.10(2)  & 0.94* $\pm$ 0.12 && &  0.56 & 0.17(4)  &  1.48 $\pm$ 0.13 & \\
Zr & 40 & 2.58 &   1.15* &  0.08(4)  &  1.40* $\pm$ 0.13  & 1.08& &   0.70  & 0.10(3) &   1.60 $\pm$ 0.13  & 1.02 & &  1.30:  & 0.10(4)  &   1.85: $\pm$ 0.13  & 1.56\\
Nb & 41 & 1.46 &    ... &      ...     &  ...   && &   0.70:    & 0.10(1)       &  2.72: $\pm$ 0.12   & &&   2.05:    & 0.05(2)       &  3.72: $\pm$ 0.12   &  \\
Mo & 42 & 1.88 & 1.40:  &  0.10(1) & 2.35: $\pm$ 0.13  & &&     0.30  &  0.10(1)   &   1.90 $\pm$ 0.13  && &   1.58::  &  0.10(1)   &   2.83:: $\pm$ 0.13  &  \\
Ba & 56 & 2.18 &   0.85  &   0.05(3)  &  1.50 $\pm$ 0.11  & 1.49& &    $-$0.02 &  0.12(2)   &  1.28 $\pm$ 0.15  & 1.17 &&   1.13  &  0.08(3)   &   2.08 $\pm$ 0.15  & 2.17  \\
La & 57 & 1.10 & 0.28:  & 0.03(2) & 2.01: $\pm$ 0.11  &1.35& &    -0.05  &  0.10(2)   &   2.33 $\pm$0.13 &2.39 & &  0.35:  &  0.05(2)   &   2.38: $\pm$0.15  & 1.74   \\
Ce & 58 & 1.58 & 1.05 &   0.05(2) &  2.30 $\pm$ 0.12  & 2.17& &   1.04:  &  0.28(3)   &   2.94: $\pm$ 0.19 & 2.32  & &   0.70 &  0.10(1)   &   2.25 $\pm$ 0.14 & 2.63  \\
Pr & 59 & 0.72 &  0.83 &   0.07(3)  &2.94 $\pm$ 0.17  &2.25& &     $-$0.20*     & 	0.10(2)      &   2.56* $\pm$ 0.13    & 2.45 &&   1.37     & 	0.13(4)      &   3.78 $\pm$ 0.13   &2.87  \\
Nd & 60 & 1.42 &  0.63  & 0.12(3) & 2.04 $\pm$ 0.12  &1.12& &    0.77  & 0.18(3)   &   2.83 $\pm$ 0.14 & 2.08  &&   1.60:  & 0.11(5)   &   3.31: $\pm$ 0.11   & 1.91 \\
Sm & 62 & 0.96 & 0.80: &  0.10(1)   &  2.67: $\pm$ 0.12 & 2.60&&     0.15  &  0.12(2)  &   2.67 $\pm$ 0.14  & 2.92 &&   0.70  &  0.10(3)   &   2.87 $\pm$ 0.14  & 2.35   \\
Eu$_{\rm LTE}$ & 63 & 0.52 &  $-$0.67  &   0.14(3)  &  1.64 $\pm$ 0.12  &1.20& &   $-$0.85 & 0.10(2) &  2.11 $\pm$0.11  &  1.26  &&   $-$0.14 &0.15(2)   &  2.47 $\pm$0.13  &1.62  \\
Eu$_{\rm NLTE}$ & 63 & 0.52 &   $-$0.54  &  0.14(3)  & 1.77$\pm$ 0.12  &&&  $-$0.80 & 0.10(2)  & 2.16 $\pm$ 0.11  &  &&  $-$0.24 & 0.10(1)  & 2.37 $\pm$ 0.13  &  \\
Gd & 64 & 1.07 &  0.91 &   0.09(2)  & 2.67 $\pm$ 0.15  &2.80& &    $-$0.15   & 	0.11(2)         &  2.26 $\pm$ 0.15  & 2.62 &&   1.20   & 	0.12(4)         &  3.26 $\pm$ 0.15  & 2.50   \\
   
Tb & 65 & 0.30 &  0.61  &  0.05(2)  & 3.14 $\pm$ 0.11 &2.64& &  $-$0.40:    & 	0.10 (2)         &  2.78:  $\pm$ 0.12   &  2.90 &&  0.65:    & 	0.10(1)   &   3.48: $\pm$   0.15 &2.69  \\
Dy & 66 & 1.10 &  0.30 &   0.10(2)  & 2.03 $\pm$ 0.15& 1.96&& 0.10  &   0.10(3)   & 	2.48 $\pm$ 0.15 &2.46 && 0.25  &   0.10(1)   & 	2.28 $\pm$ 0.13 &2.39 \\
Ho & 67 & 0.48 &   $-$0.70 &  0.10(1)  & 1.65 $\pm$ 0.12  & &&   $-$0.52  & 	0.08(3) &  2.48 $\pm$ 0.10& &&  0.01:   & 	0.14(2) &  2.66: $\pm$ 0.12 & \\
Er & 68 & 0.92 & 0.30 &  0.10(1)  &  2.21 $\pm$ 0.16     & 2.03&&    0.60:  &  0.10(1)   &   3.16:  $\pm$0.16 &2.86 &&   0.47:  &  0.15(2)   &   2.68: $\pm$ 0.16   &2.73   \\
Tm & 69 & 0.10 &   0.10 &  0.10(1)  & 2.83 $\pm$ 0.13  && &   $-$0.58   & 	0.09(2) &  2.80 $\pm$ 0.05 &2.78 & &   0.27:    & 	0.15(2) &  3.30:$\pm$ 0.14  &  \\
Yb & 70 & 0.84 &   $-$0.25 &   0.10(1)  & 1.74 $\pm$ 0.14  &&  &  -    & - & -&  & &   $-$0.10:    & 	0.10(1) & 2.19: $\pm$ 0.14  &   \\
Hf & 72 & 0.85 &  0.90*  &  0.10(1)  & 2.88* $\pm$ 0.15 & 2.72&& 0.45*     &    0.10(1)  &   3.08* $\pm$ 0.15  &2.35 & &  1.30:     &    0.10(2)  &   3.58: $\pm$ 0.16  & 3.14 \\
Ta & 73 & $-$0.12 &  1.00*  &  0.10(1)  & 3.95* $\pm$ 0.13 & && 1.00*    &   0.15(2)  &   4.60* $\pm$ 0.14   &&& 0.65:     &    0.10(1)  &   3.90: $\pm$ 0.13 &   \\
Os & 76 & 1.40 &    ...    &  ...  &     ...  & &         &   ...       &    ...    &...& &&   2.00:     &  0.10(1)      &   3.73:$\pm$ 0.15    &  \\
Ir & 77 & 1.38 &     ...   &  ...  & ... &     & &   1.50:: & 0.10(1)& 3.25:: $\pm$ 0.11& &&... &...&...&\\
Pb$_{\rm LTE}$ & 82 & 1.75 &  ...&  ... &  ... & &        &     ...   &   ...    &  ...& && 2.35:    &     0.10(1)   &   3.73:  $\pm$ 0.04  & 3.09  \\
Pb$_{\rm NLTE}$ & 82 & 1.75 &  ... &   ...  &  ... & &         &  ...   &    ...   &...&& &  2.97:     &  0.10(1)  & 4.35: $\pm$ 0.04  &   \\
\hline

\hline
\\
\end{tabular}

\tablecomments{
$^{a}$ \citet{Asplund2009} \\
$^{b}$ \citet{behara2010}\\
$:$ Uncertain abundances due to noisy/blended region\\
$::$ Very Uncertain abundances due to noisy/blended region\\
$*$ Upper limit and uncertain\\
$^{c}$ $^{12}$C/$^{13}$C ratio fom CH G band\\
$^{d}$ $^{12}$C/$^{13}$C ratio fom CN band\\}
\end{table*}
}

\setcounter{table}{4}
{\footnotesize
\begin{table*}
\caption{Elemental abundances of the objects from \citet{Karinkuzhi2021}}
\hspace*{-4cm}
\small
\renewcommand{\arraystretch}{0.9} 
\begin{tabular}{l@{\hspace{8pt}}c@{\hspace{8pt}}l@{\hspace{8pt}}r@{\hspace{8pt}}r@{\hspace{8pt}}r@{\hspace{8pt}}r@{\hspace{8pt}}r@{\hspace{8pt}}r@{\hspace{8pt}}r@{\hspace{8pt}}r@{\hspace{8pt}}r@{\hspace{8pt}}r@{\hspace{8pt}}r@{\hspace{8pt}}r@{\hspace{8pt}}r@{\hspace{8pt}}r}
\hline
\multicolumn{2}{c}{}& \multicolumn{4}{c}{CS~22887$-$048} && \multicolumn{3}{c}{CS~22891$-$171}  && \multicolumn{3}{c}{HD~145777} \\
\cline{4-6}\cline{8-10} \cline{12-14}
 &    Z  &    log$_{\odot}{\epsilon}^a$ & log${\epsilon}$&$\sigma_{l}$(N)& [X/Fe]~$\pm~ \sigma_{t}$ & &log${\epsilon}$&$\sigma_{l}$(N) & [X/Fe]~$\pm~ \sigma_{t}$  & &log${\epsilon}$&$\sigma_{l}$(N) & [X/Fe]~$\pm~ \sigma_{t}$  \\
\hline

Gd & 64 & 1.07 &  0.80: &   0.10(1)  & 1.83:$\pm$ 0.15  & &     0.30$^b$  & 	0.10(1)        &  1.73$\pm$ 0.17 & & $-$0.50$^b$ &  0.10(1)  & 0.75$\pm$ 0.17  &  \\
Tb & 65 & 0.30 &   $-$0.06*  &   0.12(3)  &  1.74* $\pm$ 0.13 & &     $-$0.77  &   0.06(2)  &  1.43 $\pm$ 0.11  &  &   $-$1.15:  &   0.10(1)  &  0.87: $\pm$ 0.11  &   \\
Dy & 66 & 1.10 &  0.55 &   0.05(2)  & 1.55 $\pm$ 0.09&& 0.00$^b$ &   0.15(2)  & 1.40 $\pm$ 0.15& & $-$0.95 & 0.08(4) & 0.27$\pm$ 0.12 &	 \\
Ho & 67 & 0.48 &   $-$0.36  &   0.10(3)  &  1.26 $\pm$ 0.11 & &   $-$0.97  &   0.04(4)  &  1.05 $\pm$ 0.08   &   &   $-$1.82:  &   0.11(2)  &  0.02: $\pm$ 0.11  &     \\
Er & 68 & 0.92 & 1.00$^b$ &  0.11(1)  & 2.18$\pm$ 0.11     & &    0.00$^b$  & 0.10(1)   &   1.58 $\pm$ 0.11  & & $-$1.20 &  0.10(1)  & 0.20$\pm$ 0.12     & \\
Tm & 69 & 0.10 &   $-$0.10  &   0.10(1)  & 1.90$\pm$ 0.13  & &   $-$0.75:  &   0.04(3)  &  1.65: $\pm$ 0.09   & &   $-$1.85  &   0.10(2)  &  0.37 $\pm$ 0.13  & \\
Yb & 70 & 0.84 &   0.70  &   0.10(1)  & 1.96 $\pm$ 0.14  & &   $-$0.50:  &   0.10(1)  &  1.16: $\pm$ 0.14  &   &   $-$1.20:  &   0.10(1)  &  0.28: $\pm$ 0.14  &    \\
Lu & 71 & 0.10 &   0.20:  &   0.10(1)  & 2.20 $\pm$ 0.10  & &   $-$0.70:  &   0.10(1)  &  1.70: $\pm$ 0.10  & &   $-$0.10*  &   0.10(1)  & 2.12* $\pm$ 0.10  &    \\
Hf & 72 & 0.85 &0.85  & 0.10(2)  &2.10$\pm$ 0.14 & &  0.45:$^b$  &   0.10(1)  &   2.10:$\pm$ 0.10  & & $-$0.65  &  0.10(1)  & 0.82$\pm$ 0.15 &\\
Ta & 73 & $-$0.12 &  0.70:  &  0.15(2)  & 2.92:$\pm$ 0.14 & & $-$0.20*     &    0.10(2)  &  2.42* $\pm$ 0.11   & &...&...&...&\\
Os & 76 & 1.40 &    1.80*     &  0.10(1) &   2.50*$\pm$ 0.15  & & 0.70 & 0.10(1)  & 1.80$\pm$ 0.15 &&    $-$0.15$^b$  & 0.10(1)  &  0.77$\pm$ 0.12    &  \\
Ir & 77 & 1.38 &    1.80*     &  0.10(1)      &   2.52*$\pm$ 0.13   & &...& ...& ...   &&    1.30::   &  0.10(1)     &  0.92::$\pm$ 0.13     & \\

\hline

\multicolumn{2}{c}{}& \multicolumn{4}{c}{CS~22942$-$019}  && \multicolumn{3}{c}{CS~30322$-$023} &&\multicolumn{3}{c}{HD~187861} \\
\cline{4-6}\cline{8-10} \cline{12-14}

Gd & 64 & 1.07 & $-$0.63$^b$ &  0.12(3)  & 0.80$\pm$ 0.16  & &   $-$2.10$^b$  & 	0.10(1)        & 0.18 $\pm$ 0.17 & &   0.15$^b$  & 	0.12(3)    & 1.68 $\pm$ 0.16 &  \\
Tb & 65 & 0.30 &   $-$0.60:  &  0.10(1)  &1.60: $\pm$ 0.11   & &    $-$2.00*  &   0.10(1)  &  1.05 $\pm$ 0.15&  &    $-$1.12:  &   0.10(2)  &  1.18: $\pm$ 0.13   &   \\
Dy & 66 & 1.10 & $-$0.65$^b$ & 0.10(1) & 0.75$\pm$ 0.14 && $-$1.45 &   0.10(1)  & 0.80$\pm$ 0.13& & $-$0.37 &   0.06(3)  & 1.13$\pm$ 0.13&  \\
Ho & 67 & 0.48 &    $-$1.30  &  0.10(1)  & 0.72 $\pm$ 0.14  & &    $-$2.30:  &   0.10(1)  &  0.57: $\pm$ 0.14 &  &     $-$1.15:  &   0.07(2)  &  0.97: $\pm$ 0.09  &        \\
Er & 68 & 0.92 & $-$0.60$^b$ &  0.12(2)  & 0.98$\pm$ 0.12     & &    $-$1.50$^b$  & 0.10(1)   &   0.93 $\pm$ 0.11 &&  0.00$^b$     & 0.10(1) & 1.68 $\pm$0.12 &    \\
Tm & 69 & 0.10 &  $-$1.30:  &  0.10(2)  &  1.10: $\pm$ 0.11  & & ...  &  ...  & ...  &&  $-$1.05  &   0.05(2)  &  1.45 $\pm$ 0.10 & \\
Yb & 70 & 0.84 &   $-$0.70:  &   0.10(1)  &  0.96: $\pm$ 0.14  & &   $-$1.90  &   0.10(1)  & 0.61 $\pm$ 0.14 &  &   $-$0.60:  &   0.10(1)  & 1.16: $\pm$ 0.14 &      \\
Lu & 71 & 0.10 &   $-$0.80*  &   0.10(1)  & 1.60* $\pm$ 0.10  & &   $-$1.60*  &   0.10(1)  &  1.65* $\pm$ 0.10  & &   $-$1.00::  &   0.10(1)  &  1.50:: $\pm$ 0.10  &       \\
Hf & 72 & 0.85 & $-$0.20$^b$  &  0.10(1)  & 1.45$\pm$ 0.09 & &   $-$1.20$^b$   & 0.10(1)     &   1.30$\pm$ 0.12   &&  0.10$^b$   &  0.10(1)    &  1.85 $\pm$ 0.10   &  \\
Ta & 73 & $-$0.12 & $-$0.70*    &   0.10(1)  &   1.92* $\pm$ 0.13  & & $-$0.76:  &  0.12(3)  & 2.71:$\pm$ 0.11 & &  $-$0.30*     & 0.10(2)  &   2.42* $\pm$ 0.11    &\\
Os & 76 & 1.40 & ...& ... &...&&  $-$0.90*    &  0.10(1)   &    1.05* $\pm$ 0.15  & &   0.60$^b$    &  0.10(1)   &  1.80 $\pm$ 0.12   &  \\
Ir & 77 & 1.38 &    1.60::     &  0.10(2)      &   2.72::$\pm$ 0.13   & &   1.40:: & 0.05(2)& 3.37:: $\pm$ 0.11 && 1.20::   & 0.10(1)&2.42::$\pm$ 0.13 &\\

\hline
\multicolumn{2}{c}{}& \multicolumn{4}{c}{HD~26} && \multicolumn{3}{c}{HD~55496} &&\multicolumn{3}{c}{HD~196944} \\
\cline{4-6}\cline{8-10} \cline{12-14}
Gd & 64 & 1.07 & 1.00:$^b$ &  0.10(1)  & 0.91:$\pm$ 0.17  & &   $-$0.70$^b$  & 	0.10(1)        & 0.33 $\pm$ 0.17 &  & $-$1.10 &  0.12(3) & 0.33 $\pm$ 0.13  & \\
Tb & 65 & 0.30 &    0.30  &  0.10(1)  &0.98$\pm$ 0.12 & &   $-$0.66:  &  0.20(2)  &1.14: $\pm$ 0.24  &  &   $-$0.60*  &   0.05(2)  &  1.60* $\pm$ 0.11  &  \\
Dy & 66 & 1.10 & 1.60: &   0.10(1)   & 1.48:$\pm$ 0.15&&  $-$0.70$^b$  & 0.10(2)  &0.30 $\pm$ 0.13 && $-$0.60$^b$ &   0.10(2)   & 0.80$\pm$ 0.15& \\
Ho & 67 & 0.48 &   0.23 &  0.06(2)  & 0.73 $\pm$ 0.11 & &    $-$1.30:  &  0.10(1)  & 0.32: $\pm$ 0.14  &  &   $-$2.00  &   0.10(1)  &  0.02 $\pm$ 0.11  &    \\
Er & 68 & 0.92 & 1.60 &  0.10(2) &  1.66 $\pm$ 0.12   & &    $-$0.48:  & 0.13(2)   &   0.70: $\pm$ 0.14  & & $-$0.50$^b$ &  0.10(1)  & 1.08$\pm$ 0.11     &  \\
Tm & 69 & 0.10 &  0.50  &  0.10(1)  &  1.38 $\pm$ 0.13  & &  $-$0.40:  &  0.10(1)  &  1.60: $\pm$ 0.13  & &   $-$1.60  &   0.10(1)  &  0.80 $\pm$ 0.13  &\\
Yb & 70 & 0.84 & ...  & ...  &  ... & &    $-$1.00:  &   0.10(1)  &  0.26: $\pm$ 0.14  &  &   $-$1.05  &   0.10(1)  &  0.61 $\pm$ 0.14  &   \\
Lu & 71 & 0.10 &   0.40::  &   0.10(1)  & 1.28:: $\pm$ 0.10  & &   0.50::  &   0.10(1)  &  2.50:: $\pm$ 0.10  &  &...&...&...&   \\
Hf & 72 & 0.85 & 2.00$^b$  &  0.10(1)  & 2.13$\pm$ 0.12 & &   $-$0.60:$^b$   & 0.10(1)     &   0.65:$\pm$ 0.09   & &  $-$0.65  & 0.05(2) & 1.00$\pm$ 0.12 & \\
Ta & 73 & $-$0.12 & ...   &  ... &   ... & & ...  &  ...  & ... & &  $-$0.60  &  0.10(1)  & 2.02$\pm$ 0.18 &\\
Os & 76 & 1.40 &    1.30$^b$  &  0.10(1) &   0.88 $\pm$0.12   & &   $-$0.40:    &  0.10(1)   &  0.30: $\pm$ 0.15   & &    0.20:    &  0.10(1)   &    1.30: $\pm$ 0.15  &  \\
Ir & 77 & 1.38 &   ...    &  ...      &   ...  & &  ... & ...& ...&&...&...&...&\\
\hline
\multicolumn{2}{c}{}& \multicolumn{4}{c}{HD~206983} && \multicolumn{3}{c}{HD~224959} &&\multicolumn{3}{c}{HD~198269}\\
\cline{4-6}\cline{8-10} \cline{12-14}
Gd & 64 & 1.07 &   0.30  & 0.10(1)     &  0.23$\pm$ 0.15 & &  0.55$^b$   & 0.10(1)    & 1.84 $\pm$ 0.17& &   $-$0.30$^b$  & 	0.10(1)    & 0.73 $\pm$ 0.17 &  \\
Tb & 65 & 0.30 &    $-$0.90  &  0.10(1)  &$-$0.20 $\pm$ 0.24  & &    $-$0.69  &   0.08(3)  &  1.37 $\pm$ 0.12  &   &   $-$1.42  &   0.15(2)  &  0.38 $\pm$ 0.21 &    \\
Dy & 66 & 1.10 &  0.40: &  0.10(1)   & 	0.30:$\pm$ 0.15 &&    0.33 & 0.07(3)&	1.59 $\pm$ 0.11& &  0.10$^b$  & 0.14(3)  & 1.10 $\pm$ 0.13 & \\
Ho & 67 & 0.48 &  ...  &  ... & ... & &    $-$0.38:  &   0.10(2)  &  1.50: $\pm$ 0.09  & &    $-$1.57:  &   0.13(4)  &  0.05: $\pm$ 0.11  &        \\
Er & 68 & 0.92 & 0.45 & 0.10(1)  &   0.53$\pm$ 0.16    & &  0.45$^b$  & 0.10(2) & 1.89  $\pm$ 0.12&  &  $-$0.15$^b$     & 0.10(1) & 1.03 $\pm$0.12 &      \\
Tm & 69 & 0.10 &  ... & ... &  ...  & &  $-$0.43  &   0.04(3)  &  1.83 $\pm$ 0.09  &&   $-$1.45  &   0.03(1)  & 0.55$\pm$ 0.09   & \\
Yb & 70 & 0.84 &  ... & ...  &  ... & &  ... &  ... &   ... &&...&...&...&     \\
Lu & 71 & 0.10 &   $-$0.35:: &  0.10(1)   & 0.55::  $\pm$ 0.10  & &   $-$0.20::  &   0.35(2)  &  2.06:: $\pm$ 0.35  &   &   $-$0.32::  &   0.35(2)  &  1.68:: $\pm$ 0.35  &     \\
Hf & 72 & 0.85 & 0.61    &   0.10(2)  &  0.76$\pm$ 0.16 & & 0.70$^b$ &   0.10(2)  &  2.21$\pm$ 0.09 &&  0.10$^b$   &  0.10(2)    &  1.35 $\pm$ 0.09   &\\
Ta & 73 & $-$0.12 & ... &  ...  &  ...& &   $-$0.25:  &  0.10(1)  & 2.23:$\pm$ 0.13    &&...&...&...& \\
Os & 76 & 1.40 &    ... & ... &...  & &  0.85$^b$  & 0.10(1)  &    1.81$\pm$ 0.12  & &   $-$0.15$^b$  &    0.10(1)   & 0.55$\pm$0.11  &\\
Ir & 77 & 1.38 &    1.30::   &  0.10(1)     &  0.92::$\pm$ 0.13     & &  1.60:: & 0.43(2)& 2.58::$\pm$ 0.43 & & 0.80::   & 0.10(1)& 1.52::$\pm$ 0.13 &\\
\hline 
\\

\end{tabular}

\tablecomments{
$^{a}$ \citet{Asplund2009} \\
$^{b}$ \citet{Karinkuzhi2021}\\
$:$ Uncertain abundances due to noisy/blended region\\
$::$ Very Uncertain abundances due to noisy/blended region\\
$*$ Upper limit and uncertain\\}

\end{table*}
}


\clearpage
\newpage

\label{Fig:Comparison}

\begin{figure*}[h]
\section{Comparison with AGB models}
\begin{minipage}[t]{0.46\textwidth}
    \includegraphics[width=0.95\textwidth]{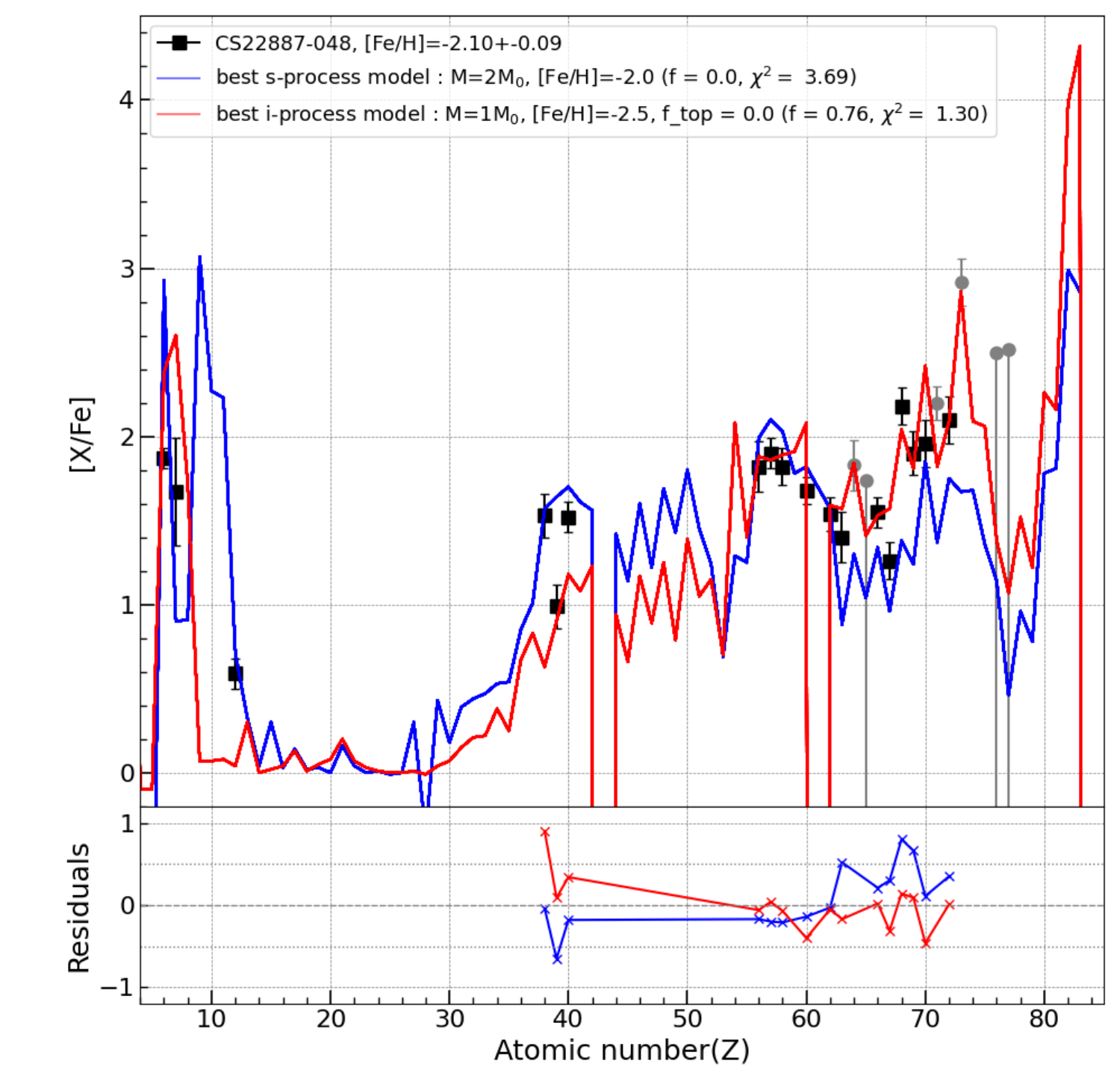}
    \label{fig:appendix-figure1}
\end{minipage}%
\begin{minipage}[t]{0.46\textwidth}
    \includegraphics[width=0.95\textwidth]{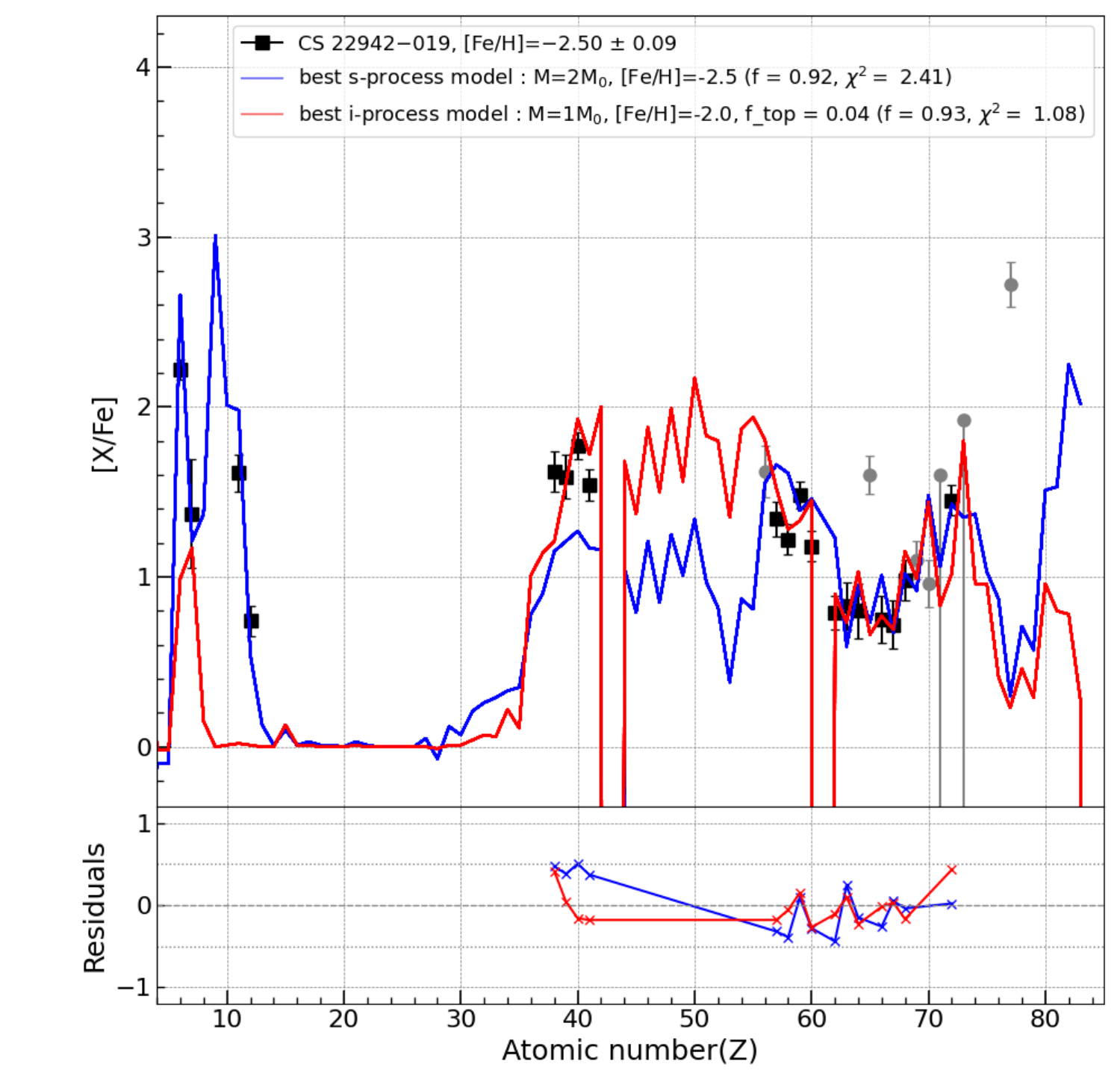}
    \label{fig:appendix-figure2}
\end{minipage}
\begin{minipage}[t]{0.46\textwidth}
    \includegraphics[width=0.95\textwidth]{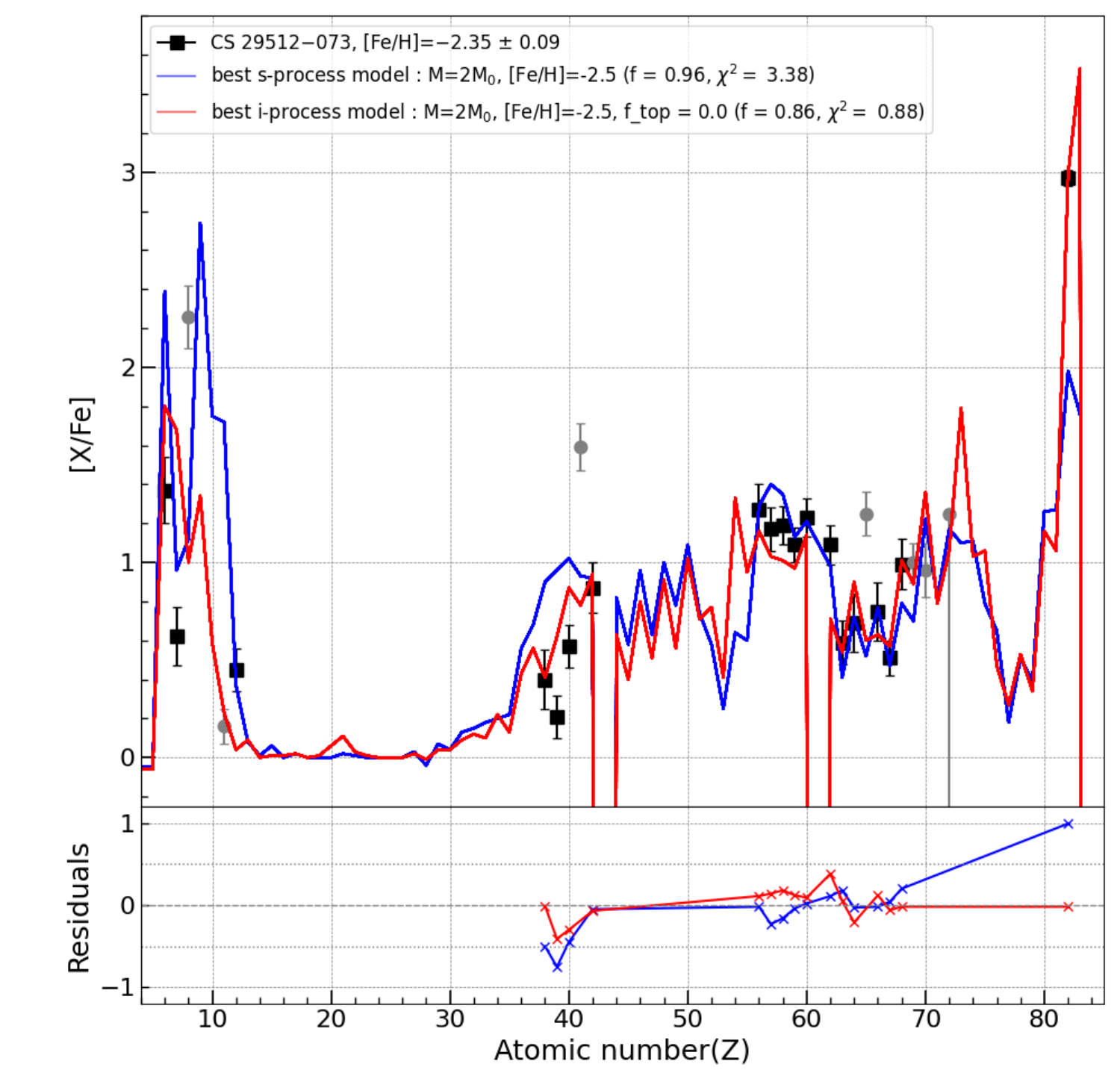}
    \label{fig:appendix-figure3}
\end{minipage}%
\begin{minipage}[t]{0.46\textwidth}
    \includegraphics[width=0.95\textwidth]{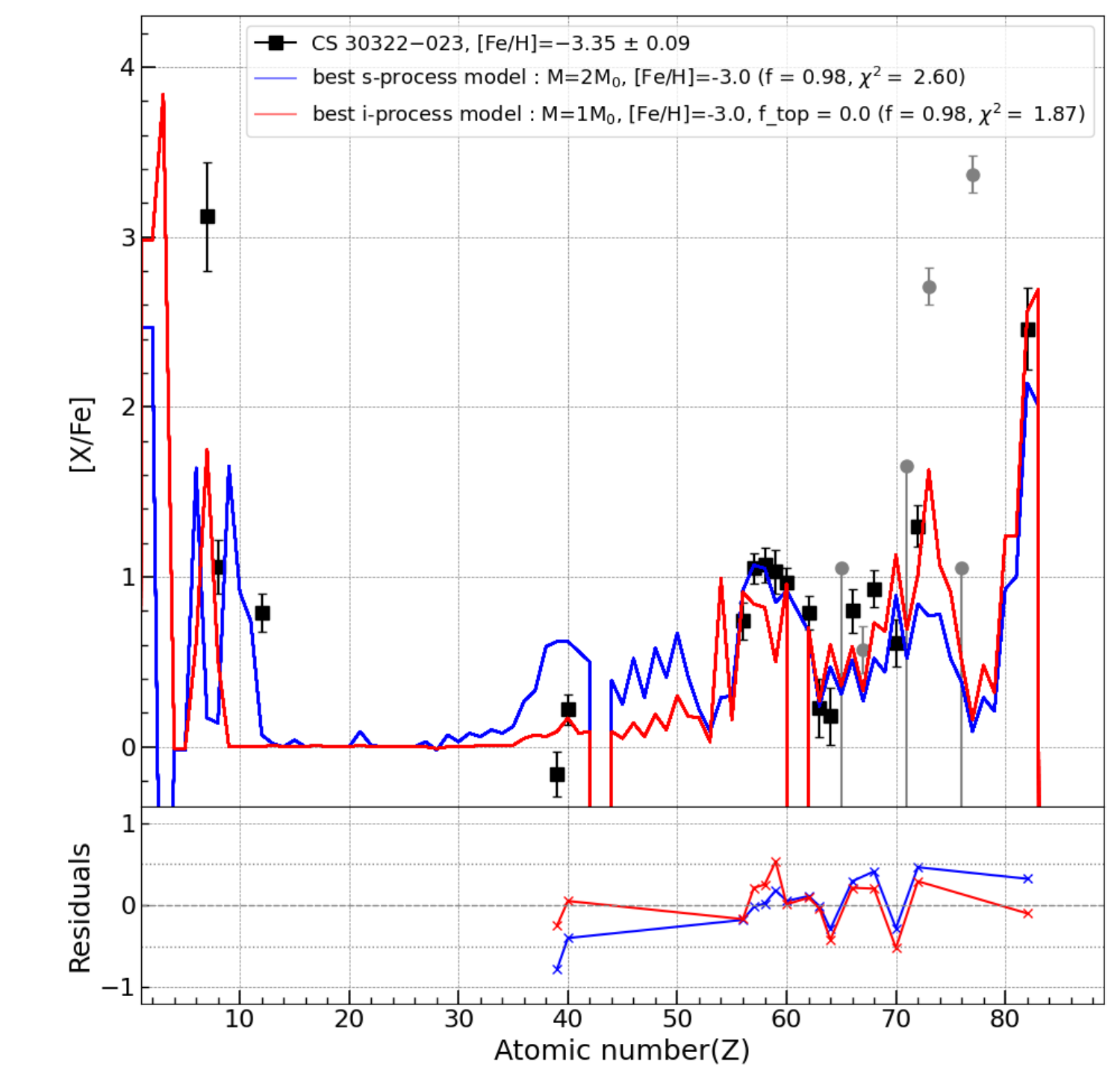}
    \label{fig:appendix-figure4}
\end{minipage}

\begin{minipage}[t]{0.46\textwidth}
    \includegraphics[width=0.95\textwidth]{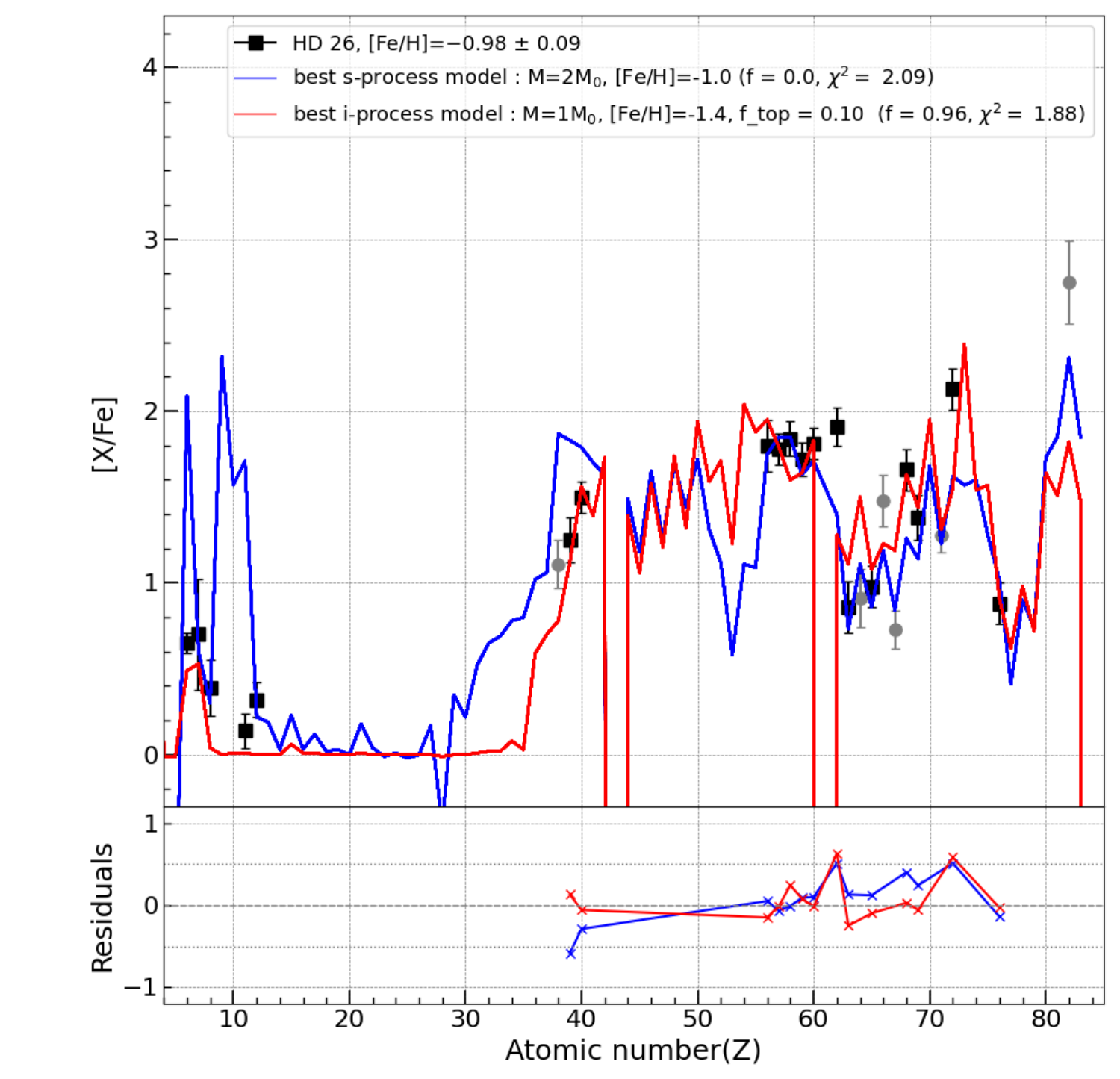}
    \label{fig:appendix-figure5}
\end{minipage}%
\begin{minipage}[t]{0.46\textwidth}
    \includegraphics[width=0.95\textwidth]{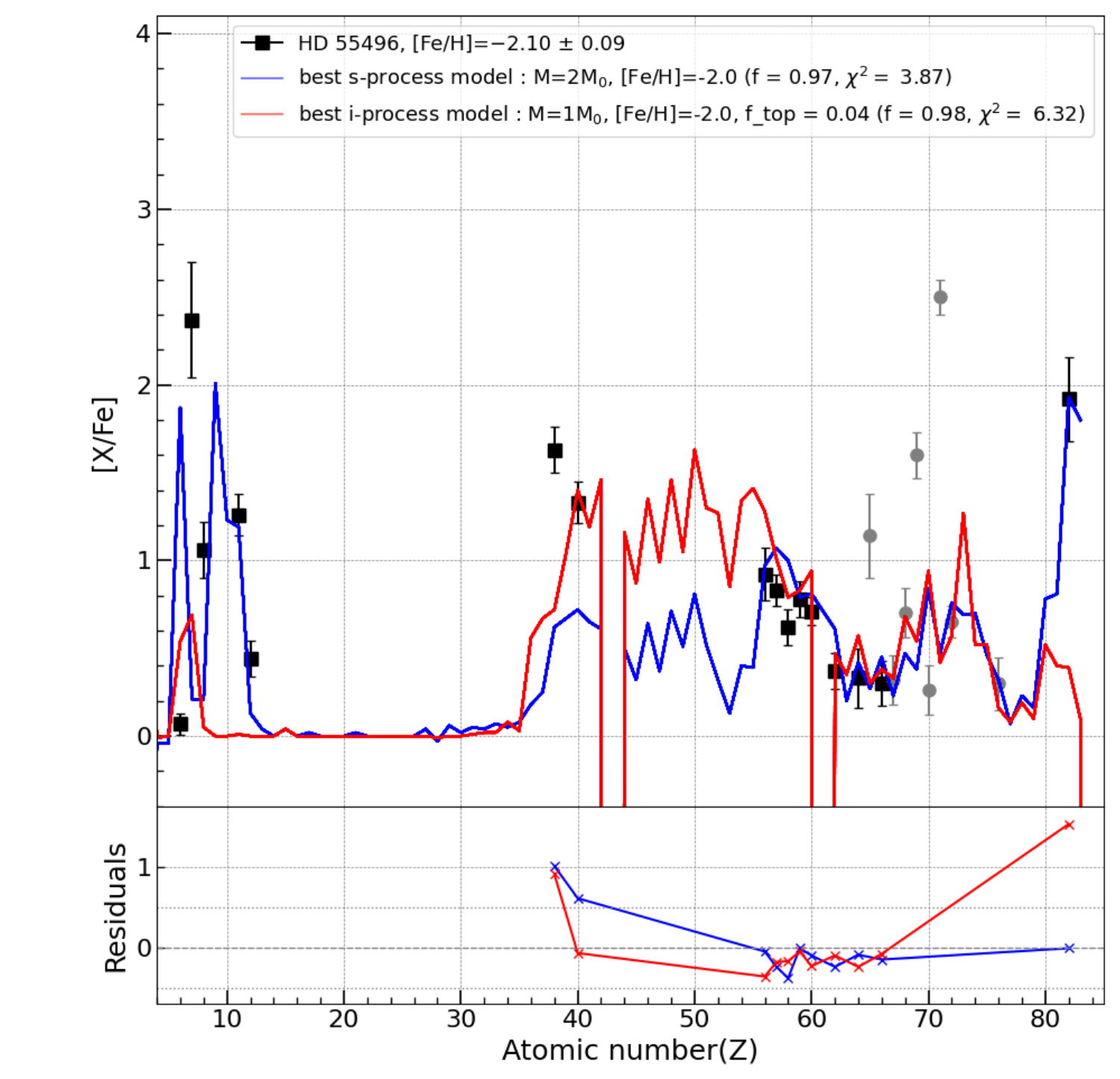}
    \label{fig:appendix-figure6}
\end{minipage}
\end{figure*}
\begin{figure*}[h]
\begin{minipage}[t]{0.46\textwidth}
    \includegraphics[width=0.95\textwidth]{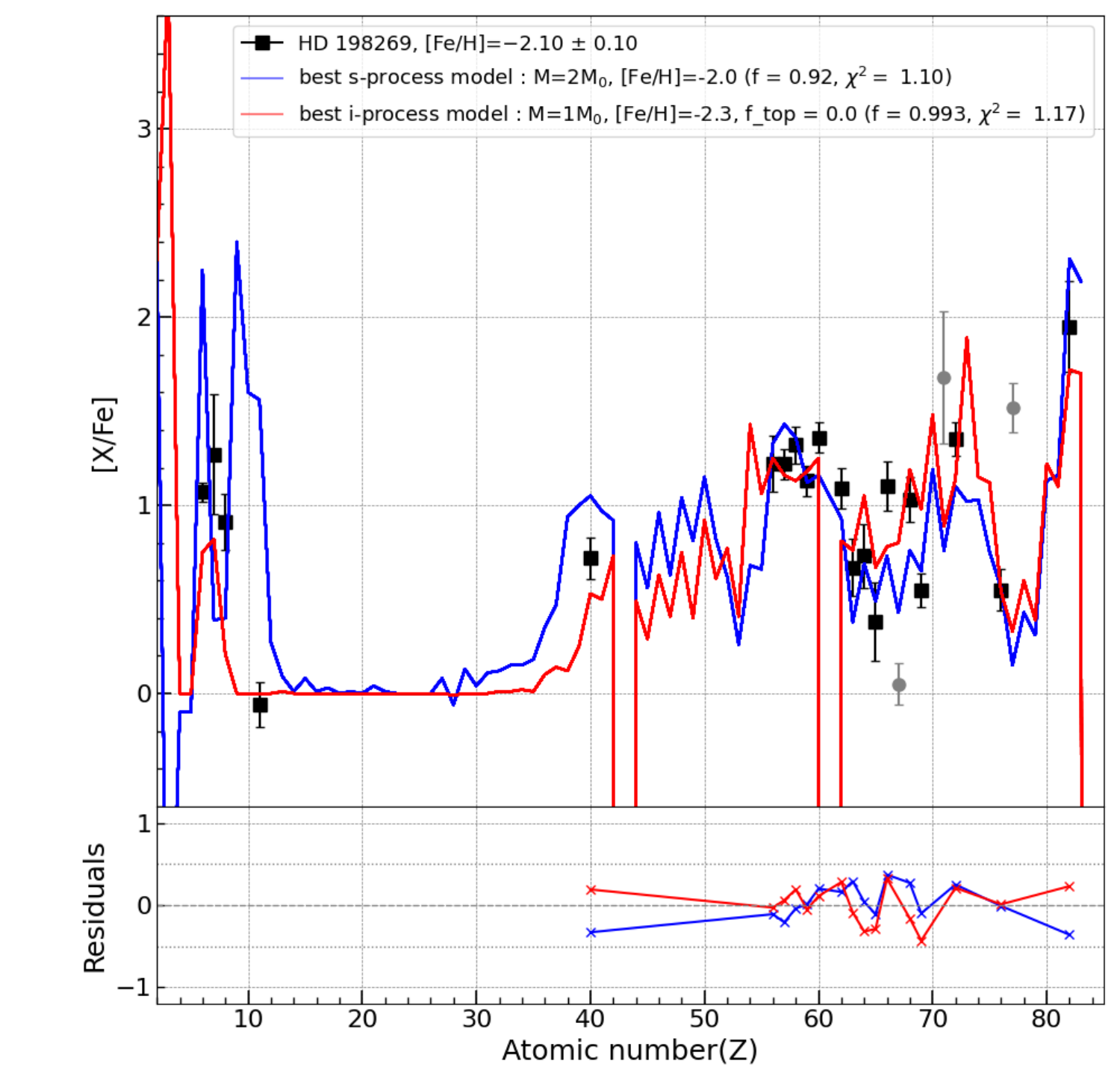}
    \label{fig:appendix-figure7}
\end{minipage}%
\begin{minipage}[t]{0.46\textwidth}
    \includegraphics[width=0.95\textwidth]{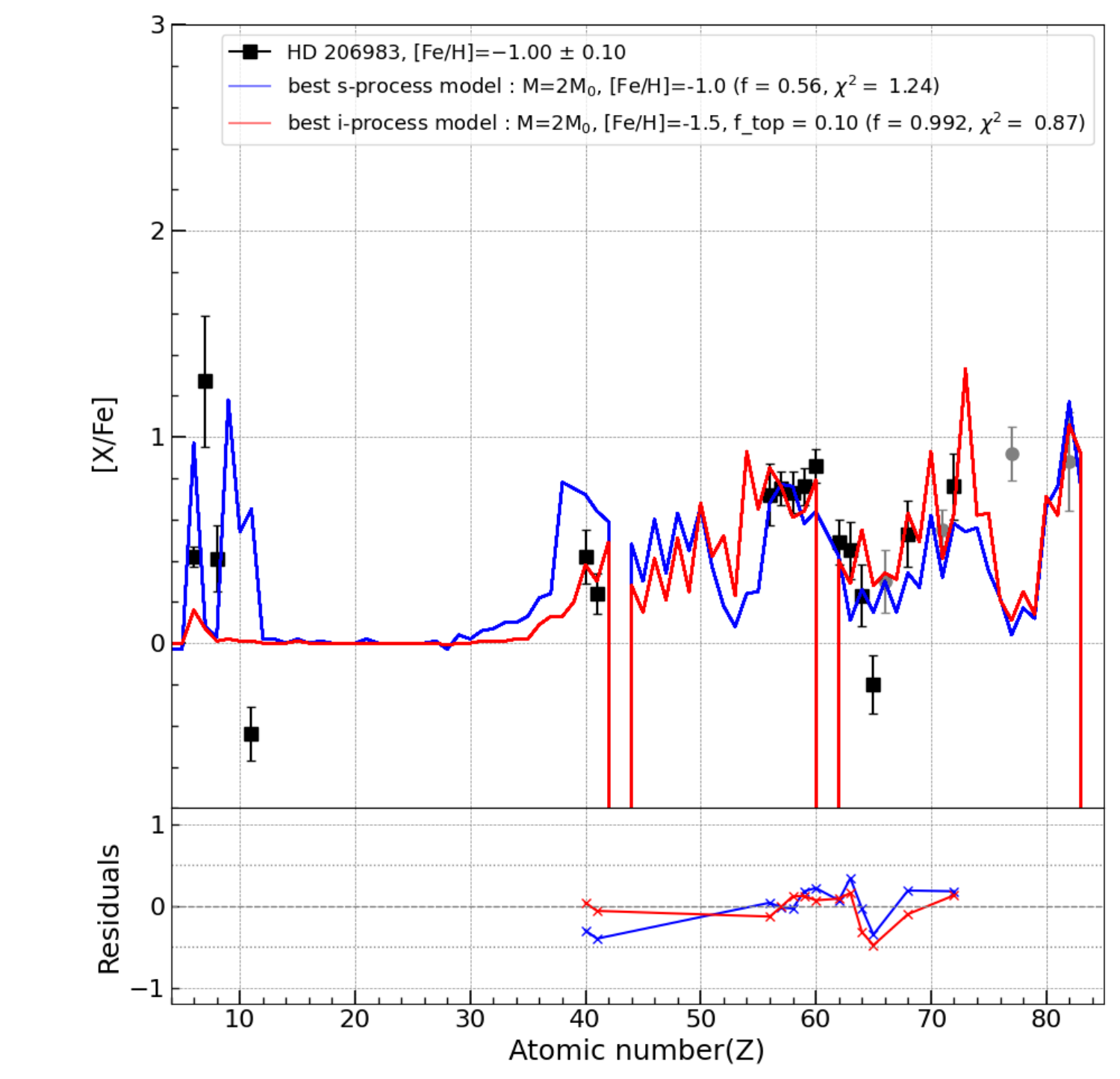} 
    \label{fig:appendix-figure8}
\end{minipage}
\begin{minipage}[t]{0.46\textwidth}
    \includegraphics[width=0.95\textwidth]{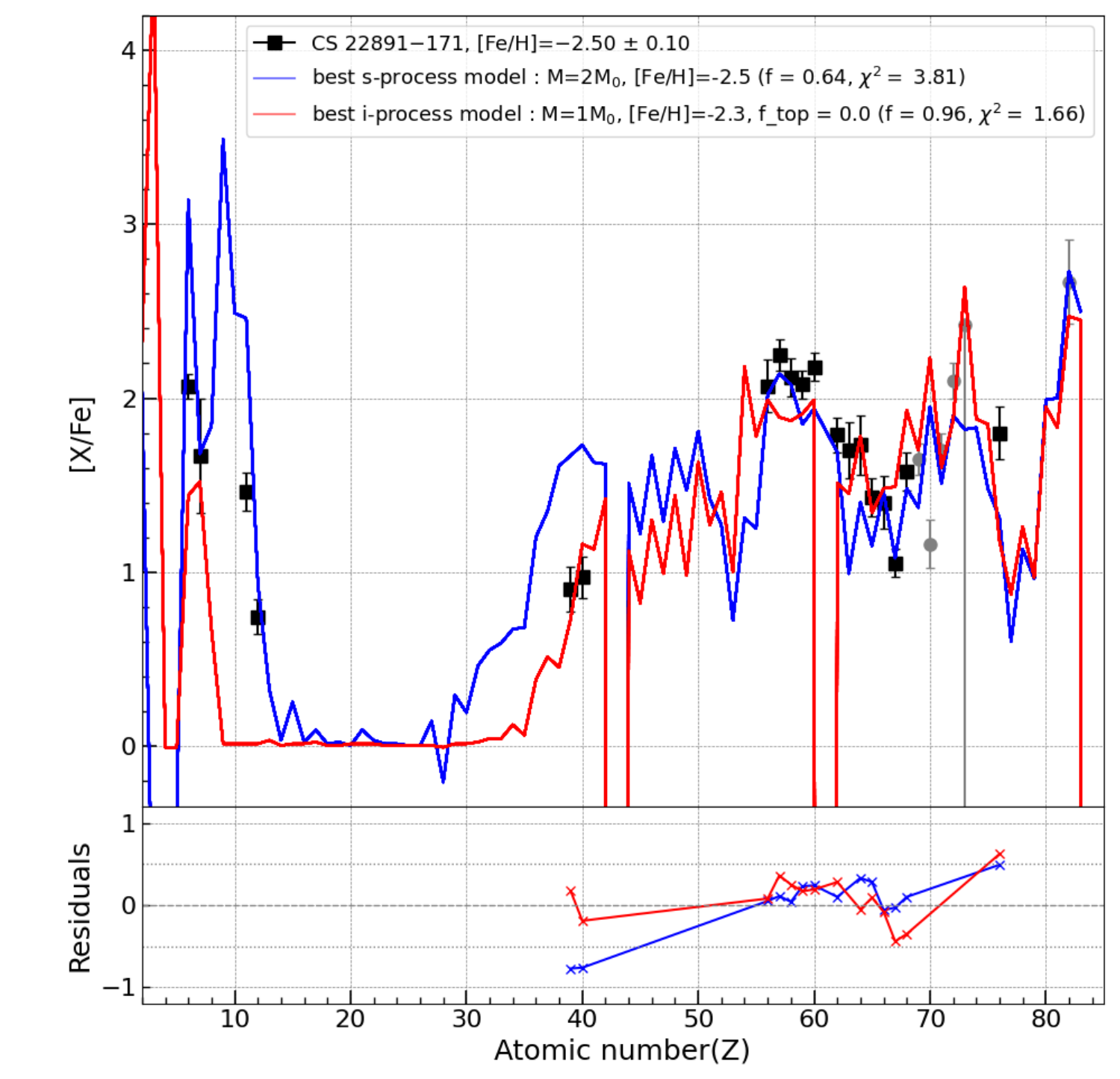}
    \label{fig:appendix-figure9}
\end{minipage}%
\begin{minipage}[t]{0.46\textwidth}
    \includegraphics[width=0.95\textwidth]{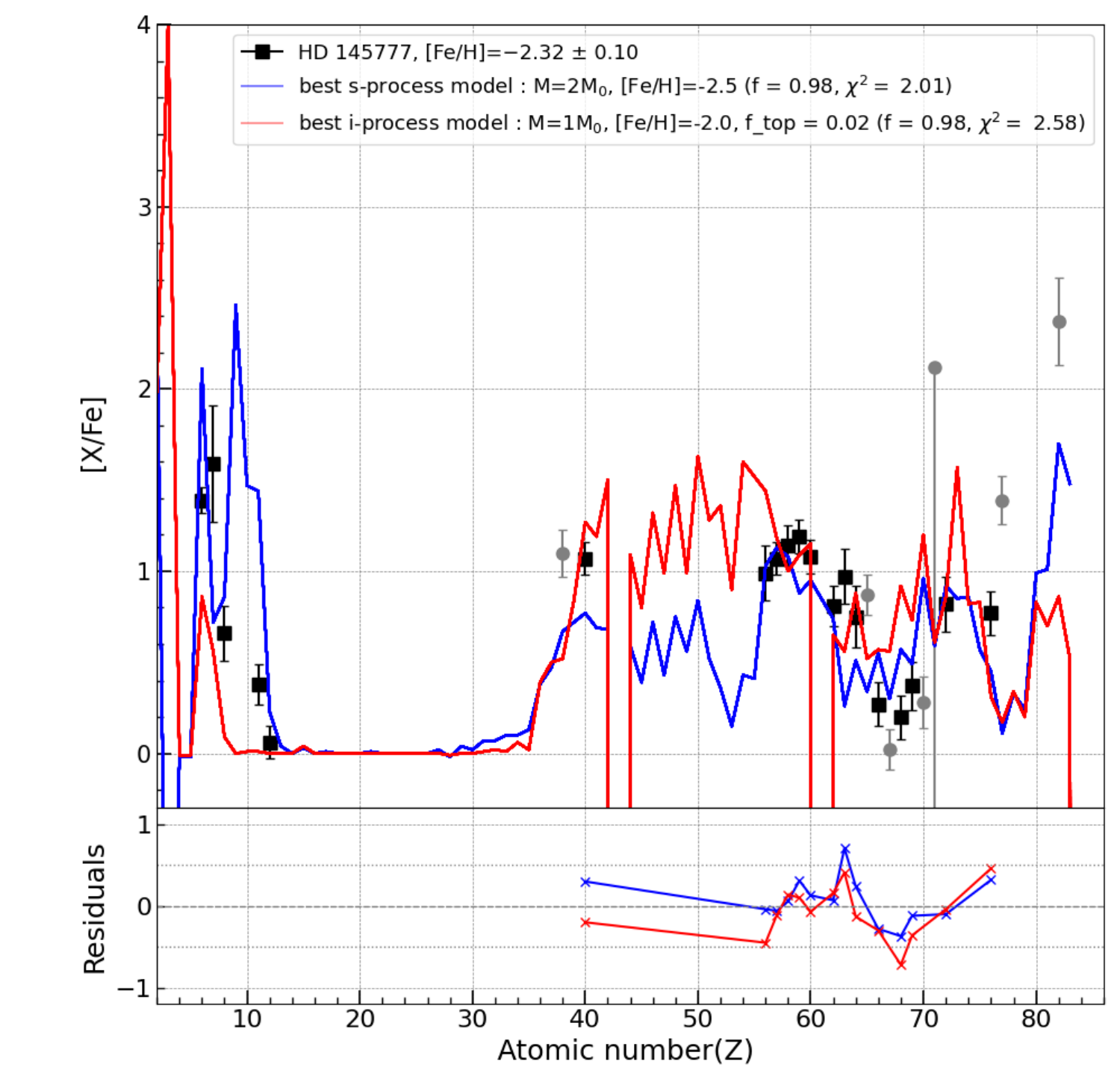}
    \label{fig:appendix-figure10}
\end{minipage}%

\begin{minipage}[t]{0.46\textwidth}
    \includegraphics[width=0.95\textwidth]{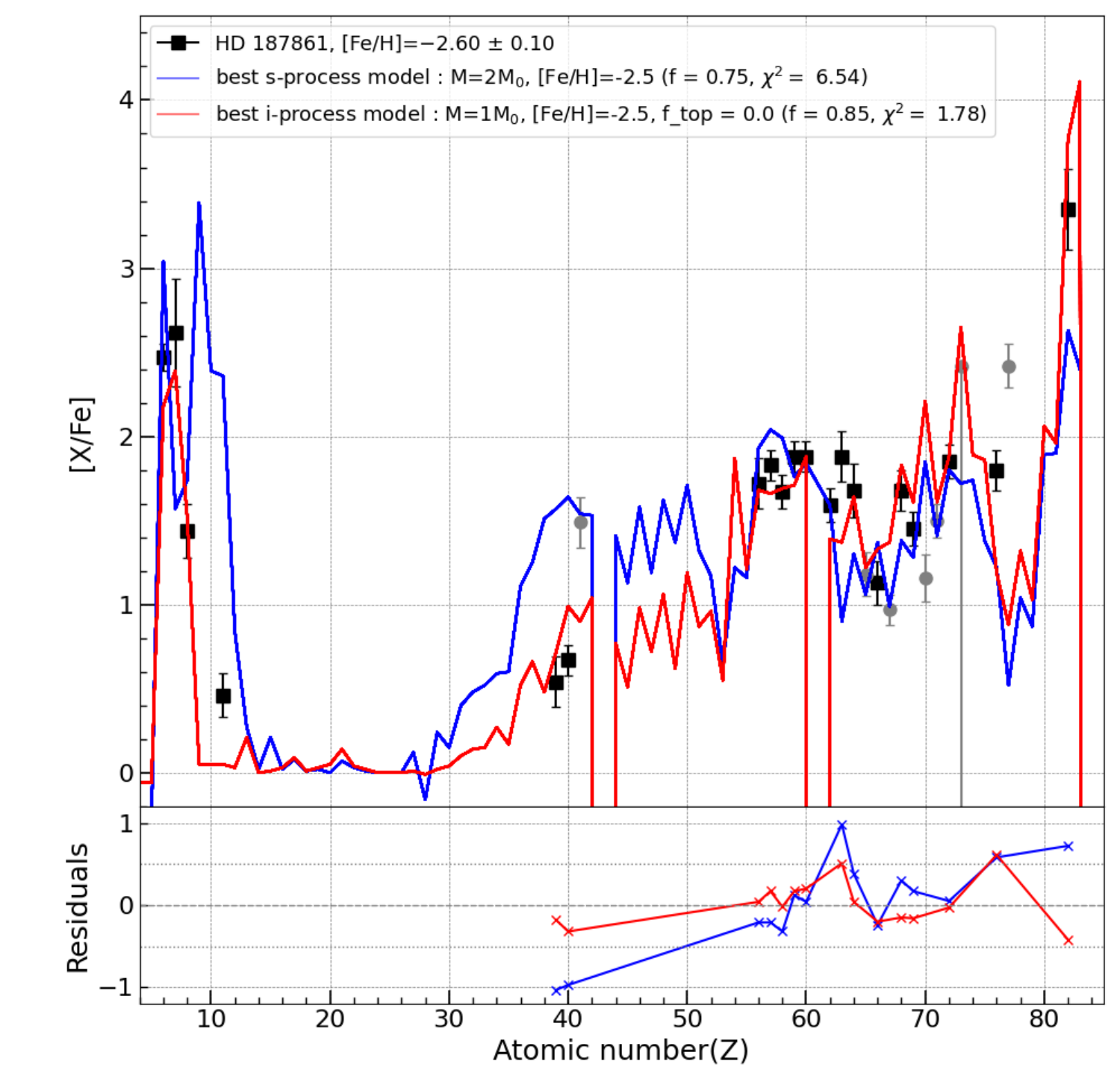} 
    \label{fig:appendix-figure11}
\end{minipage}%
\begin{minipage}[t]{0.46\textwidth}
    \includegraphics[width=0.95\textwidth]{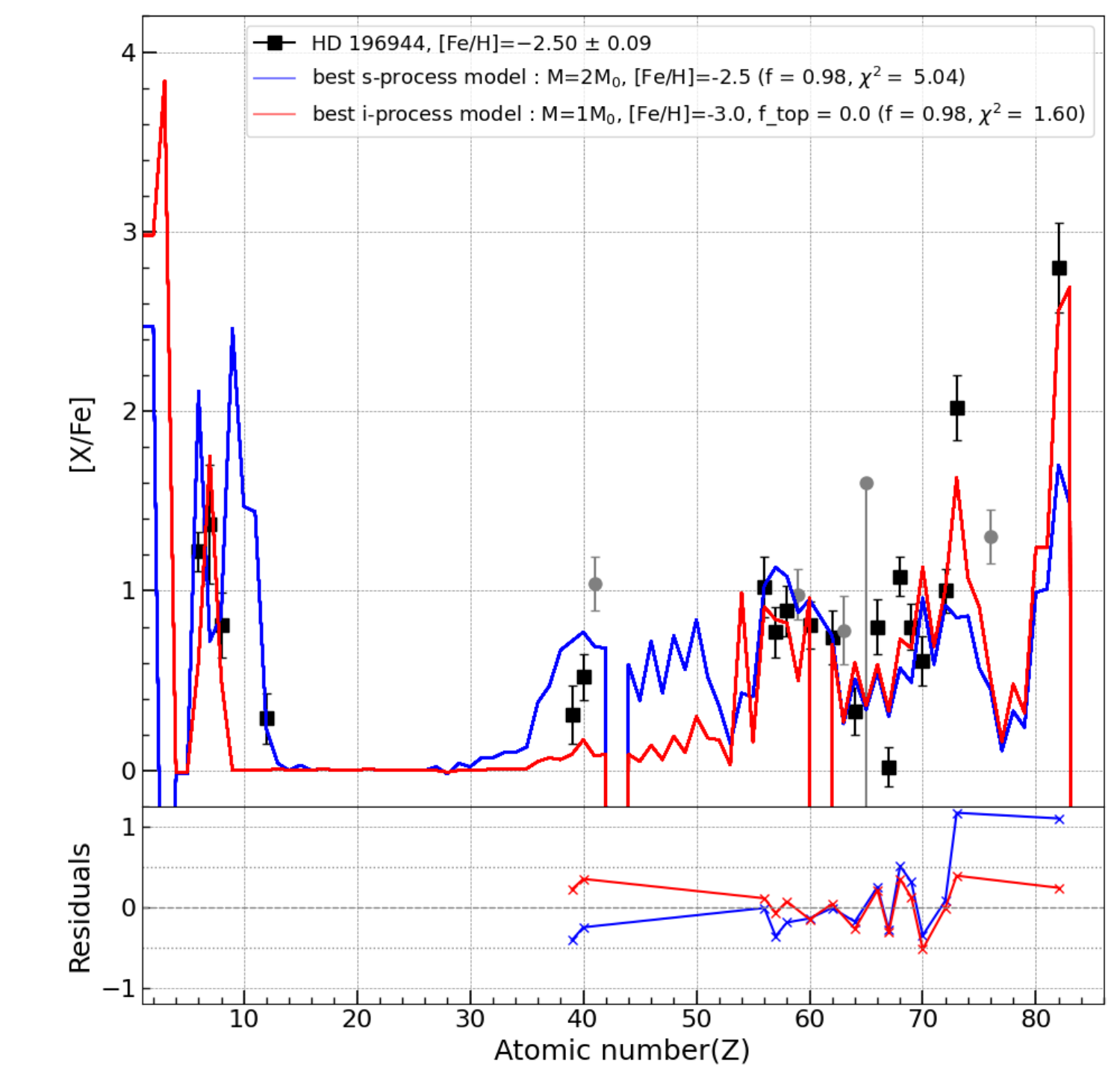} 
    \label{fig:appendix-figure12}
\end{minipage}
\end{figure*}
\begin{figure*}[h]
\begin{minipage}[t]{0.46\textwidth}
    \includegraphics[width=0.95\textwidth]{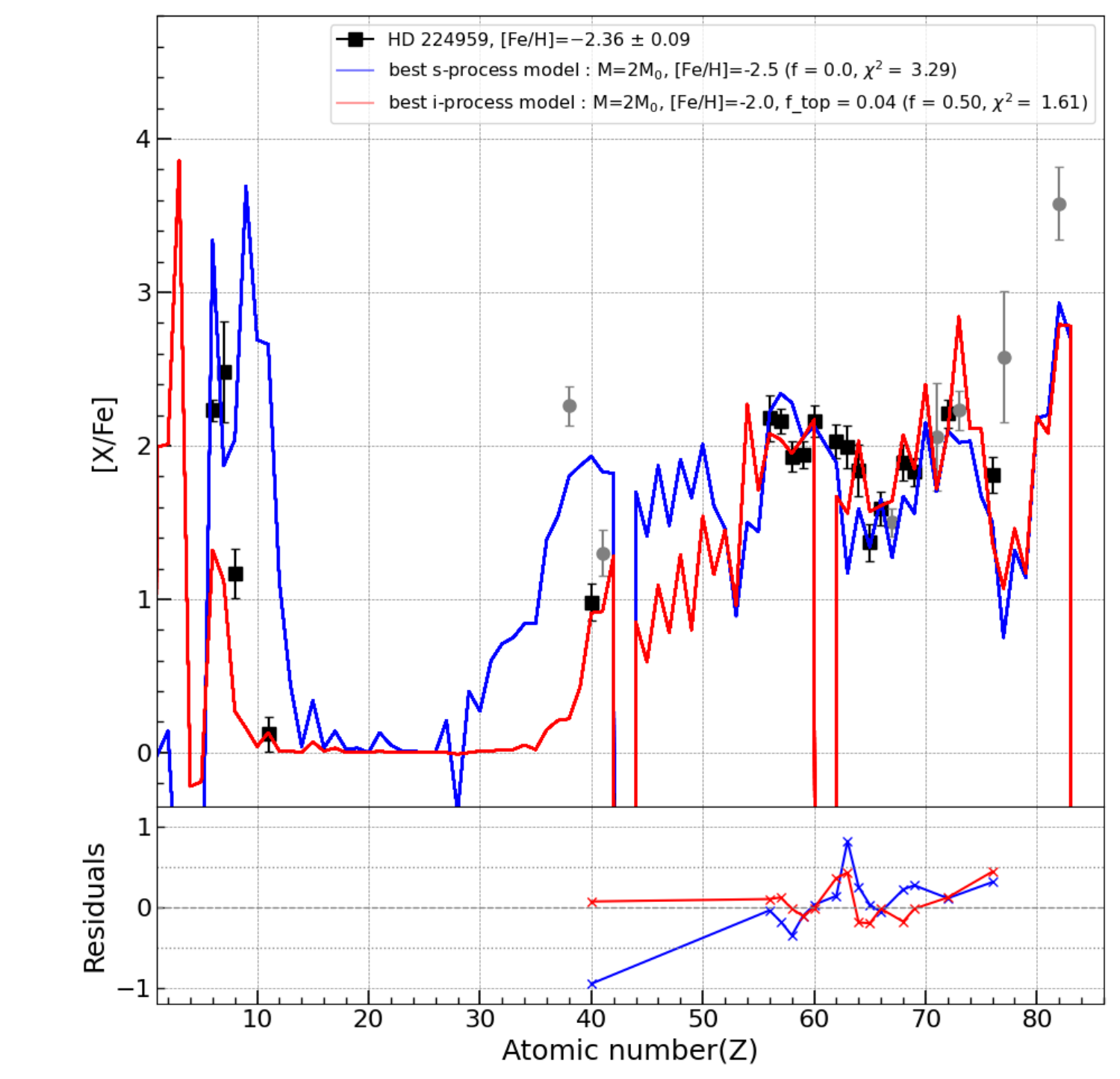}
    \label{fig:appendix-figure13}
\end{minipage}%
\begin{minipage}[t]{0.46\textwidth}
    \includegraphics[width=0.95\textwidth]{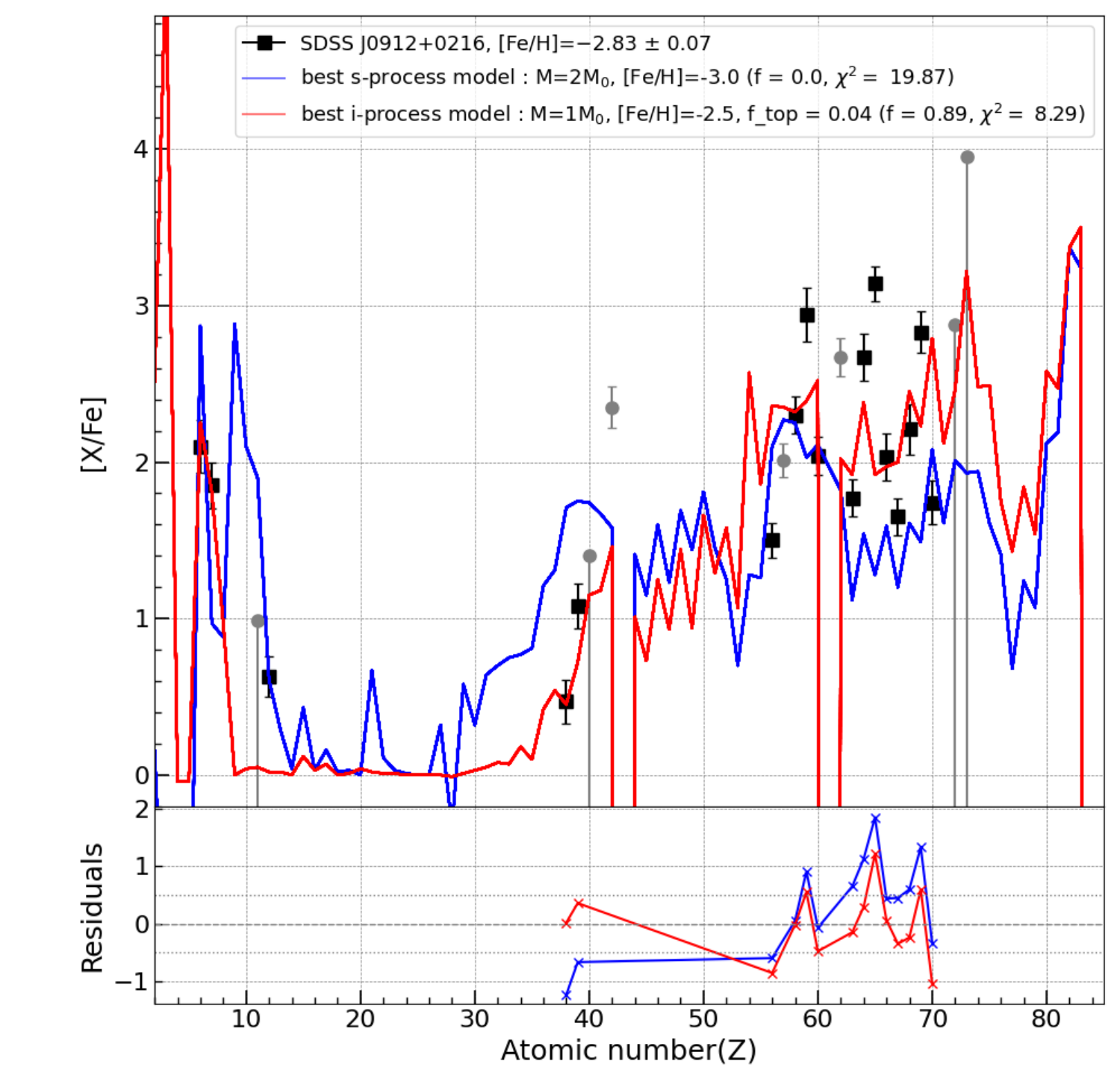}
    \label{fig:appendix-figure14}
\end{minipage}
\begin{minipage}[t]{0.46\textwidth}
    \includegraphics[width=0.95\textwidth]{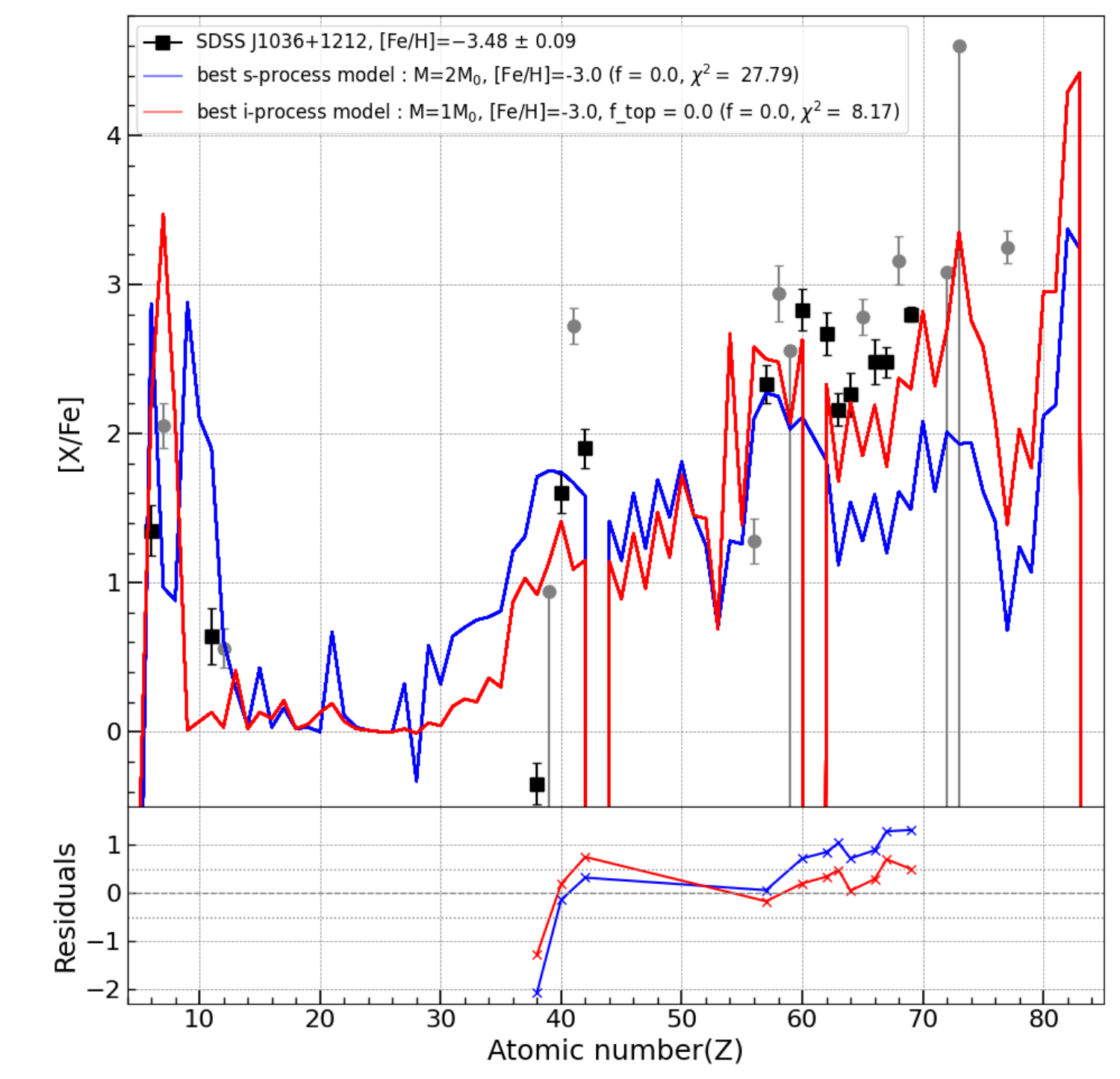} 
    \label{fig:appendix-figure15}
\end{minipage}
\begin{minipage}[t]{0.46\textwidth}
    \includegraphics[width=0.95\textwidth]{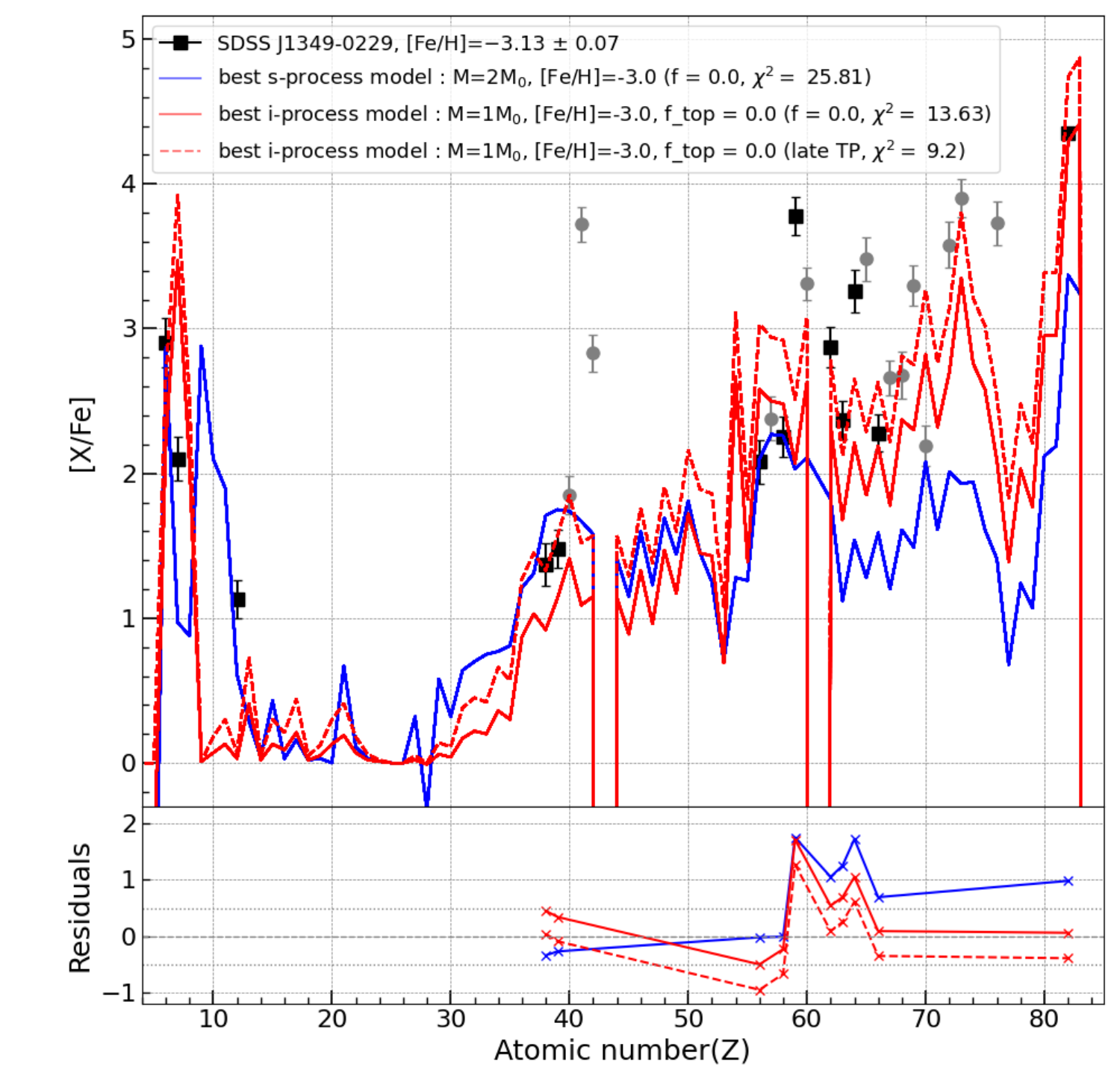} 
    \label{fig:appendix-figure16}
\end{minipage}
\caption{The abundance patterns of the 17 CEMP stars are compared with nucleosynthesis predictions from the STAREVOL code. The determined abundances are indicated by black squares, uncertain abundances by grey circles, and upper limits by grey circles with downward-pointing tails. In all cases, the best-fitting theoretical predictions for both the s-process (blue) and i-process (red) are displayed. Models are described in Sect.~\ref{Sect:nucleosynthesis}}
\label{fig:appendix-figure}
\end{figure*}

\clearpage


\bibliography{CEMP-ref}
\bibliographystyle{aasjournalv7}



\end{document}